\documentclass[12pt,preprint]{aastex}

\usepackage{graphicx}
\usepackage{lscape}

\slugcomment{To appear in ApJ}

\shorttitle{NLR and CLR in active galaxies}
\shortauthors{M\"uller S\'anchez et al.}

\begin{document}


\title{Outflows from AGN: Kinematics of the Narrow-Line and Coronal-Line Regions in Seyfert Galaxies\footnotemark[1] $^,$\footnotemark[2]
}


\author{F. M\"uller-S\'anchez$^{3,4}$, M.~A. Prieto$^{3,4}$, E.~K.~S. Hicks$^{5}$, H. Vives-Arias$^{6}$, R.~I. Davies$^{7}$, M. Malkan$^{8}$, L.~J. Tacconi$^{7}$, R. Genzel$^{7, 9}$}

\affil{$^3$ Instituto de Astrof\'isica de Canarias, V\'ia L\'actea s/n, La Laguna, E-38205, Spain}

\affil{$^4$ Departamento de Astrof\'isica, Facultad de F\'isica, Universidad de la Laguna, Astrof\'isico Fco. S\'anchez s/n, La Laguna, E-38207, Spain}

\affil{$^5$ Department of Astronomy, University of Washington, Box 351580, U.W., Seattle, WA 98195-1580, USA}

\affil{$^6$ Departamento de Astronom\'ia y Astrof\'isica, Universidad de Valencia, Dr. Moliner 50, Burjassot (Valencia), E-46100, Spain}

\affil{$^7$ Max Planck Institut f\"ur Extraterrestrische Physik, Postfach 1312, 85741, Garching, Germany}

\affil{$^8$ Department of Physics and Astronomy, University of California, Los Angeles, CA 90095-1562, USA}

\affil{$^9$ Department of Physics, 366 Le~Conte Hall, University of 
        California, Berkeley, CA 94720-7300, USA}

\footnotetext[1]{Based on observations at the European Southern Observatory VLT (60.A-9235, 070.B-0649, 070.B-0664, 074.B-9012, 075.B-0040, 076.B-0098 and 083.B-0332).}

\footnotetext[2]{Based on observations at the W. M. Keck Observatory, which is
operated as a scientific partnership among the California Institute
of Technology, the University of California and the National Aeronautics
and Space Administration. The Observatory was made
possible by the generous financial support of the W.M. Keck Foundation.}





\begin{abstract} 

As part of an extensive study of the physical properties of active galactic nuclei (AGN) we report high spatial resolution near-IR integral-field spectroscopy of the narrow-line region (NLR) and coronal-line region (CLR) of seven Seyfert galaxies. These measurements elucidate for the first time the two-dimensional spatial distribution and kinematics of the recombination line Br$\gamma$ and high-ionization lines [Si~{\sc vi}], [Al~{\sc ix}] and [Ca~{\sc viii}] on scales $<300$ pc from the AGN. The observations reveal kinematic signatures of rotation and outflow in the NLR and CLR. The spatially resolved kinematics can be modeled as a combination of an outflow bicone and a rotating disk coincident with the molecular gas. High-excitation emission is seen in both components, suggesting it is leaking out of a clumpy torus. 
While NGC 1068 (Seyfert 2) is viewed nearly edge-on, intermediate-type Seyferts are viewed at intermediate angles, consistent with unified schemes.
A correlation between the outflow velocity and the molecular gas mass in $r<30$ pc indicates that the accumulation of gas around the AGN increases the collimation and velocity of the outflow. 
The outflow rate is 2--3 orders of magnitude greater than the accretion rate, implying that the outflow is mass loaded by the surrounding interstellar medium (ISM). 
In half of the observed AGN the kinetic power of the outflow is of the order of the power required by two-stage feedback models to be thermally coupled to the ISM and match the $M_{BH}-\sigma^*$ relation. In these objects the radio jet is clearly interacting with the ISM, indicative of a link between jet power and outflow power.

\end{abstract}



\section{Introduction} \label{int} 

The narrow-line region (NLR) in active galactic nuclei (AGN)
plays a crucial role in attempts to understand the AGN structure and evolution, as well as the interaction between the nuclear engine and the circumnuclear interstellar medium (ISM) of the host galaxies. Among the fundamental AGN components, the NLR is the largest observable structure in the UV, optical and near-IR that is directly affected by the ionizing radiation and dynamical forces from the active nucleus. 
The size and properties of the NLR has been scrutinized intensely over the past few decades. Ground-based and $Hubble$ $Space$ $Telescope$ ($HST$) [O~{\sc iii}] $\lambda5007$\r{A} (hereafter [O~{\sc iii}]) and H$\alpha$ $+$ [N~{\sc ii}] narrow-band imaging of Seyfert galaxies (Pogge 1988a, 1988b; Haniff et al. 1988; Mulchaey et al. 1996; Evans et al. 1991, 1993; Schmitt et al. 1996, 2003) revealed extended emission in several of these objects. The overall morphology of the NLR is varied, but there is a trend in the sense that the NLRs in Seyfert 2 galaxies are more elongated than those in Seyfert 1 galaxies. 
The fact that several Seyfert 2s presented conically shaped NLRs (e.g., NGC 1068 and NGC 4388) indicates that the source of radiation ionizing the gas is collimated, a result that gave strong support to the torus version of the Unified Model of AGN (Antonucci 1993; Urry \& Padovani 1995). However, when comparing ground-based  [O~{\sc iii}] images of Seyfert galaxies from Mulchaey et al. (1996) with the $HST$ snapshot survey of Schmitt et al. (2003), the latter reveal, on average, four times smaller NLR sizes, probably due 
to low sensitivity (quick exposures). Furthermore, one needs to be cautious when talking about ``NLR size'' as  [O~{\sc iii}], or other low-ionization lines, can be produced by other physical processes not related to the AGN such as star formation or shock-ionized gas.

The physical conditions in the NLR have also been studied extensively,
and it is now widely accepted that photoionization by the central continuum source plays a dominant role in exciting the gas (e.g. Kraemer \& Crenshaw 2000a; Kraemer et al. 2000; Bennert et al. 2006) although shocks may play an important role in localized regions (Kraemer \& Crenshaw 2000b; Dopita et al. 2002; Contini et al. 2009).

Ground-based long-slit spectroscopic studies attempted to determine the kinematics of the NLR, but a general consensus on the velocity flow pattern was not reached; cases were made for infall, rotation, outflow, etc. (e.g. Veilleux 1991; Moore \& Cohen 1996; Cooke et al. 2000; Fraquelli et al. 2000; Ruiz et al. 2001; Fraquelli et al. 2003). 
Spectroscopic studies with the Faint Object Camera (FOC) on board of $HST$ also fail in this regard. Several models were advocated on the basis of these data, including rotation, radial acceleration by radio jets, and lateral expansion of gas around the jets (Winge et al. 1997, 1999; Axon et al. 1998; Capetti et al. 1999). 
Observations with the Space Telescope Imaging Spectrograph (STIS) on $HST$ of a number of key objects have suggested that radial outflow is the dominant component of motion in the NLR. 
The first reliable evidence for outflow in the NLR came from $HST$/STIS long slit observations of NGC 1068 (Crenshaw \& Kraemer 2000) and  NGC 4151 (Hutchings et al. 1998; Crenshaw et al. 2000). Since then, detailed constraints on kinematic models of the NLR in Seyfert galaxies became possible (see, e.g., Crenshaw et al. 2010 and references therein). 
Although the comparison of the radial velocity curves with biconical models provide a good fit to the overall flow pattern, there exist significant local variations that remain unexplained. 
Furthermore, the dynamics, origin and energetics of the outflows, as well as the role of the radio jet and the presence of other types of motion in the NLR kinematics, are not fully understood.
Two-dimensional velocity fields with high spatial resolution are needed to get a more complete picture of the kinematics in the NLR. Recent studies based on integral-field spectroscopy (a few of them done with Adaptive Optics in the near-IR) have started to find evidence for radial outflows in the NLR of Seyfert galaxies (Garc\'ia-Lorenzo et al. 2001; Barbosa et al. 2009; Storchi-Bergmann et al. 2010; Riffel et al. 2010). 

 



The high-ionization lines, or coronal lines, observed in the spectra of many AGNs, are forbidden fine structure transitions in the ground level of highly ionized atoms which have threshold energies above 100 eV. 
In contrast to [O~{\sc iii}] or H$\alpha$, which are commonly used to study the NLR, the coronal lines are free of potential contributions from star formation.
Therefore, these lines provide direct insights into the structures and dynamical forces associated with the AGN. 
Ground-based spectroscopic studies revealed that these lines are present with comparable strength in AGN spectra regardless of their class (Penston et al. 1984; Erkens et al. 1997; Prieto \& Viegas 2000; Rodr\'iguez-Ardila et al. 2002; Reunanen et al. 2003), tend to be broader than NLR lines ($400<$FWHM$<1000$ km s$^{-1}$), and their centroid velocity is often blueshifted with respect to the systemic velocity of the galaxy (e.g. Rodr\'iguez-Ardila et al. 2006). These facts have been interpreted as evidence for a location of the coronal-line region (CLR) closer to the AGN than the NLR, but outside the broad line region (BLR), and probably associated with outflows (e.g. M\"uller-S\'anchez et al. 2006, hereafter MS06). 
Because of their high ionization potential (IP$>100$ eV), 
the lines can be produced by either a hard UV to soft X-ray continuum, fast shocks, or a combination of both. The success of photoionisation models to reproduce, within a factor of few, the measured line ratios points towards photoionization as the main excitation mechanism of the CLR 
\citep{kraemer00, groves04, mullaney08, mazzalay10}, although an additional contribution from fast shocks must be invoked to fully explain the line ratios in several cases (MS06; Rodr\'iguez-Ardila et al. 2006; Contini et al. 2009). However, the morphology, size and kinematics of the CLR remain uncertain because in most observational studies the coronal lines are not spatially resolved. 

Detailed studies of the physical conditions across the CLR were not possible before the advent of the $HST$ and ground-based Adaptive Optics (AO). Recently, \citet{prieto05} determined, for the first time, the size and morphology of the CLR in four Seyfert galaxies. These authors found that the CLR as traced by the [Si~{\sc vii}] $\lambda 2.48 \micron$ line extends out to $r=30–-200$ pc, and is aligned preferentially with the direction of the lower ionization cones seen in [O~{\sc iii}] or H$\alpha$. 
The morphology and size of the CLR indicate that the nuclear radiation is either intrinsically collimated or that some additional excitation mechanism (such as fast shocks) contributes to the formation of coronal gas at those observed distances from the ionizing source. 
With the long-slit capability of $HST$ (STIS), \citet{mazzalay10} attempted to study the CLR kinematics in a sample of ten Seyfert galaxies. They found that the radial velocity curves of the coronal lines follow the same pattern as those of  [O~{\sc iii}] emission and concluded that neither the strength nor the kinematics of the CLR scale in any obvious and strong way with the radio jet.

There is great current interest in NLR and CLR kinematics, to resolve the much debated question of feedback from AGN via outflows of ionized gas, now recognized as a potentially crucial input to the $M_{BH}-\sigma^*$ relation (e.g. Gebhardt et al. 2000): AGN outflows provide energy into the ISM of the host galaxy, controlling the star formation and accretion processes, thus regulating the black hole and bulge growth. Recent results from Krongold et al. (2007) and Storchi-Bergmann et al. (2010) indicate that the total kinetic energy released by the AGN outflow is comparable to the energy required to disrupt the hot phase of the ISM, which could then disrupt star formation. These important topics had not previously been widely studied because high signal-to-noise (S/N) spectra over a full two-dimensional (2D) field with high spatial resolution are required, and these are still scarce. 
Particularly, as the coronal lines are exclusively driven by the AGN power, the CLR is an ideal laboratory for investigating the physical processes within $<300$ pc from the central black hole. 

In an effort to draw general conclusions about the kinematics of the NLR and CLR in active galaxies, 
we have carried out a program of high spatial resolution, near-IR integral field spectroscopy of Seyfert galaxies. The measurements discussed here were obtained with the use of AO and the instruments SINFONI on the European Southern Observatory (ESO) 8.2 m Very Large Telescope (VLT) and OSIRIS on the 10 m Keck II Telescope. In this paper, 
which is the first of a series devoted to investigate the role of NLR and CLR kinematics in the co-evolution of supermassive black holes (SMBHs) and their host galaxies, 
we discuss the NLR and CLR distribution and kinematics of seven Seyfert galaxies using the Br$\gamma$ $\lambda 2.16 \micron$ recombination line as tracer of the NLR and the near-IR coronal lines:  [Si~{\sc vi}] $\lambda 1.96\micron$ (IP $= 167$ eV),  [Al~{\sc ix}] $\lambda 2.04\micron$ (IP $= 285$ eV) and [Ca~{\sc viii}] $\lambda 2.32\micron$ (IP $= 127$ eV). 
The existing studies of the kinematics of the NLR are mostly based on long-slit spectroscopy obtained in the optical with $HST$. Our goal is to extend these
studies to the near-IR, a region which is less affected by dust, and where AO permits diffraction-limited integral field spectroscopy of various emission lines that are a vital addition to the lines observed at optical and UV frequencies. Furthermore, this work serves as the basis for further studies of the NLR with future near-IR telescopes and interferometers, such as, e.g. LBT, ELT and $JWST$. We aim to determine for the first time the 2D kinematics of the CLR, examine previous claims of radial outflow along the NLR, as well as investigate possible interactions between the NLR/CLR and other components of the AGN or the host galaxy in the central $<300$pc. When a significant outflowing structure is present, the 2D kinematics are characterized with appropriate kinematic models and the mass outflow rates are calculated.

The paper is structured as follows. In Section 2 we describe the sample characteristics, observations and the methods used for the data reduction and analysis. The general properties of the NLR and CLR are described in Section 3. In Section 4 we analyze in detail the observed kinematics and perform kinematic modeling. The implications of the results are discussed in Section 5, including the origin of the ionized gas motions and possible interactions with other components of the unified model of AGN and the host galaxy. The overall conclusions of the study are outlined in Section 6, and in the Appendix a more detailed description of the individual objects is presented. Throughout
this paper, $H_0 = 75$ km s$^{-1}$ Mpc$^{-1}$ is assumed.


\section{Sample, Observations and Data Processing}\label{observations}

\subsection{Galaxy Sample}\label{sample} 

The seven galaxies studied in this paper were observed with AO and the instruments VLT/SINFONI and Keck/OSIRIS. Source selection of the SINFONI and OSIRIS subsamples was driven principally by technical considerations, being the primary criteria for selecting AGN that (1) the nucleus should be bright and compact enough for good AO correction, used either as natural guide star (NGS) or as a tit-tilt guide for the laser guide star (LGS) AO system, (2) the galaxy should be close enough ($D < 70$ Mpc) that small spatial scales can be resolved at the near-IR diffraction limit of an 8-m class telescope, and (3) the galaxies should have been observed at different wavelengths so that a multiwavelength approach can be used when analyzing the integral field data. 
The only exception is NGC~6814 in the SINFONI subsample which was observed without the AO module.

The resulting sample of seven galaxies is listed in Table~\ref{table1}. Both the OSIRIS and SINFONI subsamples contain
NGC~6814 and NGC~7469. The measured and derived properties from both datasets are consistent, and a comparison of the data from each instrument is presented in the Appendix. Although the OSIRIS data have a better spatial resolution than the SINFONI data of these two galaxies, we decided to perform the kinematic analysis using the SINFONI datacubes, since the SINFONI FOV covers better the extension of the NLR and CLR (see Section~\ref{results}), making the fitting procedure more accurate (Section~\ref{kinematics}). As can be seen in Table~\ref{table1}, there exists bias towards intermediate type 1 objects. This bias is unintenional, as there is no preference for a particular type of Seyfert nucleus for the study presented here. Narrow Br$\gamma$ and coronal lines have been detected in the optical and infrared spectra of type 1 and type 2 Seyfert galaxies in approximately equal fractions 
(Rodr\'iguez-Ardila et al. 2011). As can be seen in Table~\ref{table1}, we detect broad Br$\gamma$ (FWHM $> 1000$ km s$^{-1}$) in all galaxies classified as type 1.5 and 1.9.  
In general, we only consider an emission-line detected and ``useful'' for kinematic analysis when its power is a factor of 3 above the rms scatter of the spectrum $(S/N > 3)$. Narrow Br$\gamma$ and [Si~{\sc vi}] fulfill this requirement in all galaxies of the sample, and therefore will be used as tracers of the NLR and CLR, respectively.

\subsection{Observations and Data Reduction}\label{reduction} 

SINFONI and OSIRIS are AO-assisted, cryogenic near-IR
integral-field spectrographs, commissioned at the VLT and the W.~M. Keck Observatory, respectively. They deliver spectra simultaneously over a contiguous two-dimensional field of view in the wavelength range from $1.1-2.45 \mu$m at a resolving power of $1500-4000$. Details about the SINFONI instrument can be found in Eisenhauer et al. (2003) and Bonnet et al. (2004), and a more complete discussion on the integral field spectrometer OSIRIS can be found in \citet{larkin06}.


The SINFONI data presented here were taken in nine separate runs between July 2004 and June 2009. The spectral range was similar for all the runs (1.95--2.45 $\mu$m) with typical spectral resolution R$\sim$4000 (FWHM 70 km s$^{-1}$). The pixel scale was either $0.0125\arcsec\times0.025\arcsec$ or $0.05\arcsec\times0.1\arcsec$, resulting in a field of view (FOV) of $0.8\arcsec\times0.8\arcsec$ or $3.2\arcsec\times3.2\arcsec$, respectively. The only exception is NGC~6814 which was observed without the AO module and the largest pixel scale of $0.125\arcsec\times0.25\arcsec$ (FOV=$8\arcsec\times8\arcsec$). For each galaxy, the nucleus was used as NGS for a near-diffraction-limited correction by the AO module \citep{bonnet03}. 
The exposure times for Circinus, NGC~3783, NGC~1068 and NGC~6814 were 30, 40, 110, and 140 minutes, respectively, and for the remaining galaxies 60 minutes of
on-source data were obtained. 
Data reduction was accomplished using the SINFONI custom reduction package SPRED \citep{abuter05}. This performs all the usual steps needed to reduce near-IR spectra, but with the additional routines for reconstructing the data cube. The reduction steps include trimming, bias and background subtraction, flat fielding,
cosmic rays cleaning, correction for nonlinearity, alignment and interpolation of the data, extraction of the spectra, wavelength calibration, sky subtraction and coaddition of different exposures.
Telluric correction and flux calibration were performed using a standard star (A- or B-type) observed next in time to the science frames. 
In addition, flux calibration was cross-checked with VLT/NACO data in $1\arcsec-3\arcsec$ apertures. In some objects the flux calibration was further cross-checked in larger $3\arcsec-5\arcsec$ apertures using 2MASS data. 
Agreement between the different data sources was consistent to 20\%.

The OSIRIS galaxies were observed between April 2006 and January 2008 using both the NGS and LGS AO systems (van Dam et al. 2004; van Dam et al. 2006; Wizinowich et al. 2006). The data have $0.035\arcsec$ pixel$^{-1}$, a FOV of $0.56\arcsec \times 2.24\arcsec$ and a spectral resolution of R$\sim3000$ in the wavelength range $1.965-2.381 \micron$. Details on the observations of individual objects can be found in Hicks et al. (2009, hereafter H09). The OSIRIS data was reduced using the OSIRIS final data reduction pipeline (DRP), which performs all the usual steps needed to reduce near-IR spectra, but with the additional routines for reconstructing the data cube (see above). 

The spatial resolution of each data cube was estimated from both the broad Br$\gamma$ emission and the non-stellar continuum, the two always unresolved in 8-m class telescopes. 
The only exception was Circinus. 
In this galaxy the point spread function (PSF) was derived by comparison to higher resolution VLT/NACO data (see MS06 for details on the fitting procedure). 
A summary of the data, including the specific spatial resolution achieved for each galaxy, is given in Table~\ref{table1}.

\subsection{Extraction of the Emission-Line Gas Properties}\label{analysis} 

In order to draw general conclusions about the properties of the ionized gas in AGNs, all of the galaxies in the sample are evaluated using a consistent method to
minimize differences due to spatial resolution and the shapes of the spectral profiles. 
Before discussing our findings, we first describe the methods used to fit the spectral line profiles and visualize the results.


\subsubsection{Two-dimensional Flux, Velocity, and Dispersion Maps}\label{linefit} 

We derived emission-line flux distributions, radial velocity and velocity dispersion maps of the NLR (narrow Br$\gamma$ emission) and the CLR using the IDL  code LINEFIT. This method, described in Davies et al. (2007, hereafter D07), fits the emission lines by convolving a Gaussian with a spectrally unresolved template profile (an OH sky emission line) to the continuum-subtracted emission line profile at each spatial pixel in the data cube. The continuum is estimated using a polynomial function fitted to the line-free regions adjacent to the emission line. 
The continuum and line intervals must be set as input when executing the program. To allow for the best possible extraction of the line properties, the width of the fitting window is introduced manually after visual inspection of the line profile with the visualization tool QFitsView\footnote{see http://www.mpe.mpg.de/$\sim$ott/QFitsView/}. The fitting window should cover the complete spectral range of the line and enough range of continuum to the left and right of the line profile. We estimate that the uncertainty in the continuum subtraction is on the order of 5\%. This method allowed us to obtain uncertainties for the kinematic maps in the range of $(5-20)$ km s$^{-1}$. The dispersion extracted by this fitting procedure is already corrected for instrumental broadening. 

In the central regions of Seyfert 1 galaxies, the Br$\gamma$ line profile consists of a narrow component (FWHM $< 400$ km s$^{-1}$) superposed on a broad component (FWHM $> 1000$ km s$^{-1}$, see e.g. Fig.~1 of D07). In these cases the narrow Br$\gamma$ properties were extracted using the same procedure as described above but utilizing the spectral region of the rather flattened top part of the broad component as the interval where the continuum is fitted. Since this study is focused on the NLR in Seyfert galaxies, hereafter we refer to the narrow component of Br$\gamma$ just as Br$\gamma$. 



As it was shown in MS06 for the case of the Circinus galaxy, the [Ca~{\sc viii}] line sits on top of the stellar $^{12}$CO(3-1) $\lambda 2.323\mu$m absorption bandhead. This occurs in all AGNs with [Ca~{\sc viii}] detection, and must be corrected in order to restore the true emission-line profile. This correction was performed by convolving stellar templates of red (super)giants with a Gaussian broadening function, and varying its
parameters to minimize the $\chi^2$, which was measured across the $^{12}$CO(2-0) bandhead (free of emission lines).
The $^{12}$CO(3-1) bandhead of the template was then convolved with the broadening
function and subtracted from the galaxy's spectrum (see Fig.~8 of MS06).

\begin{figure}
\epsscale{.99}
\plotone{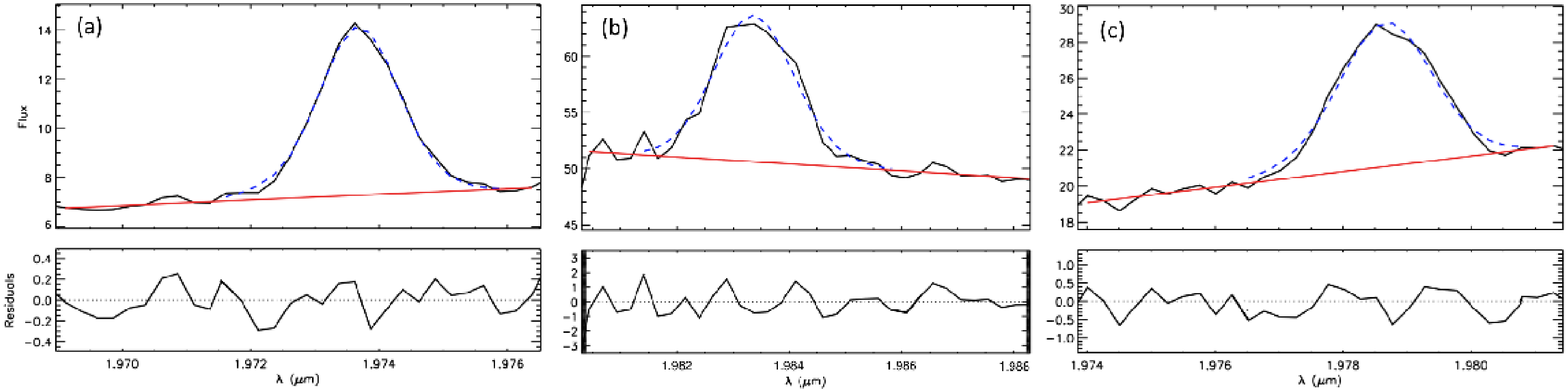}
\caption{Example [Si~{\sc vi}] line profiles extracted from the central regions of three sample galaxies, chosen to represent the main types of profile shapes obseved over the fields. The data are shown by a solid black line, individual Gaussian fits by dashed blue lines and the continuum level fit in red. Below each spectrum is the residual plot. Fluxes and residuals are given in W m$^{-2}$ $\mu$m$^{-1}$. (a) An example line fit from NGC~6814 showing a symmetric profile. Similar profiles are observed over the fields of the majority of the galaxies (except NGC~1068); (b) a spectrum from NGC~3783 showing a slightly redshifted tail. These profiles are typical of NGC~3783 and NGC~6814; (c) example of a spectrum of NGC~2992 showing a slightly blueshifted tail. These profiles are typical of Circinus (see MS06), NGC~2992 and NGC~7469.  
\label{fig1a}}
\end{figure}
\begin{figure}
\epsscale{.99}
\plotone{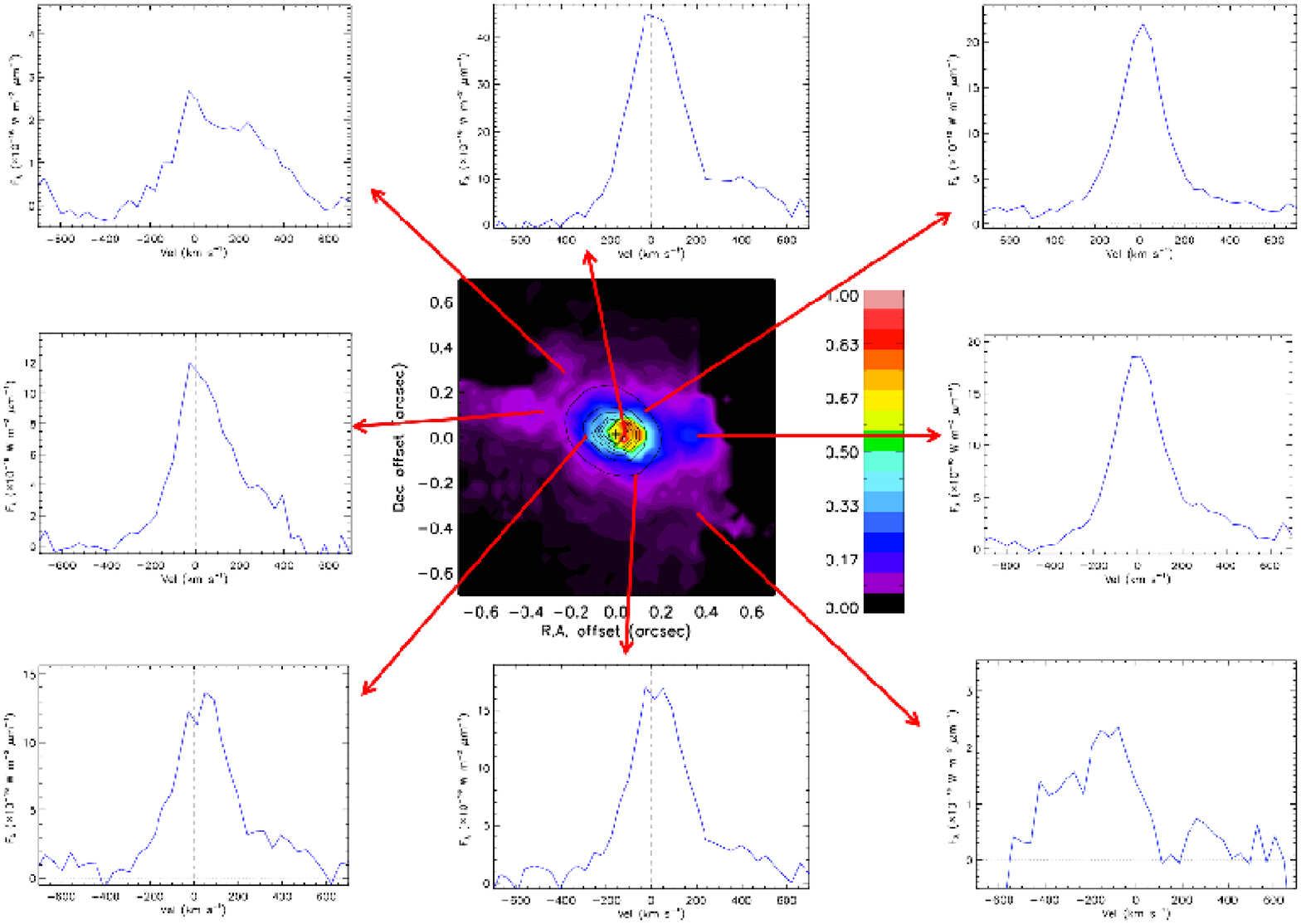}
\caption{[Si~{\sc vi}] emission line profiles from different regions in the central $1.2\arcsec\times1.2\arcsec$ ($80\times80$ pc) of NGC~4151. The spectra have been extracted from apertures of $0.35\arcsec$ (23 pc) diameter. The locations of these regions are indicated by the arrows. The central panel shows the flux map of [Si~{\sc vi}] emission. 
\label{fig1}}
\end{figure}

Despite some asymmetries observed in the emission line profiles, we were successful in fitting a single Gaussian component to Br$\gamma$ and the coronal lines (except in NGC~1068, see Section~\ref{tomography}). Typical fits to the [Si~{\sc vi}] emission line profile are shown in Figure~\ref{fig1a}. The instrumental spectral profile is implicitly taken into account by convolving the assumed emission line profile (the Gaussian) with a template line (tracing the effective instrumental resolution). As a result, small asymmetries in the emission line profiles can be well matched. For more details on the key features of LINEFIT see \citet{davies09}. 
This sample represents the variety of different profile shapes we find across the IFU fields, and demonstrates the high quality of the data and the accuracy of the line-fitting. 
As can be seen in Figure~\ref{fig1a}, redshifted or blueshifted tails are common features in the coronal line profiles. An extreme special case is observed in the CLR of NGC~4151 (see Figure~\ref{fig1}). While fitting a single Gaussian with LINEFIT allows us to reliably measure the [Si~{\sc vi}] properties in the major part of the field of view, despite the apparent presence of a broad component in the very central pixels which has a considerably lower S/N making its characterization rather uncertain, the NE and SW regions (regions A and H) exhibit complex profiles in which it is rather difficult, if not impossible, to correctly identify the ``true'' number of components (Crenshaw et al. 2000 and Storchi-Bergmann et al. 2010 present studies of the NLR in NGC~4151 based on two-Gaussian fits to the emission-line profiles). For this reason, and for reasons of consistency and simplicity, we decided to limit ourselves to one single main component, as in the rest of the galaxies. Furthermore, this approach 
facilitates the modeling of the velocity maps and 
permits a direct comparison with the fits to the Br$\gamma$ line profiles, which are normally well approximated by a single Gaussian. 
These fits, in combination with the velocity channel maps, which are more sensitive to asymmetries and faint components (see Section~\ref{tomography}), allowed us to develop a general understanding of both the kinematics and the origin of the ionized gas emission. 


In addition, 
integrated intensity maps of Br$\gamma$ and coronal lines were constructed by summing up all spectral channels containing line emission in each spatial pixel, and subtracting from each the continuum level. These measurements result in similar flux distributions as those obtained with LINEFIT, confirming that the single Gaussian fits are good approximations. 

\subsubsection{Velocity channel maps}\label{tomography} 

\begin{figure}
\epsscale{.99}
\plotone{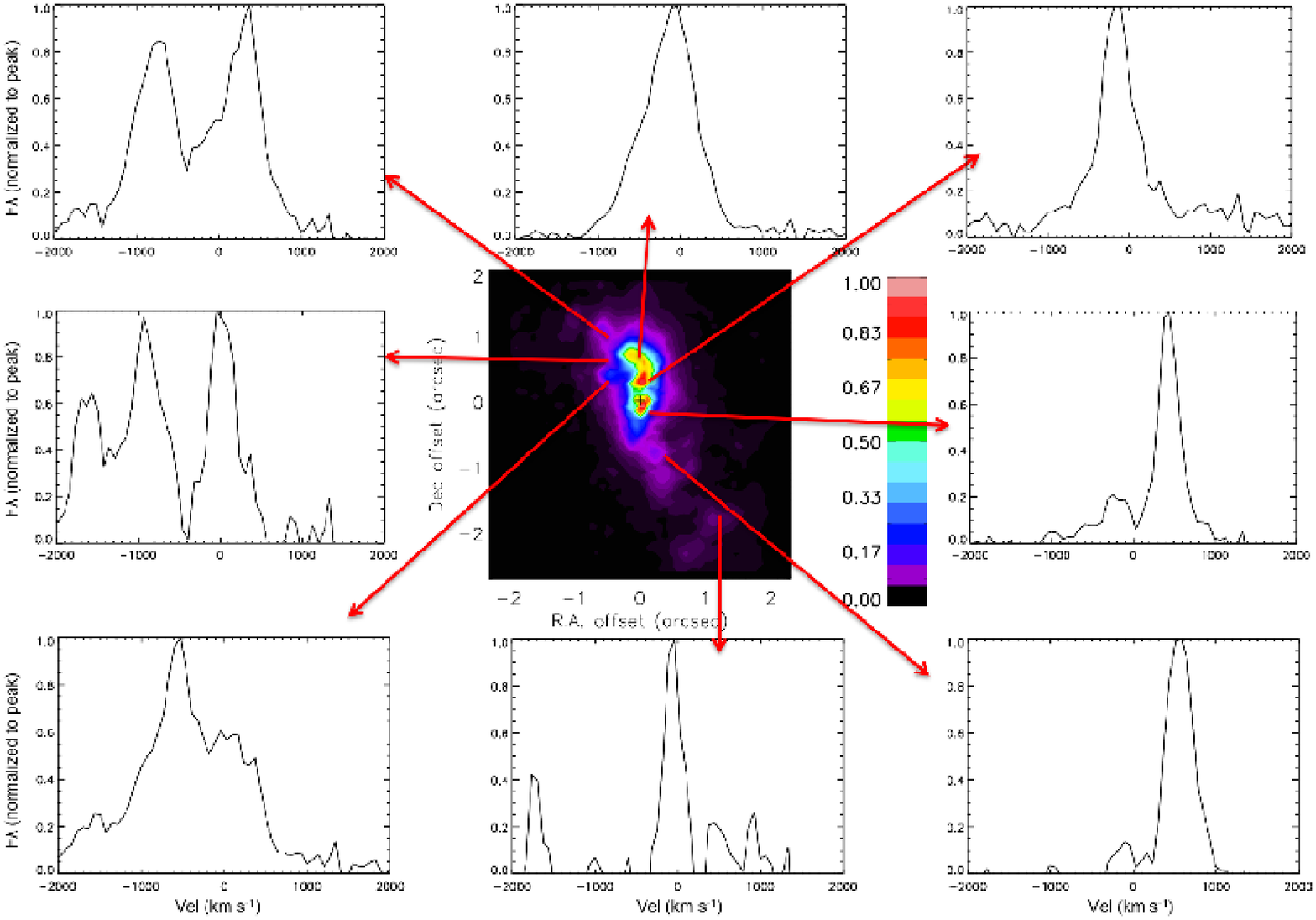}
\caption{[Si~{\sc vi}] emission line profiles from different regions in the central $4\arcsec\times4\arcsec$ ($280 \times 280$ pc) of NGC~1068. The spectra have been extracted from apertures of $0.5\arcsec$ (35 pc) diameter. The locations of these regions are indicated by the arrows. The central panel shows the flux map of  [Si~{\sc vi}] emission. 
\label{fig2}}
\end{figure}

The NLR and CLR of NGC~1068 are unique with respect to the other galaxies observed, presenting multiple spectral components (see Figure~\ref{fig2}). In consequence, a single-Gaussian fit would be insufficient to derive the properties of the ionized gas, and a multi-Gaussian approach would be difficult and uncertain due to the large number of degrees of freedom to constrain in such fits. In order to take into account the asymmetries (and the multiple components as in NGC~1068) in the coronal lines spectral profiles in a consistent manner, a velocity tomography of these lines was performed. 

Flux images in different velocity channels were constructed in a similar way to that used to create the integrated line flux image (see previous Section). The data cubes were ``sliced'' along the spectral coverage of each emission line into small wavelength ranges to create each channel map. The slices were obtained after the subtraction of the continuum determined as averages of the fluxes from
both sides of the emission line. In most of our data, each channel corresponds to 
$\sim80$ km s$^{-1}$, 
approximately the spectral resolution of the data. 
The number of channels depends on the width of the emission line for each galaxy.
The zero velocity is adopted as the one corresponding to the peak wavelength of H$_2$ 1-0S(1) within an integrated spectrum centered at the nucleus. In all cases, this corresponds to approximately the systemic velocity of the galaxy. 

The velocity channel maps provide different information from that of the 2D velocity maps. The latter 
is dependent on the relative intensities of different kinematic components eventually present to form an emission line showing velocities which  
usually correspond to the components with the highest fluxes. In the velocity channel maps, the emission in the wings of the profiles is fully mapped, allowing the detection of gas with higher radial velocities, but with lower emission intensity. 





\section{General Properties of the Ionized Gas}\label{results} 

This section contains a discussion of the general results of our near-IR observations, along with comparisons to relevant measurements at other wavelengths. 
More detailed analyses and interpretations can be found in the following Sections.
Specific details and analyses for individual objects can be found in the Appendix.

\subsection{Spatial Distribution and Kinematics of the Narrow Line Region}\label{nlr} 

Figures~\ref{fig3} to~\ref{fig4} present the Br$\gamma$ flux, velocity and dispersion maps for the full galaxy sample. 
In all panels, the contours delinate the $K-$band continuum emission 
and the position of the AGN, defined as the peak of non-stellar continuum emission at 2.2 $\mu$m, is marked with a cross. 
Regions in white correspond to pixels where the line properties are uncertain and thus were masked out. These rejected pixels in the velocity and dispersion maps are those with a flux density lower than $5\%$ of the peak of Br$\gamma$ emission. 
The Br$\gamma$ fluxes were measured by integrating
the flux inside a rectangular region that included all the visible emission, F(Br$\gamma$)$_{\mathrm{int}}$, 
as well as within a circular aperture equivalent to two times the spatial resolution achieved in each galaxy ($2\times$FWHM) and centered at the nucleus, F(Br$\gamma$)$_{\mathrm{nuc}}$. In NGC~6814 and NGC~7469 these quantities were measured in both the OSIRIS and SINFONI datasets, and the results are consistent within $5\%$. 
The measurements are presented in Table~\ref{table2}.

Despite the diversity of the observed morphologies, most of the Br$\gamma$ emission consists of a marginally resolved core (except in NGC~3783 which is unresolved) and extended emission out to several tens of pc. The emission in all cases is diffuse or filamentary, and it is difficult to determine whether it further breaks down into compact knots or blobs such as those found in H$\alpha$ or [O~{\sc iii}] images, even though the resolutions are comparable. The only two cases where the nuclear emission shows detailed structure (narrow-line clouds) are NGC~1068 and NGC~4151. In order to quantify the extent of the Br$\gamma$ emission in all galaxies, 
we have plotted in Figure~\ref{fig5} the azimuthally averaged surface brightness radial profiles of the Br$\gamma$ flux distribution. The azimuthal average of the Br$\gamma$ emission is consistent with a full-width at half-maximum (FWHM$_{\mathrm{Br\gamma}}$) less than 60 pc in all of the AGN, with a mean FWHM$_{\mathrm{Br\gamma}}$ for the sample of 22 pc (Figure~\ref{fig5} and Table~\ref{table2}). These measurements reflect the size of the compact and bright core in each galaxy. 
We also measured the extent of the photometric semimajor axis of the Br$\gamma$ emission, R$_{\mathrm{Br\gamma}}$, as well as its PA$_{\mathrm{Br\gamma}}$ (Table~\ref{table2}). The measurements of R$_{\mathrm{Br\gamma}}$ and PA$_{\mathrm{Br\gamma}}$ were done directly on the images, using as reference the contours corresponding to $5\%$ of the peak of Br$\gamma$ emission. A comparison of the measured Br$\gamma$ sizes with measurements from [O~{\sc iii}] images is presented in Table~\ref{table2}. 
The size of Br$\gamma$ emission in two objects (NGC~3783 and NGC~6814) is consistent with the sizes of the NLR estimated from [O~{\sc iii}] images. The $HST$ [O~{\sc iii}] image of NGC~7469 shows a compact bright core ($r<180$ pc), also seen in UV images \citep{munoz}, and extended emission up to $r\sim1300$ pc. The OSIRIS Br$\gamma$ flux map of this galaxy shows a similar tendency (see Figure~\ref{fig4}).    
While the compact emission is likely associated with the AGN, the extended emission can be attributed to circumnuclear starformation \citep{heckman86, genzel95}. Therefore, the NLR in this galaxy must be very compact and its size is then consistent with the measurements of R$_{\mathrm{Br\gamma}}$ ($130<r<180$ pc).  
In Circinus, NGC~1068, NGC~4151 and NGC~2992, the observations are limited by the relatively small FOVs of SINFONI and OSIRIS, but the Br$\gamma$ flux maps at these scales show very similar morphologies to those of the NLR as traced by [O~{\sc iii}] or H$\alpha$ images. All these facts confirm the validity of using Br$\gamma$ as tracer of the NLR. 

As can be seen in the velocity maps of Figures~\ref{fig3} to~\ref{fig4}, there is a great deal of diversity in the kinematics of the NLR at these scales in the sample galaxies. Three cases are clearly identified: (1) velocity fields dominated by rotation as in Circinus and NGC~7469, (2) disturbed rotational patterns (NGC~2992, and NGC~6814), and (3) velocity fields dominated by non-circular motions (NGC~3783, NGC~4151 and NGC 1068). 
As it was discussed in Section~\ref{tomography} the Br$\gamma$ and [Si~{\sc vi}] line profiles of NGC~1068 are composed of multiple spectral components. This is clearly affecting the velocity and dispersion maps and therefore they cannot be used for analysis. 
The kinematics of the NLR and CLR in NGC~1068 will be analyzed in Section~\ref{outflows} by means of the velocity channel maps.


The velocity dispersion maps, or $\sigma$ maps, show a wide range of values ranging from 30 km s$^{-1}$ to $\sim300$ km s$^{-1}$. Interestingly, in each galaxy, there exist regions where the velocity dispersion clearly increases. This will be further discussed in Section~\ref{outflows}. Table~\ref{table2} lists the mean and the highest $\sigma$ measured for each of the galaxies.

\subsection{Spatial Distribution and Kinematics of the Coronal Line Region}\label{clr} 

In general, the S/N of the spectra is high enough to reliably measure the coronal gas flux distributions in all galaxies of the sample. As it can be seen in Table~\ref{table1}, coronal line emission was detected in the seven Seyfert galaxies of the sample. However, the kinematics can only be accurately measured where the emission line is strong. 
Because [Si~{\sc vi}] is generally the most prominent coronal line in the $K-$band, it was used for the measurements of the size and kinematics of the CLR. Figures~\ref{fig6} to~\ref{fig7} show the [Si~{\sc vi}] flux, velocity and dispersion maps for the SINFONI and OSIRIS subsamples. In all panels the contours delinate the $K-$band continuum emission and the position of the AGN is marked with a cross. 
Once again, in the kinematic maps we only display velocities for regions with a flux density higher than $5\%$ of the peak of [Si~{\sc vi}] emission. 
The 2D flux, velocity and dispersion maps of the second strongest coronal line in our spectra [Ca~{\sc viii}] look very similar to the equivalent [Si~{\sc vi}] maps, so they are not shown here. Measurements of the [Al~{\sc ix}] line also result in similar flux, $\sigma$ and velocity distributions. Figure~\ref{fig8} demonstrates the similarities between the morphologies of [Si~{\sc vi}], [Al~{\sc ix}] and [Ca~{\sc viii}] in the three galaxies where we detected the three coronal lines. 
The quantitative analysis of the line fluxes over the entire CLRs, F([Si~{\sc vi}])$_{\mathrm{int}}$, as well as within a circular aperture equivalent to two times the spatial resolution achieved in each galaxy ($2\times$FWHM) and centered at the nucleus, F([Si~{\sc vi}])$_{\mathrm{nuc}}$, is presented in Table~\ref{table3}. Once again, in NGC~6814 and NGC~7469 these quantities were measured in the OSIRIS and SINFONI datacubes, and the results from both datasets are consistent within $5\%$. 

As can be seen in Figures~\ref{fig6} to~\ref{fig7}, in all cases the CLR is resolved (except in NGC~3783) and the peak of emission is located at the position of the AGN (or very close to it), being spatially coincident with the peak of Br$\gamma$ emission. This supports the hypothesis that most of the Br$\gamma$ in these objects is produced by the AGN, rather than star-formation (probably except in Circinus and NGC~7469, see the Appendices \ref{circinus} and \ref{7469}). 
Despite the diversity of the observed morphologies, most of the CLRs consist of a bright nucleus and extended diffuse emission along a preferred position angle, which usually appears to coincide with that of the Br$\gamma$ gas. In NGC~1068 and NGC~4151 the emission is enhanced at $\sim0.35\arcsec$ north and west of the nucleus, respectively. In each of these galaxies, this bright knot is spatially coincident with the jet/cloud interaction region (the ``northern tongue'' of molecular gas in NGC~1068, M\"uller-S\'anchez et al. 2009), just as it occurs with the secondary peak of Br$\gamma$ (see Section~\ref{radio}). 

In order to quantify the extent of the CLR, we have constructed azimuthal averages within circular radial annuli of the flux distribution of [Si~{\sc vi}]. The resulting spatial profiles are shown in Figure~\ref{fig9}. The FWHM of the CLR is less than 60 pc in all of the AGN, with a mean FWHM$_{\mathrm{[Si~{vi}]}}$ for the sample of 24 pc (Figure~\ref{fig9} and Table~\ref{table3}). 
These measurements reflect the sizes of the compact and bright [Si~{\sc vi}] cores, which are equivalent to the sizes of the Br$\gamma$ cores. 
We also measured the extent of the photometric semimajor axis of the [Si~{\sc vi}] emission (R$_{\mathrm{[Si~{vi}]}}$), as well as its PA$_{\mathrm{[Si~{vi}]}}$, in the same way as it was done for the NLR. The results shown in Table~\ref{table3} indicate that the CLRs and the NLRs are extended along the same PA, and that the CLRs are slightly more compact than the NLRs.   

The fact that the flux distributions of [Si~{\sc vi}], [Ca~{\sc viii}] and [Al~{\sc ix}] are resolved (or marginally resolved) and extended, support an origin for these coronal lines in the inner part of the NLR or in the transition region between the BLR and the NLR, as suggested by previous authors (e.g. Prieto et al. 2005). As can be seen in Figure~\ref{fig8}, [Ca~{\sc viii}] is always less extended than [Si~{\sc vi}], and [Al~{\sc ix}] is always the most compact of the three coronal lines.  Recent studies (Rodr\'iguez-Ardila et al. 2006) support the formation of the observed stratified CLR through photoionization by the central source, with higher ionization potential species located closer to the AGN. Our measurements of [Al~{\sc ix}], which has the highest ionization potential of the three, are consistent with this hypothesis. However, it is surprising the smaller extent of the [Ca~{\sc viii}] emission region when compared to that of the [Si~{\sc vi}], which has higher ionization potential. 
This fact can be attributed to depletion of gas-phase calcium as it is deposited in dust grains, which can alter its gas-phase abundance by large factors (e.g. Groves et al. 2004). 
Still, the apparent differences in size might be merely an artifact of the lower S/N of the weaker lines.    

As can be seen in the velocity maps of Figures~\ref{fig6} to~\ref{fig7}, the kinematics of the CLR in the majority of the AGNs appears to be dominated by non-circular motions, except for Circinus, NGC~2992 and NGC~6814, in which it appears to be dominated by rotation. When comparing the velocity fields of [Si~{\sc vi}] and Br$\gamma$, one notices that the kinematics of the NLR and CLR in five of the seven Seyfert galaxies shown in Figures~\ref{fig6} to~\ref{fig7} are very similar, suggesting that the mechanism responsible for the gas motions is the same in the two regions. As it will be seen in Section~\ref{outflows}, this is also true for the NLR and CLR in NGC~1068. The only exception is NGC~7469, in which the NLR appears to be rotating and the CLR exhibits a completely different velocity field, probably dominated by radial motions. A detailed kinematic analysis of the NLR and CLR will be presented in Section~\ref{kinematics}.   

The velocity dispersion maps shown in Figures~\ref{fig6} to~\ref{fig7} confirm the fact that the coronal lines tend to be broader than the lines in the NLR, but narrower than those in the BLR ($400<$FWHM$<1000$ km s$^{-1}$). Surprisingly, the velocity dispersion variations of the gas in the NLR are very similar to those observed in the CLR for the majority of the galaxies in the sample, providing further evidence for a common mechanism responsible of the gas motions in the two regions. Table~\ref{table3} lists the mean and the highest $\sigma$ measured for each of the galaxies.




\section{Kinematic Analysis}\label{kinematics} 

\subsection{Evidence for rotation and outflow in the NLR and CLR of AGN}\label{outflows} 

We have extracted Br$\gamma$ and [Si~{\sc vi}] kinematics in the nuclear region of seven Seyfert galaxies, allowing for the first time a spatially resolved two-dimensional study of the low- and high-ionization gas motions on scales $<300$ pc around the AGN. 
At these scales, the Br$\gamma$ velocity fields of four galaxies (Circinus, NGC~2992, NGC~6814 and NGC~7469) appear rather regular, despite some deviations from pure rotation. A similar situation occurs in three velocity maps of [Si~{\sc vi}] emission (Circinus, NGC~2992 and NGC~6814). This fact allows us to use a simple axisymmetric disk model as a first approximation to characterize the NLR and CLR kinematics. 

This first characterization was done using the kinemetry method, developed and described in detail by \citet{krajnovic06}. Briefly, kinemetry is an extension of surface photometry to the higher-order moments of the velocity distribution. Kinemetry was originally designed for use with very high S$/$N ($> 100$) and relatively extended stellar kinematic data.
To apply this method to our lower S$/$N emission line data covering small spatial scales, the breadth of the analysis must be
somewhat restricted. We therefore employ kinemetry in
a more limited capacity; rather than using kinemetry to
measure and interpret subtle kinematic features of a velocity
field, we instead use it to determine the strength of
deviations of the observed velocity fields from the ideal rotating disk case. This is identical to
assuming that any deviations from the ideal case that
might occur in a disk (e.g. in- and outflows, warps, spiral
structure) induce similar power in the higher Fourier coefficients
than those caused by the noise, and therefore the analysis of these coefficients would be difficult and uncertain. We then considered the entire velocity field at once and solved for a global position angle (PA) and inclination $i$. The kinematic center is assumed to be spatially coincident with the AGN defined as the peak in the $K-$band nonstellar emission. The PA of the kinematic major axis is measured counterclockwise from North ($0\degr$) to East ($90\degr$) and have values in the range $[-180, 180]$ degrees. The gradient of velocities, from blueshifts to redshifts, determine the sign of the PA.  


The PA and $i$ of the kinematic major axes as listed in Table~\ref{table4} were obtained by minimizing the difference between
the SINFONI and OSIRIS velocity fields and the ``circular'' velocity field (the 2D image reconstructed from only the cosine term B1 in the kinemetric expansion, Krajnovi\'c et al. 2006). The quality of the fit is represented in each case as the average of the absolute values of the velocity residuals map ($\mu$), obtained after subtracting the rotational component to the Br$\gamma$ and [Si~{\sc vi}] velocity maps. Residuals greater than the estimated kinematic errors ($\sim20$ km s$^{-1}$, see Section~\ref{linefit}) represent either deviations from circular motion (bars, inflows/outflows, etc.), or artifacts induced by the noise. 
For comparison purposes, we include in Table~\ref{table4} the best fit PA and $i$ of the molecular gas disks derived by H09 for the majority of the sample galaxies.  These values are in addition consistent with those determined from the stellar kinematics in the central $<300$ pc and the large scale disk, as checked by H09 and \citet{friedrich10} for all the galaxies in the sample. Three cases are clearly identified in Table~\ref{table4} (note that NGC~1068 is excluded from this analysis). 
Regions with residuals consistent with the estimated kinematic errors ($\mu<20$ km s$^{-1}$) and best-fit PA and $i$ in good agreement with those determined from stellar and/or H$_2$ kinematics (within a typical uncertainty of $12\degr$) are interpreted 
to be consistent with co-planar disk rotation (NLRs in Circinus, NGC~2992 and NGC~7469).  
The regions with $\mu>20$ km s$^{-1}$ are likely to be dominated by non-circular motions (NLR and CLR in NGC~3783, NGC~4151 and NGC~6814; and CLR in NGC~7469). Finally, in two CLRs (Circinus and NGC~2992), we note inconsistency between the geometric parameters of the [Si~{\sc vi}] disk and those of the H$_2$ disk (PA differences greater than the measurement error of $\sim12\degr$, see Table~\ref{table4}), but the residuals indicate a good fit ($\mu<20$ km s$^{-1}$). Since the derived kinematic parameters of the CLR in the Circinus galaxy could be affected by the instrumental artifacts present at $\sim0.25\arcsec$ and $\sim-0.25\arcsec$ in the FOV, we consider this result as uncertain. In consequence, a detailed modeling of the [Si~{\sc vi}] kinematics in this galaxy was not attempted. The particular case of NGC~2992 will be discussed in detail in Section~\ref{model}.

Now we concentrate on the objects with residuals greater than the kinematic errors ($\mu>20$ km s$^{-1}$), indicating the presence of non-circular motions. 
A second characterization was done keeping the best fit parameters of the rotating molecular gas disk constant during the kinemetry expansion. However, in this case the measured errors are even larger than in the previous characterization 
indicating that rotation alone cannot explain the kinematics of these objects. This can be clearly seen in Figure~\ref{fig10} were we show the velocity fields of H$_2$, Br$\gamma$ and [Si~{\sc vi}] in NGC~3783, NGC~4151, NGC~6814 and NGC~7469 (the four galaxies with $\mu>20$ km s$^{-1}$). The velocity maps of the NLR and CLR in these objects are very similar (except in NGC~7469), and present redshifted and blueshifted velocities in the direction opposite to the sense of rotation as traced by the H$_2$ velocity maps. In NGC~6814 there is a region north of the AGN in which the Br$\gamma$ and [Si~{\sc vi}] velocities are redshifted but the H$_2$ velocities appear approximately at the systemic velocity of the galaxy, or even blueshifted, showing a complete different behaviour. 
As the direction of motion (radial velocity changing from blueshift to redshift) occurs along a single PA, there is no evidence of a significant warp
(e.g. twisting of the kinematic major axis PA and/or
inclination angle). Thus, any approach to reproduce the observed kinematics by a rotating or warped disk model can be excluded. The only exception is NGC~6814 in which there appears to be a twist of the kinematic major axis, which could be a signature of either a warped disk or a second kinematical component superimposed on co-planar disk rotation.
   
In order to better understand such motions, we make use of the velocity channel maps described in Section~\ref{tomography}. Figures~\ref{fig11}--\ref{fig14a} show the slices around the [Si~{\sc vi}] line through the data-cubes of NGC~1068, NGC~3783, NGC~4151, NGC~6814, and NGC~7469, respectively. For completeness, Figures~\ref{fig14b} and \ref{fig14c} show the velocity channel maps of [Si~{\sc vi}] in NGC~2992 and Br$\gamma$ in Circinus, respectively. Since the Circinus datacube is affected by instrumental artifacts around the [Si~{\sc vi}] wavelength, we do not attempt to produce velocity channel maps of this line. 
It should be noticed that in the rest of the galaxies the velocity slices of Br$\gamma$ are very similar to the equivalent [Si~{\sc vi}] maps, so we do not show them here. The only exception is NGC~7469, but since the Br$\gamma$ velocity map of this galaxy is exceptionally regular (see Figure~\ref{fig10}), the velocity slices do not add significant and useful information to this work. 

As can be seen in Figures~\ref{fig11}--\ref{fig14a}, the velocity channel maps reveal blueshifted and redshifted velocity components on each side of the nucleus, whose speed appears to increase with radius to a value that exceeds 200 km s$^{-1}$. Note that in NGC~3783 this occurs mostly in the north part of the galaxy. These approaching and receding regions are in fact the main kinematical components in the radial velocity maps, but the magnitude of the velocity vectors is smaller. These radial motions indicate that a considerable amount of ionized gas must be moving either towards or outwards the nucleus rather than orbiting it on circular paths. The non-circular velocities show values in the range 200-500 km s$^{-1}$, except for NGC~1068 which exhibits a maximum velocity of $\sim1400$ km s$^{-1}$ at $\sim1\arcsec$ from the nucleus. 
The velocities of these bright clouds at $\sim100$pc from the nucleus are too high to be explained by any sort of reasonable gravitational potential (including contributions from the SMBH, nuclear stellar cluster, and bulge; Das et al. 2007), which rule out infall or orbital motions as being dominant. Rotational or gravitational infall models would require unreasonably high masses in the central $\sim50$ pc (greater than $10^{9}$ M$_\sun$) to produce the high velocities that are seen ($>300$ km s$^{-1}$). In the case of NGC~6814, these high velocities suggest that a radial component superimposed on disk rotation is responsible for the apparent change of the kinematic major axis PA. Furthermore, gas falling towards the nucleus is expected to increase its velocity as it approaches the AGN \citep{mueller09} or present defined structure (signatures of bars or inner spirals) in the radial velocity maps or the velocity residuals \citep{schoenmakers97, davies09}, properties that are absent in our data. The observed velocity variations as a function of distance provide support for a non-gravitational force which is accelerating the gas from the nucleus and therefore can only be explained in terms of radial outflow.

Several other pieces of evidence suggest that the non-circular motions observed in the kinematics of the NLR and CLR correspond to outflows (see the Appendix for a discussion of the individual galaxies):
\begin{itemize}
\item For the galaxies with existing radio images (NGC~1068, NGC~4151 and NGC~6814), we found that the collimated radio emission is spatially coincident with the NLR/CLR gas (or at least the two are aligned, see Figure~\ref{fig15}). The direction of motion seen in [Si~{\sc vi}] and Br$\gamma$ is blueshifted in the jet pointing toward us and redshifted in the counterjet. This is a clear indication of a bipolar outflow of material from the nucleus.
	\item Interestingly, in most of the galaxies there exist regions where the velocity dispersion clearly increases. These increments are observed at the nucleus and/or along the directions of the blueshifts and redshifts, providing evidence for radial outflow, as the combination of the line-of-sight (LOS) velocities from clouds located at the front and back parts of the outflow increment the width of the spectral profiles. Gas moving on circular orbits or falling into the nucleus along eccentric orbits or under the influence of a bar or nuclear spiral is expected to show a rather uniform dispersion map (H09), or even a low dispersion along its trajectory towards the nucleus, since the gas is expected to be gravitationally bound \citep{mueller09}. 
	\item Some galaxies show evidence for radial acceleration to a projected distance, followed by deceleration (NGC~1068, NGC~3783 and NGC~7469). 
For example, in NGC~1068, the maximum blueshift of $\sim-1400$ km s$^{-1}$ in Br$\gamma$ and [Si~{\sc vi}] occurs at $\sim80$ pc NE from the nucleus, but there are low velocity components at further distances (see e.g velocity channel with -1000 km/s), indicating that the NLR/CLR gas accelerates outward from the nucleus to a turnover point, and subsequently appears to decelerate. Radial acceleration plus deceleration was first noticed in $HST$/STIS observations of NGC 1068 and NGC 4151 (Crenshaw et al. 2000a, 2000b), being interpreted as radial outflow. 
\end{itemize}  

The integral-field observations are revealing kinematic signatures of rotation and/or outflow in the NLR and CLR of Seyfert galaxies. Therefore, kinematic models incorporating these two types of motions were used to reproduce the observed velocity fields. The fitting method and the derived model parameters will be discussed in Section~\ref{model}.

\subsection{Kinematic modeling}\label{model} 

To quantify the kinematics of the NLR and CLR 
in the galaxies with significant non-circular motions, 
we have modeled the observed velocities as extra-planar motions with biconical geometry, or biconical outflows, superimposed on co-planar disk rotation. The biconical geometry was selected because of its adaptability to the present investigation as demonstrated by previous morphological and kinematical studies of the NLR in Seyfert galaxies with $HST$ \citep{crenshaw00, schmitt03b, das06}. Our observations also provide support for the use of such geometry. As can be seen in Figure~\ref{fig10} and Figures~\ref{fig11} to~\ref{fig14a}, the outflows appear to be bipolar and are following a preferred direction with a certain opening angle, characteristics that can be well matched with a bicone.

The disk is modeled as a set of rotating coplanar annuli following the cosine law approximation \citep{krajnovic06}. The kinematic center is assumed to be coincident with the AGN defined as the peak in the nonstellar emission. The disk model has three free parameters: the magnitude of the circular velocity $v_c$, the position angle of the major axis of the disk PA$_{\mathrm{disk}}$, and the inclination of the disk $i_{\mathrm{disk}}$. Fitting the velocity fields with this model would be equivalent to use kinemetry to derive the best-fit disk parameters as it was done in Section~\ref{outflows}. 

The second component of the model is a biconical outflow. The modeling of this structure is based on the method described in \citet{das06}. The model consists of two hollow cones having interior and exterior walls placed apex to apex at the position of the AGN. This point corresponds to the apex of the bicone. The bicone is symmetrical in geometry about the apex. Each cone extends from this point to a maximum distance $z_{\mathrm{max}}$ measured in parsecs along the bicone axis. The radius of the inner and outer walls of the bicone are defined by the inner and outer half-opening angles ($\theta_{\mathrm{in}}$ and $\theta_{\mathrm{out}}$), respectively. The orientation of the bicone in the sky is defined by a pair of angles: PA$_{\mathrm{bicone}}$, and $i_{\mathrm{bicone}}$. PA$_{\mathrm{bicone}}$ is the position angle of the bicone axis in the plane of the sky measured counterclockwise from North ($0\degr$) to East ($90\degr$), and $i_{\mathrm{bicone}}$ is the inclination of the bicone axis with respect to the line-of-sight (LOS), being zero when the bicone axis is perpendicular to the LOS, positive when the northern part of the bicone (or eastern part if PA$_{\mathrm{bicone}}=90\degr$) is inclined towards the observer, and negative otherwise. The velocity at each point is a  three-dimensional vector moving radially away from the AGN. Since the non-circular (outflow) velocities appear to increase with distance $r$ from the nucleus, we tested velocity laws of the form $v\propto r^n$, and found that a simple proportionality (n=1), or $r$-law ($v_a=k_1r$), provided a good match to the data. Some galaxies show evidence for radial acceleration to a projected distance, followed by deceleration that approaches the systemic velocity (NGC~1068, NGC~3783 and NGC~7469). 
We therefore incorporated a deceleration term to the velocity law of the form $v_d=v{_\mathrm{max}} - k_2r$, where $v{_\mathrm{max}}$ is the maximum velocity reached by the outflow at a distance $r_t$ from the nucleus. The velocities along the bicone walls will be zero at $z_{\mathrm{max}}$. Note that $v_d$ does not have any impact on the cases where deceleration is not observed (as in NGC~4151), since this term is only used at distances $r>r_t$, and therefore it occurs outside the observed emission. In these cases, the exact value of $z_{\mathrm{max}}$ is not constrained. 

The bicone model was programmed in IDL as a three-dimensional array filled with velocities for a given parameter set and velocity law. The velocity value of each element of the array corresponds to the component of the velocity vector in the direction of the LOS, as would be measured by the observer. 
An example of such model following an $r$-law of acceleration and deceleration is shown in Figure~\ref{fig16}. The three-dimensional structure and front and back projections of two test models with PA$_{\mathrm{bicone}} = 0\degr$ and two inclinations, $i_{\mathrm{bicone}} = 90\degr$ and $i_{\mathrm{bicone}} = 0\degr$ are shown in this Figure. For the purposes of illustration, the velocities are assigned a color based on the amount of redshift or blueshift. The dark purple color is the highest blueshifted (or negative) velocity, and the light red is the highest redshifted (or positive) velocity, with intermediate velocities given in gradients. Green represents approximately zero velocity. 

Since the position in the LOS direction of the more luminous clouds in the NLR/CLR is not known,
for each galaxy we performed three fits: (1) using the LOS velocities in the front part of the bicone, (2) using the LOS velocities in the back part of the bicone and (3) using the average of the LOS velocities at each pixel. 
With each of these methods it is possible to carry out a two-dimensional kinematic fit of the bicone model plus rotation to the velocity maps. 
In addition, for each galaxy we performed two fits, one using only the bicone model, and a second one adjusting simultaneously the disk and bicone parameters. 
The best-fit parameters were found by minimizing the sum of the squared differences between the model velocity field and that extracted from the data, taking into account the measured errors in each spaxel ($\chi^2$). 
The best-fit corresponds to the model presenting the minimum $\chi^2$, and the confidence intervals are determined by a range of $\chi^2$ values greater than ${\chi^2}_{min}$, defined as $\Delta \chi^2 = \chi^2 - {\chi^2}_{min}$ \citep{lampton76}. Before computing confidence levels we have renormalized $\chi^2$ following a procedure already adopted by \citet{barth01} and \citet{marconi06}, which consists in rescaling the error bars by adding a single constant in quadrature to the velocity uncertainties. This constant is chosen for each galaxy such that the best fit model has $\chi^2/\mathrm{dof}\approx1$. 

The results of applying this fitting procedure to the four galaxies with significant radial motions in their CLR ($\mu>20$ km s$^{-1}$) are summarized in Table~\ref{table6}. Since the velocity maps of the NLR and CLR in these objects are very similar (except in NGC~7469, see Figure~\ref{fig10}), the fitting procedure yielded best-fit parameters which are equivalent for the two regions. 
Figures~\ref{fig17}--\ref{fig20} show the best-fitting model velocity fields of NGC~3783, NGC~4151, NGC~6814 and NGC~7469, respectively. In all cases, the best fits were obtained using the average of the LOS velocities. 
In NGC~3783 we adjusted a kinematic model consisting of one single cone in the north part of the galaxy plus rotation, as part of the southern cone is hidden by the plane of the galaxy (see Section~\ref{orientation}). 
For comparison purposes, we present in Table~\ref{table6} the minimum $\chi^2$ values achieved for three different types of models (without renormalization): (1) pure rotation ($\chi^2_{\mathrm{rot}}$), (2) pure biconical outflow ($\chi^2_{\mathrm{out}}$) and (3) rotation plus biconical outflow ($\chi^2_{\mathrm{rot+out}}$). Although the number of fitted parameters is different in each of these three models, the number of degrees of freedom remains practically unchanged, since the number of data points in the velocity maps ($N$) is large. Thus, $N$ was used as the number of degrees of freedom in each case (Table~\ref{table6}). 
In addition, we show in Figure~\ref{fig20b} the best fit with pure rotation for a rotation-dominated case (NGC~6814, $\chi^2_{\mathrm{rot}}<\chi^2_{\mathrm{out}}$), and the best fit with pure outflow for an ouflow-dominated case (NGC~3783, $\chi^2_{\mathrm{out}}<\chi^2_{\mathrm{rot}}$). In the two cases, the quality of the fit, represented by the velocity residuals map and the $\chi^2$ values (see Table~\ref{table6}), decreases considerably when compared to the best fits incorporating rotation plus outflow. As can be seen in Figure~\ref{fig20b} specific flow patterns are noticeable in the velocity residuals maps. Particularly interesting are the residuals in an outflow-dominated case, which reveal signatures of rotation as traced by the H$_2$ disk.  
The NLR/CLR kinematics of NGC~1068 was modeled using a bicone structure similar to the one described here. However, the fitting method was slightly different, as in this galaxy we aimed to reproduce the velocity channel maps and PV diagrams \citep{storchi09}. The modeling of the NGC~1068 kinematics is described in a separate publication (M\"uller-S\'anchez et al. in prep.), and here we only present the main results (Table~\ref{table6}), which will be used in the discussion presented in Section~\ref{discussion}. 


Our kinematic models of biconical outflow plus rotation provide a good match to the overall velocity patterns, despite slight local variations in the velocity fields. No obvious regular patterns are noticeable in the residuals maps, suggesting that the features observed in the residuals are only due to noisy pixels in the data (Figures~\ref{fig17}--\ref{fig20}). Furthermore, the residuals show mostly values between $[-20, 20]$ km s$^{-1}$ consistent with the measured errors of the velocity maps. 
As can be seen in Table~\ref{table6}, in two galaxies 
the best-fit is obtained for models incorporating only biconical outflow
(NGC~1068 and NGC~7469) and in other two galaxies there are no signatures of deceleration (NGC~4151, NGC~6814). Note that in NGC~4151 further observations with larger FOVs are required to conclude on this topic. 
For NGC~1068 the best-fit model parameters are in agreement with those from previous $HST$ spectroscopic studies of the NLR in this galaxy \citep{crenshaw00, das06}. However, in NGC~4151 our estimates of PA$_{\mathrm{bicone}}$, $v_{\mathrm{max}}$ and $r_t$ are significantly different to the ones derived in previous kinematic studies (Das et al. 2005). We attribute these differences to the fact that we are incorporating a rotational component and to the shape of the FOV of OSIRIS. Unfortunately, our observations do not fully cover the regions where the outflows are observed, so we are probably seeing only the starting and lateral parts of the outflow. However, the rest of the model parameters are in good agreement with previous kinematic models of the NLR in this galaxy. In all cases where a rotational component is present, the PA$_{\mathrm{disk}}$ and $i_{\mathrm{disk}}$ values are consistent with those determined from the stellar or H$_2$ kinematics (Table~\ref{table4}) and the large scale galactic disk. This will be further discussed in Section~\ref{disks}. 

The case of NGC~2992 is now revisited. Since the [Si~{\sc vi}] velocity map of Circinus is uncertain (see Section~\ref{outflows}), this is the only galaxy in which we note inconsistency between the geometric parameters of the putative [Si~{\sc vi}] disk and those of the H$_2$ disk (PA differences greater than the measurement error of $\sim12\degr$, see Table~\ref{table4}), but $\mu<20$ km s$^{-1}$ (Section~\ref{outflows} and Table~\ref{table4}). The orientation of the galactic disk in this galaxy agrees with that of the H$_2$ kinematics (Friedrich et al. 2010), 
suggesting that although purely co-planar disk rotation matches reasonably well the [Si~{\sc vi}] velocity field, the CLR kinematics is probably more complex. NGC~2992 exhibits blue-wing asymmetric coronal line profiles in the whole FOV, reaching values up to $\sim-300$ km s$^{-1}$. This situation is also observed in Circinus and NGC~7469 among the sample galaxies (see Figure~\ref{fig1a} and Figure 8 of MS06), and has been interpreted as evidence for radial outflow (MS06). Further evidence for outflows comes from the increments of velocity dispersion with radius, and from the H$_2$ kinematics, which are explained by incorporating an ouflowing component (Friedrich et al. 2010).     
Therefore, we interpret the variations in the geometrical parameters of the disk as evidence for radial outflow, and decided to use a biconical outflow model superimposed on co-planar disk rotation to explain the CLR kinematics in this galaxy. 
The results of applying this fitting procedure to the CLR of NGC~2992 are presented in Table~\ref{table7} and Figure~\ref{fig22}. This model provides also a better fit to the Br$\gamma$ velocity field when compared with the pure rotation model. Furthermore, the best-fit parameters of the bicone are consistent with previous studies of the NLR in this galaxy \citep{veilleux01, garcialorenzo01}. 

Table~\ref{table7a} summarizes the main results of the kinematic analysis. In the galaxies where the best fit incorporates rotation and radial outflow, the dominant component is considered to be the one exhibiting a smaller $\chi^2$ when fitting the velocity field just with that type of motion. A visual inspection of the residuals maps was also used to identify the dominant component in the NLRs and CLRs. The results of these two approaches are consistent with each other. 
As it was mentioned in Section~\ref{nlr}, in four galaxies (Circinus, NGC~1068, NGC~2992 and NGC~4151) the size of the NLR, as traced by [O~{\sc iii}] emission, is larger than our FOVs. However, the model parameters of NGC~1068, NGC~2992 and NGC~4151 are consistent with those derived in previous kinematic studies of the NLR in these galaxies with larger FOVs. All CLRs are fully covered by the SINFONI and OSIRIS FOVs, except in the Circinus galaxy. Since we have not been able to characterize a putative outflowing component in Circinus on scales of less than 16 pc from the nucleus, although there exist spectroscopic and morphological evidences of its presence (Marconi et al. 1994; Prieto et al. 2005; MS06), this galaxy is excluded from the discussion in the following sections, 
except for Section~\ref{disks}, where the presence of rotational components in all galaxies is discussed.

\section{Discussion}\label{discussion}

\subsection{Inner disks of ionized gas}\label{disks}

In general, the Br$\gamma$ velocity fields in the sample galaxies exhibit co-planar rotational components, being the only exception NGC~1068 (see Table~\ref{table7a}). A similar situation is observed in the [Si~{\sc vi}] velocity maps, this time being NGC~1068 and NGC~7469 the only exceptions. Furthermore, the best-fit parameters to the NLR and CLR rotational components are consistent with those derived from the stellar$/$H$_2$ kinematics. Another important fact is that the magnitudes of the rotational velocity vectors in each galaxy are consistent with the values found by H09 for the molecular gas, within $\pm 20$ km s$^{-1}$. Therefore, the dynamical masses in the central $r\sim50$ pc estimated from the rotational components of ionized gas (considering $\sigma_{\mathrm{mean}}$ of Br$\gamma$ as the dispersion in the disk, Table 2) are in good agreement with those found from H$_2$ kinematics (between $2-20\times10^7$ M$_\odot$). 
In addition, the consistency of the nuclear kinematics with the large-scale disk orientation reported in the literature for the sample galaxies indicates that these inner disks are well aligned with the host galaxy. 

Evidence for an underlying rotational component in the NLR is also found in ground-based and $HST$ studies of several other objects, such as NGC~1365 (Hjelm \& Lindblad 1996), NGC 2992 \citep{veilleux01}, and NGC 3516 (Arribas et al. 1997), and recent $HST$ spectroscopic studies of the CLR \citep{mazzalay10} with a general trend for the low-ionization gas to be a better tracer of the disk component, while the high-ionization gas presents more deviant behavior, usually associated with outflow. Our integral-field data confirm this hypothesis and provide accurate measurements of the physical properties of the disk and outflow components.

The presence of inner disks of low- and high-ionization gas provides additional constraints to the classical picture of the NLR in unification models (e.g. Antonucci 1993), which postulate that an optically and geometrically thick toroidal structure produces a bicone of radiation that ionizes the NLR. In this model, the ionized gas is co-spatial with the bicone of radiation and is outflowing. Within this context, rotating disks of ionized gas may be observed if the ionization bicone intersects the disk plane, causing NLR and CLR emission with a rotation signature. An example of such interaction has been observed in Mrk 573 with $HST$/STIS \citep{fischer10}. This hypothesis can be tested quantitatively with our kinematic models. If the angle between the axis of the bicone and that of the host galaxy ($\beta$) is smaller than the half openig angle of the bicone ($\beta<\theta_{\mathrm{out}}$), the ionizing radiation will intersect the disk of the host galaxy, and a rotational component will be observed. Because the disk has a finite thickness, when $\beta>\theta_{\mathrm{out}}$ radial motions are expected to dominate, but a rotational component may be present. As can be seen in Table~\ref{table6}, the two galaxies presenting only radial outflow in their CLR (NGC~1068 and NGC~7469) have $\beta>\theta_{\mathrm{out}}$, and therefore rotational signatures are absent. In NGC~2992, NGC~3783 and NGC~6814 $\beta<\theta_{\mathrm{out}}$, and as a result co-planar disk rotation is observed. Finally, in NGC~4151 $\beta>\theta_{\mathrm{out}}$ and a rotating disk is present, suggesting that either the disk height is high enough to provide a large intersection area with the ionization bicone, or there exist hard UV$/$soft X-ray radiation located outside of the ionization bicone in the disk plane. 
Although the first explanation seems plausible, the second should not be ruled out. 
First, in all four galaxies with rotational and outflowing components (NGC~2992, NGC~3783, NGC~4151 and NGC~6814; see Table~\ref{table7a}) the low- and high-ionization gas emission extends beyond the borders of the bicone as represented by the models (Figures~\ref{fig17}--\ref{fig20} and Figure~\ref{fig22}). Additionally, in all our sample galaxies, the disks of ionized gas are spatially coincident with the disks of molecular gas, 
suggesting that the two types of gas are spatially mixed. 
The presence of NLR/CLR gas outside the ionization cones seems to contradict the predictions of the classical torus model of AGN, as it would imply the escape of radiation through the walls of the obscuring structure. 
One way to reconcile this discrepancy is to assume that the molecular gas within this nuclear region is in a clumpy distribution such that a fraction of ionizing photons escape in different random directions outside the bicone 
and ionize the gas in the disk producing the observed flux and velocity maps. 
Recently, \citet{mor09} analyzed mid-IR spectra of AGN using clumpy torus models and found that for a mean number of 5 clouds along any equatorial radial line (number of clouds varying between 1 to 10) there is a reasonable probability ($>\sim1\%$) to have a LOS that does not encounter any clouds, i.e. UV radiation can escape unhindered.   
The clumpy distribution is also supported by theoretical models \citep{schartmann05, elitzur06, nenkova08}, and observations of thermal dust emission (e.g. Jaffe et al. 2004), silicate absorption (e.g. Deo et al. 2007) and molecular gas (MS06; H09). 

\subsection{Collimation and acceleration of the bipolar outflows}\label{acceleration}

High degree of collimation (full opening angle generally not more than $15\degr$) is a conspicuous feature of relativistic jets \citep{lada96}. The ionized gas outflows studied here have full opening angles ($2\theta_{\mathrm{out}}$) $>50\degr$. In this regime, we consider an outflow collimated when its full opening angle is $<70\degr$ and less-collimated when it has a value $>90\degr$. 

The outflows observed in the velocity maps of NGC~3783 and NGC~4151 show values  which increase with distance from the nucleus up to velocities $>200$ km s$^{-1}$ and are collimated, having half-opening angles ($\theta_{\mathrm{out}}$) in the range $29-34\degr$ (see Table~\ref{table6}). The velocity channel maps of NGC~1068 (Figure~\ref{fig11}) reveal a similar behaviour, but with a maximum velocity which is $4-9$ times higher than the values estimated in other CLRs and the smallest $\theta_{\mathrm{out}}$. These results are consistent with what has been observed previously in kinematic studies of the NLR of NGC~1068 and NGC~4151 \citep{das05, das06}. In NGC~2992, NGC~6814 and NGC~7469 the outflows are less collimated with $\theta_{\mathrm{out}}$ between $45-57\degr$ and are not experiencing high accelerations reaching $v_{\mathrm{max}}<200$ km s$^{-1}$. These results suggest the existence of two types of outflows: fast and collimated, and slow and less-collimated. We find an anticorrelation between the  [Si~{\sc vi}] outflow velocity ($v_{\mathrm{max}}$) and $\theta_{\mathrm{out}}$ of the bicone in the sense that more collimated outflows have higher velocities. This relationship is shown in Figure~\ref{fig23}. Linear fits to the maximum velocities with respect to $\theta_{\mathrm{out}}$ indicate that there is a significant anticorrelation, with Pearson's correlation coefficient of $-0.77$. The significance of this anticorrelation will be investigated further in Section~\ref{origin}.

Interestingly, the velocity channel maps of NGC~4151 and NGC~6814 (Figures~\ref{fig13} and~\ref{fig14}), provide evidence for a combined type of flow: the highest velocities are not only observed at the edges of the CLR (or the FOV in the case of NGC~4151), but also at the nucleus. These components do not appear in the radial velocity fields because, as it was explained in Sect.~\ref{tomography}, the Gaussian fits probe the brightest emission, while the channel maps also probe fainter emission (in the wings of the line profiles).     
This is consistent with the rotational component measured in the velocity fields of these two galaxies, as in the vicinity of the nucleus the brightest component is the one originating in the galactic disk, while away from the AGN the brightest component corresponds to the outflowing one. 
In the velocity channel maps we see high velocity gas at the nucleus, suggesting that the outflow does not leave the nucleus at zero velocity. Thus, these observations are probably not revealing acceleration of a single wind, but instead the interaction of an already accelerated wind with the surrounding medium. This scenario is consistent with the one proposed by \citet{everett07} and \citet{storchi09} who suggested that the interaction of an outflowing nuclear wind with the ISM could explain the observed velocities in NGC~4151. A more detailed discussion on the origin of the outflows will be presented in Section~\ref{origin}.





\subsection{Correlation of the outflowing gas and the molecular gas properties}\label{correlation} 

Unified schemes of AGNs \citep{antonucci93, urry95} require an optically and geometrically thick structure of gas and dust around the central engine to explain the great diversity of AGN classes. The compact sizes (only a few parsecs) determined in recent mid-IR interferometric observations (e.g. Jaffe et al. 2004; Beckert et al. 2008; Burtscher et al. 2009) require that the obscuring matter be located inside the region where the black hole gravity dominates over the galactic bulge. Three fundamental properties characterize this structure: (1) anisotropic obscuration so that sources viewed face-on are recognized as type 1 objects, those observed edge-on are type 2, (2) dust reemission in the IR of the AGN obscured radiation, and (3) collimation power responsible for producing a bicone of radiation that ionizes the NLR. 
In the following we refer to this small structure as
torus, keeping in mind that its detailed morphology and kinematics are rather uncertain, might not resemble closely a rotating smooth torus. As in principle the torus could be the structure responsible for the collimation, in this Section we investigate possible correlations between the outflowing gas and the molecular gas properties.


On scales of a few tens of pc from the nucleus, H09 found that the molecular gas in the sample Seyfert galaxies (except for NGC~2992) is in a generally rotating, geometrically thick clumpy disk that is associated with the smaller scale obscuring torus. We find an anticorrelation between $\theta_{\mathrm{out}}$ of the bicone and the molecular gas mass ($M_{\mathrm{gas}}$) in $r<30$ pc in the sense that galaxies with greater amount of gas have higher degree of collimation. This relationship can be seen in Figure~\ref{fig24}. The mass of molecular hydrogen is approximated using the typical gas mass fraction of $10\%$ suggested by four independent estimates (H09).
The anticorrelation coefficient is $-0.85$. 
This result confirms that the molecular gas structure seen on scales of a few tens of pc has a strong impact on physical processes directly associated with the AGN (Davies et al. 2006; M\"uller-S\'anchez et al. 2009; H09). Furthermore, it indicates that the torus is indeed the structure responsible for the collimation of outflows on small spatial scales. The accumulation of gas clouds towards the nucleus (higher $M_{\mathrm{gas}}$) increment the collimation power of the torus (stops the outflow expanding sideways). 


A correlation is also found between the [Si~{\sc vi}] outflow velocity $v_{\mathrm{max}}$ and the molecular gas mass in the sense that galaxies with greater amount of gas have higher outflow velocities. This relationship, shown in Figure~\ref{fig25}, results from the fact 
that these two properties are anticorrelated with $\theta_{\mathrm{out}}$ (Figures~\ref{fig23} and~\ref{fig24}). The significance of this correlation will be investigated further in Section~\ref{origin}.

\subsection{Orientation effects}\label{orientation}

The geometry of the bipolar outflow can be used to test the predictions of the unified model of Seyfert galaxies. The basic idea is that the torus outside of the BLR produces a bicone of radiation that ionizes the NLR and that Seyfert type depends on viewing angle with respect to the bicone axis and half-opening angle of the bicone ($i_{\mathrm{bicone}}$ and $\theta_{\mathrm{out}}$).
The bipolar outflow in Seyfert 1 galaxies (or intermediate type Seyferts) is expected to satisfy the condition 
$\left| i_{\mathrm{bicone}} \right| + \theta_{\mathrm{out}} > 90\degr$, so that a view to the BLR is possible, whereas in Seyfert 2 galaxies the condition is 
$\left| i_{\mathrm{bicone}} \right| + \theta_{\mathrm{out}} < 90\degr$. According to this definition, an AGN with, e.g $\left| i_{\mathrm{bicone}} \right|=65\degr$, would be a type 2 nucleus as long as we are not looking into the bicone ($\theta_{\mathrm{out}} < 25\degr$). Assuming that the plane of the small scale torus is perpendicular to the bicone, such an hypothetical AGN would have $\left| i_{\mathrm{torus}} \right|=25\degr$, nearly pole-on. 
We wanted to point out this fact to remark the importance of $\theta_{\mathrm{out}}$ in the standard classification of AGN. 


The inclination of the bicone axis in the only Seyfert 2 galaxy studied here, NGC~1068, is $9\degr$, nearly perpendicular to the LOS, and  $\theta_{\mathrm{out}}=27\degr$. The sum of these two angles is $<90\degr$ consistent with the prediction of the torus model of AGN for a type 2 object. Since $\theta_{\mathrm{in}} > \left| i_{\mathrm{bicone}} \right|$, the front part of the bicone is always blueshifted and the back part always redshifted, showing blueshifts and redshifts on either side of the AGN. 
However, blueshifted velocities are expected to slightly dominate the velocity vectors in the NE simply because there are more clouds in the LOS moving towards us in this region (the distance across the front part of the NE cone is higher than in the back part, due to the inclination). Based on the same argument, redshifted clouds should dominate in the SW part of the galaxy. 
As it can be seen in Figure~\ref{fig11}, the bipolar outflow in NGC~1068 is fully consistent with this scenario (see e.g velocity channels with $-200$ and $+400$ km s$^{-1}$), despite some obscuration in the SW produced by the plane of the galaxy (see below). 

Surprisingly, the five intermediate type Seyfert galaxies in our sample (four Sy 1.5 and one Sy 1.9, see Table 1) are seen at $\left| i_{\mathrm{bicone}} \right| + \theta_{\mathrm{out}} < 90\degr$ (outside the bicone), which apparently contradicts the predictions of the unified model. This puzzling result has also been observed previously in kinematic studies of the NLR of NGC~2992 and NGC~4151 \citep{veilleux01, das05}. Our results, however, are consistent with the unified model within the uncertainties. The angle between the outer edge of the bicone and the LOS is defined as $\alpha=90\degr - \left| i_{\mathrm{bicone}} \right| - \theta_{\mathrm{out}}$. Tables~\ref{table6} and \ref{table7} show that in most intermediate type Seyferts we have a view to the bicone that is very close to the outer edge, being NGC~3783 the only exception. Considering a typical uncertainty of $15\degr$ in $\alpha$ (corresponding to the mean of the sum of the uncertainties in $i_{\mathrm{bicone}}$ and $\theta_{\mathrm{out}}$) a direct view to the BLR is available in most cases (except in NGC~3783). It is interesting to point out that other available mesurements of $\alpha$ in three intermediate type AGN (NGC 2992: Veilleux et al. 2001, NGC 4051: Fischer et al. in prep., and NGC 4151: Das et al. 2005), indicate that the sight line is very close to the edge of the cone, rather than ``down the middle'' (in NGC 4051 the view is actually close to pole-on and inside the cone). This is also supported by previous calculations of the inclination angles of Seyfert 1 galaxies using results from reverberation mapping and X-ray data \citep{wu01, bian02}. Such studies find mean inclination angles between $32-36\degr$, which translate to $i_{\mathrm{bicone}}=54-58\degr$. Considering the mean half-opening angle of the galaxies in our sample $\theta_{\mathrm{out}}=40\degr$ as a typical value, it is reasonable to conclude that most Seyfert 1s are actually seen close to the edge of the cone, particularly intermediate type Seyfert galaxies (like the ones presented in this study, see Table~\ref{table1}) which are expected to have $i_{\mathrm{bicone}}\sim30-45\degr$ \citep{curran00}. However, it is intriguing that, according to these calculations, many Sy 1.5s would have sight lines that are outside of the bicone, i.e. the BLR is not directly seen. Although the uncertainty in our measurements could place most of the intermediate type Seyferts in our sample marginally inside the bicone, this may not work in the same direction for each one. Further integral-field observations of a larger sample of Sy 1s and Sy 1.5s are needed to fully test the idea that intermediate type Seyferts are viewed at intermediate angles and determine the probability for sight lines in and around the cone. 

Since $\alpha$ is rather small in our galaxies, it seems plausible that, in order to get a direct view to the BLR, there exist also small geometric distortions in the torus-bicone system. One possibility would be a small misalignment between the plane of the torus and the bicone axis such that they are not completely perpendicular. In this case the condition to observe the BLR needs to be modified by adding a correction term: $\left| i_{\mathrm{bicone}} \right| + \theta_{\mathrm{out}} + \delta > 90\degr$, where $\delta$ is the angle between $i_{\mathrm{bicone}}$ and the true perpendicular to the plane of the torus. Another possibility would be that the aperture of the outer part of the torus, where the ionization cone escapes, is slightly larger than $\theta_{\mathrm{out}}$, which in fact is determined by the most inner part of the torus. In this case the correction term $\delta$ can be understood as an enlargement to $\theta_{\mathrm{out}}$. Even if these two apertures are equal, $\theta_{\mathrm{out}}$ can be considered always slightly larger due to the existance of a LOS which allows us to see the far side of the BLR. 
Whatever plausible these scenarios may seem, they cannot account for the case of NGC~3783, where our line of sight is far outside of the bicone ($\alpha=26\degr$).  


Another way to reconcile the apparent discrepancy between our models and the LOS to the BLR (particularly in NGC~3783) is to assume that the collimating structure of molecular gas is in a clumpy distribution such that our view to the BLR is almost independent of the inclination and opening angles of the bicone. In the clumpy torus scenario, the classification of a Seyfert galaxy as type 1 or type 2 depends more on the intrinsic properties of the torus rather than on its inclination. Recently, \citet{ramos-almeida11} fitted models of clumpy tori to the spectral energy distributions (SEDs) of seven Seyfert galaxies including NGC~4151, NGC~6814 and NGC~7469. Their results do not show a clear trend in the inclination of the torus in Sy 1.5 and Sy 2 galaxies (although views of Sy 2s are slightly more inclined than those of the Sy 1.5 galaxies), but it is remarkable that for NGC~7469 they found $i_{\mathrm{torus}}=83\degr$. 
Their measurements of the inclination angles in the other two galaxies (NGC~4151 and NGC~6814) are consistent with ours within the uncertainties. Analytical models of clumpy accretion disks in combination with SINFONI data were used by \citet{vollmer09} to explain the properties and evolution of the torus in six Seyfert galaxies including NGC~3783 and NGC~7469. These authors found that while NGC~7469 shows a moderately transparent torus, NGC~3783 has the most transparent torus of the galaxies in their sample, a result that gives support to our observations.

The five intermediate type Seyfert galaxies studied here have intermediate inclinations, $\theta_{\mathrm{in}} < \left| i \right|$, and $\left| i_{\mathrm{bicone}} \right| + \theta_{\mathrm{out}} \lesssim 90\degr$ (or slightly larger than $90\degr$ considering the uncertainties). Therefore, they show almost entirely blueshifts on one side and redshifts on the other side of the central AGN, simply because there are more clouds with blueshifted (or redshifted) velocities at each side of the AGN (one external face and two internal faces of the bicone). The intermediate inclination is consistent with unified schemes and may explain the properties of intermediate type Seyferts. Recently, Nagao, Taniguchi \& Murayama (2000) compared several line strengths and line ratios of Sy 1.5s with those of type 1 and 2, and found that although the Sy 1.5s have an intermediate value in the line strengths between the Sy 1s and Sy 2s, the relative line strengths cover the whole observed ranges of both the Sy 1s and the Sy 2s. These results suggest that the three types of AGN are basically the same physical phenomenon and the differences may be due to inclination effects. In Sy 1.5s a significant part of the BLR is obscured by the torus, resulting in composite profiles consisting of both the NLR and BLR emission. This is supported by the fact that a spectrum with the appearance of an intermediate Seyfert galaxy can be obtained by adding a Seyfert 2 galaxy spectrum to a Seyfert 1 galaxy spectrum (Cohen 1983). Alternatively, a clumpy torus could also explain the properties of intermediate type Seyfert galaxies. In this case when a cloud is passing the line of sight to the central engine, the BLR can be (partially) obscured, resulting in a composite profile as well. 

In Figure~\ref{fig26}, we show the geometry of the biconical outflow and the host galaxy disk for all galaxies exhibiting radial outflow. For simplicity, the midplane of the galaxy and the outer surface of the bicone are portrayed, even though the bicone extends to an inner opening angle in the models. The color blue indicates that the majority of the radial velocities in the cone are blueshifted, although the back part of the cone in some cases may be redshifted as explained above. Accordingly, a red cone indicates that the majority of the radial velocities are redshifted. 
In the case of NGC 1068, the SW side of the galaxy is closer to us, and the SW cone therefore experiences more extinction (the red cone). This is consistent with the Br$\gamma$ and  [Si~{\sc vi}] images in Figures~\ref{fig3} and~\ref{fig6}, which shows weaker, and in some places absent, emission SW of the
location of the AGN. 
In NGC~2992 the majority of the NW cone is blueshifted and the SE cone is redshifted. The SE part of the disk is closer to us and therefore the blue cone experiences less extinction, on average, than the red cone. 
In the case of NGC 4151, the SW cone is entirely blueshifted and the NE cone is entirely redshifted. The host galaxy disk is close to the plane of the sky, and the SW cone shows less extinction than the NE cone shows. This picture is also applicable to the case of NGC~7469 (see Table~\ref{table6}). 
All the above geometries lead to integrated emission-line profiles that are asymmetric to the blue, in agreement with the observed profiles of these galaxies (see e.g. Figure~\ref{fig1a}) and the majority of previous spectroscopic studies of the CLR based on integrated line profiles. In the case of NGC~3783, the PA and inclination of the galactic disk are such that the blueshifted cone in the south is in its majority occulted by the disk and the redshifted cone in the north experiences less extinction. Thus, this model explains the slightly redshifted tails in the emission-line profiles of this galaxy, as well as its radial velocity field. A similar geometry is applicable to the case of NGC~6814 (see Table~\ref{table6}), which also presents asymmetric redshifted profiles (see Figures~\ref{fig1a} and~\ref{fig14}). These results are consistent with the analysis of the effects of extinction by a disk on the NLR profiles presented in \citet{crenshaw10}.

\subsection{Origin of the outflows} \label{origin} 

Recently, studies have shown that most of the emission lines seen in the NLR and CLR can be accounted for by assuming a central ionization source illuminating a multicomponent gas \citep{kraemer00, groves04, mullaney08, mazzalay10}. 
Since it is now generally accepted that photoionization is the dominant source of 
ionization of the NLR and CLR gas, with shocks possibly contributing in localized regions \citep{viegas02, prieto05, rodriguez06}, we focus our discussion on dynamical properties to gain more insight into the sources of energy that may affect the acceleration of the gas.

\subsubsection{Comparison between the gas kinematics and the radio jet} \label{radio} 


Figure~\ref{fig15} shows comparisons of the Br$\gamma$ linemaps to maps at radio wavelengths. The outflows are observed in approximate alignment with compact radio emission only in the central parsecs of NGC~1068 and NGC~4151, and possibly also in NGC~6814. Note that in NGC~6814 the OSIRIS Br$\gamma$ flux map has three times better spatial resolution than the radio image.
In the case of NGC~7469, in addition to the diffuse radio emission associated with the nuclear starburst ring, an East-West (E-W) extended nuclear source is also detected \citep{orienti09}, which, at high spatial resolution, separates into five sources along an E-W direction within $0.1\arcsec$ of the nucleus (Lonsdale et al. 2003). Thus, the outflow of coronal gas in this galaxy seems to follow the radio distribution at very small spatial scales, but a direct comparison between the two data sets could not be made. The 6 cm radio emission in NGC~2992 forms a figure-of-eight structure along a PA$=160\degr$ (Ulvestad \& Wilson 1984). In the central $4\arcsec$, the radio continuum is elongated along a PA=$38\degr$ presenting a good correlation with the H$_2$ and Br$\gamma$ morphology, but not with the [Si~{\sc vi}] emission, as this is very compact (total size of $\sim1.2\arcsec$) and the radio emission presents a cavity in this region. In NGC~3783 there exists only an unresolved nuclear radio source \citep{orienti09} and in Circinus there is no radio image of comparable resolution available in the literature (see Davies et al. 1998 for a comparison of near-IR and radio images in the central $30\arcsec \times30 \arcsec$ of this galaxy).     

Thus, a perfect spatial correlation between the radio images and the NLR/CLR outflows is not seen, only an alignment between them and signatures of interaction between the radio jet and the emitting gas in two galaxies: NGC~1068 and NGC~4151. Mundell et al. (2003) and Gallimore et al. (2004) suggest that some NLR clouds in NGC~1068 and NGC~4151, respectively, are actually responsible for bending the radio jet. 
Our observations are consistent with this interpretation. In these two objects there is a knot of enhanced Br$\gamma$ and coronal-line emission at $\sim0.35\arcsec$ north of the nucleus in NGC~1068, and west of the AGN in NGC~4151. In each of these galaxies, this bright knot is spatially coincident with one of the components of the nuclear radio jet (NGC1068: Component C, Gallimore et al. 2004; NGC~4151: Component C, Mundell et al. 2003). The channel maps of Figures~\ref{fig11} and~\ref{fig13} reveal line flux enhancements at the locations of radio knots with relatively high redshifted and blueshifted velocities (speeds in the range 200-600 km s$^{-1}$). 
These knots can possibly be explained by the lateral expansion of hot
gas away from the radio jet, as claimed by Axon et al. (1998) and
Capetti et al. (1997), and indicate that the radio jet has been deflected at this location. This flow, however, is restricted to clouds within a few parsecs from the nucleus, along the region of the radio jet. 
The large-scale outflow of the bright clouds cannot be accounted for by the radio jet alone and needs a different acceleration mechanism. 
This can be clearly seen in NGC~4151 where the flux distribution at $r<20$ pc from the nucleus is spatially coincident with the radio jet, but at distances greater than this the radio jet is oriented at PA=$77\degr$ and the bicone at PA=$50\degr$ (Figure~\ref{fig15}). The distinct orientations of the two structures suggest that the radio jet is not driving the outflow of ionized gas. A similar situation is observed in the rest of the sample galaxies and in NGC~3783 the coronal gas clouds seem to accelerate even in the absence of any radio emission. However, at small spatial scales, the radio jet may have a contribution to the formation of the outflows, transfering momentum to the circumnuclear gas (see Section~\ref{agn_wind}).

\subsubsection{An accelerated wind from the AGN}\label{agn_wind} 

In the kinematic models proposed for the sample galaxies, the NLR/CLR gas first accelerates up to a typical distance of $\sim180$ pc (except in NGC~2992, where $v_{\mathrm{max}}$ is reached at $r=900$ pc) from the nucleus and
then decelerates. While the deceleration phase is observed in NGC~1068, NGC~3783 and NGC~7469, this stage is not present in the rest of our data. 

Dynamical models including radiative forces, gravitational forces and a drag force due to the NLR clouds interacting with a hot ambient medium have been developed \citep{das07}. While these models fail to identify the acceleration mechanism, they do demonstrate that drag forces can produce the required deceleration. 
The possible acceleration mechanisms, radiative pressure, magnetic fields and thermal winds, fail in explaining the gradual acceleration over the inner 100 pc, reaching terminal velocities at distances $<10$ pc from the central source \citep{everett05, chelouche05, das07, everett07}. By analogy with the bipolar outflows observed in young stellar objects (Lada \& Fich 1996; Matzner \& McKee 1999), \citet{everett07} suggested that perhaps the observed $v\propto r$ AGN outflows result from the interaction of an already accelerated nuclear wind with the ISM, instead of acceleration of a single wind. In protostellar outflows, a wide-angled magnetized wind expands like a bubble and sweeps up the ambient medium in a momentum conserving manner into a shell at the wind-bubble ambient medium interface. The observed velocity law is a general property of momentum-conserving winds blowing into $1/r^2$ density profiles \citep{bally83, shu91}. More recently, Matzner \& McKee (1999) have argued that this velocity relation is recovered for a wide variety of ambient density distributions. 
Our observations provide evidence for such a two-stage process involving the local interaction of high velocity gas (an accelerated wind from the AGN) with low velocity gas from the disk (see Section~\ref{acceleration}). The AGN winds (e.g. hydromagnetic and radiatively driven winds), and probably also the radio jet in some cases, transfer momentum to the circumnuclear gas and push the gas from the disk, causing the observed emission-line profiles and velocity structure along the bicone. 

In this scenario, the collimation power of the torus and the relative angle between the outflow axis and the disk of the galaxy ($\beta$), are crucial to explain the diverse NLR$/$CLR kinematics observed.  
The correlation between molecular gas mass and $v_{\mathrm{max}}$ (Figure~\ref{fig25}) indicates that the collimation power of the torus influences considerably the physical properties of the outflows. When the collimation power of the torus is high (high $M_{\mathrm{gas}}$) and $\beta>\theta_{\mathrm{out}}$, we observe fast outflows with no rotational component as in NGC~1068. When the collimation power is high and $\beta<\theta_{\mathrm{out}}$ we see relatively fast outflows and rotation as in NGC~3783. In the cases where the collimation power is low and $\beta<\theta_{\mathrm{out}}$, rotation and outflows with low accelerations are present as in NGC~2992 and NGC~6814, and when the collimation power is low and  $\beta>\theta_{\mathrm{out}}$, we only see a slow outflow with no rotational component as in NGC~7469. In addition, as it was discussed in Section~\ref{disks}, 
a small fraction of the NLR/CLR gas rotating in the disks may be attributed to ionizing photons which have travelled through a clumpy distribution of molecular gas. This is probably the case of NGC~4151, in which fast outflows (high collimation power) and rotation are observed but $\beta>\theta_{\mathrm{out}}$.  

D07 showed that there is considerable emission from young stars (with ages in the range 10-300 Myr) within the central few parsecs of an AGN-hosting galaxy. It is thus reasonable to consider that the kinematics of the NLR/CLR can be affected by these (post-)starbursts at radii approaching a few tens of parsecs from the central engine. 
Models of starburst–-driven galactic outflows whose dynamics depends
on both radiation and thermal pressure from supernova (SN) explosions have been succesful in explaining large-scale outflows of ionized gas in starburst and ultraluminous galaxies \citep{nath09}. Using models that synthesize the evolution of populations of massive stars we can estimate the mass and energy returned from starbursts (e.g. Leitherer et al. 1999, Starburst99). 
In the solar-metallicity case where the mass-loss rate and mechanical luminosity are constant beyond $\sim40$ Myr and scale with SFR,
$\dot{E} = 7.0 \times 10^{41} (SFR/M_\odot$ yr$^{-1})$ erg s$^{-1}$ \citep{veilleux05}. 
We note that for instantaneous star formation the mechanical luminosity is approximately constant with a value $\sim 10^{40}$ erg s$^{-1}$. For a typical SFR of $1$ M$_\odot yr^{-1}$ in the central 200 pc of these galaxies (D07), the estimated injection of energy from the post-starbursts is in the range $10^{40}-7 \times 10^{41}$ erg s$^{-1}$, which is on the order of the estimated kinetic power of the observed outflows in four galaxies (see Table~\ref{table8}), indicating that they are in the limit of kinetic power to be able to drive the outflows. Note that in actively starforming galaxies the measured SFRs are $>5$ M$_\odot $yr$^{-1}$. 

Two considerations argue against starburst-driven outflows: timescale and morphology. The dynamical timescales of the outflows ($t_{\mathrm{dyn}}= r_t/v_{\mathrm{max}}$) range from 0.1 to 4 Myr (see Tables~\ref{table6} and \ref{table7}). Even considering a longer active phase of the outflow, tipically 10 Myr, these timescales are inconsistent with the derived starburst ages but in agreement with the typical timescale of the active phase of an AGN. Regarding their morphology, starburst-driven outflows observed in nearby starburst galaxies (M82: Westmoquette et al. 2009, NGC~253: M\"uller-S\'anchez et al. in prep.) have a different geometry to the one observed in Seyfert galaxies, which consists basically of truncated hollow cones. This reflects the fact that they originate in the central hundred pc of the starburst galaxy. As the geometry of the outflows in our sample Seyfert galaxies is a bicone having a sharp apex, the collimation is occuring on very small spatial scales, supporting an AGN origin. We therefore conclude that if the central post-starbursts have a contribution on the acceleration of ionized gas in Seyfert galaxies, these contributions are modest, being the AGN the major driving mechanism. 

\subsection{Mass outflow rates and energetics}\label{feedback}

Recent simulations of mergers and quasar evolution have invoked AGN induced
outflows to explain the correlations between the masses
of black holes and the galaxy bulge properties (e.g. Silk \& Rees
1998; di Matteo et al. 2005). The majority of these models require that a significant fraction of the radiated luminosity $L_{\mathrm{bol}}$ of the AGN couples thermodynamically to the surrounding gas 
($\sim5\%$, di Matteo et al. 2005), although recent work by \citet{hopkins10} demonstrate that a two-stage feedback model, similar to the scenario proposed in Section~\ref{origin}, requires a much lower energy injection ($\sim0.5\% L_{\mathrm{bol}}$)   
All current simulations have been developed for galaxies in which the black hole (BH) is growing rapidly via merger-induced accretion, i.e. the feedback is driven by quasars or radio galaxies. Seyfert galaxies are generally less luminous than quasars and they are likely to be fuelled via secular processes associated with disk evolution \citep{davies10}. 
Although the mechanism responsible for triggering accretion on to massive BHs in Seyfert galaxies and quasars may be different, the two lead to the formation of an AGN. The subsequent physical processes associated to the AGN phenomenon are expected to be similar in local (Seyfert) and more distant (quasars) galaxies. 
Therefore, it is important to investigate whether the properties of the ionized gas outflows in these local AGNs are consistent with the assumptions of the quasar
feedback models.

The feedback from the AGN –- in the form of the mass
outflow rate ($\dot{M}_{\mathrm{out}}$) and its kinetic power ($\dot{E}_{\mathrm{out}}$) –- can be estimated using the ionized gas properties and geometry of the outflow. 
We concentrate on CLR gas, in contrast to previous studies in which these quantities were estimated using NLR lines (e.g. Holt et al. 2006; Storchi-Bergmann et al. 2010), because the outflow properties can be better constrained in the kinematics of the high-ionization species (see Figure~\ref{fig10} and Section~\ref{model}), 
but remark that the analysis also applies for the NLR gas as the geometry and kinematics of the outflow are very similar in both regions, except for NGC~7469 in which only the CLR is in an outflow (see Table~\ref{table7a}). 


For a steady-state outflow, the mass outflow rate is given by
\begin{equation}
\dot{M}_{\mathrm{out}} = 2\, m_\mathrm{p}\, N_e\, v_{\mathrm{max}}\, A\, f \nonumber
\end{equation} 
where $m_\mathrm{p}$ is the proton mass, $N_e$ is the ionized gas density, $A$ is the lateral surface area of the outflowing region (one cone) and $f$ is the filling factor. The factor 2 is introduced to take into account the bicone geometry. 
We assumed typical values of $N_e=5000$ cm$^{-3}$ and $f=0.001$, based on previous studies of the CLR which indicate $10^3 < N_e < 10^4$ cm$^{-3}$ and $f<0.01$ \citep{morwood96, oliva97, rodriguez06, schneider06}. It is important to point out that since the critical density of the CLR is $2-3$ orders of magnitude higher than that of the typical NLR gas (Fergusson et al. 1997; Rodr\'iguez-Ardilla et al. 2011), $N_e$ could be up to 200 times higher. However, since $N_e \propto f^{-1/2}$ \citep{oliva97}, the filling factor would be very much reduced in those cases. 
If this calculation was performed for the NLR, $N_e$ would be smaller $10^2 < N_e < 10^3$ cm$^{-3}$ and $f$ would be larger $0.01 < f < 0.1$ \citep{taylor03, bennert06, storchi09}. 
$A$ and $v_{\mathrm{max}}$ are obtained directly from our kinematic models, both measured at the position of $r_t$. Therefore, we are confident that the uncertainties in these two last parameters (the measurements errors) are negligible compared to those in the assumed values of gas density and filling factor.



The results for the sample galaxies are presented in Table~\ref{table8}. Our calculations yield mass outflow rates ranging from 1 to 10 M$_\odot$ yr$^{-1}$, except for NGC~2992 which has $\dot{M}_{\mathrm{out}} \approx$ 120 M$_\odot$ yr$^{-1}$. Due to uncertainties in the assumed parameters ($N_e$ and $f$) the derived mass outflow rates are probably accurate to no better than about $\pm0.5$ dex. 
The high $\dot{M}_{\mathrm{out}}$ in NGC~2992 reflects the fact that this galaxy has the largest lateral surface area of the outflow. However, it is unlikely that the radius of the outflow region has been overestimated ($r_t=900$ pc) since it is consistent with the one of the figure-of-8 observed in radio observations \citep{ulvestad84} and shorter than the one measured in morphological and spectroscopic studies of H$\alpha$ and [O~{\sc iii}] emission \citep{veilleux01, garcialorenzo01}. If anything, the outflowing region may be slightly larger. Therefore, the outflow in this galaxy presents different characteristics to the ones observed in the rest of the sample galaxies, probably having an hybrid origin (stellar and the AGN, see Appendix~\ref{2992}). In fact, the estimated $\dot{M}_{\mathrm{out}}$ in this galaxy is perfectly consistent with those detected in powerful starburst sources such as ULIRGs (10 to 1000 M$_\odot$ yr$^{-1}$, Veilleux et al. 2005). 

These values can be compared with those reported in the literature.  
\citet{crenshaw07} obtain $\dot{M}_{\mathrm{out}}\approx 0.16$ M$_\odot$ yr$^{-1}$ 
for NGC~4151 using blueshifted absorption lines in the UV spectrum
of this galaxy. However, \citet{storchi09} report $\dot{M}_{\mathrm{out}}\approx 2.4$ M$_\odot$ yr$^{-1}$ for NGC~4151 based on AO-assisted near-IR integral-field observations obtained with the instrument NIFS at the Gemini North Telescope. \citet{veilleux05} report
values of $\dot{M}_{\mathrm{out}}$ calculated from warm ionized gas masses
in the NLR in the range 0.1 to 10 M$_\odot$ yr$^{-1}$. However, \citet{barbosa09} obtained mass outflow rates 10-100 times smaller for a sample of Seyfert galaxies observed with seeing-limited integral field spectroscopy. 

We can also compare the measured outflow rates with the mass accretion rates $\dot{M}_{\mathrm{acc}}$ necessary to feed the central massive black holes in the galaxies of the sample. The mass accretion rate at scales down to one Schwarzschild radius $R_S$ can be estimated from the mass-to-luminosity conversion efficiency of a black hole $L_{\mathrm{bol}}=\eta\,c^2\,\dot{M}_{\mathrm{acc}}$, where $\eta$ is the accretion efficiency and $L_{\mathrm{bol}}$ is the bolometric luminosity of the AGN. A typical accretion efficiency value for a standard geometrically thin accretion disk 
is $\eta=0.1$ \citep{shakura73}. The values of $L_{\mathrm{bol}}$ and the resulting mass accretion rates are listed in Table~\ref{table8}. We can conclude that, on average, there is $\sim10^2-10^3$ times more mass outflowing from the nuclear region than being accreted into the black hole. A larger outflow rate can be understood if the
outflowing gas does not originate in the nucleus, but is actually gas
from the circumnuclear ISM being pushed away by a nuclear wind, consistent with the two-stage feedback model. 
\citet{veilleux05} and \citet{storchi09} report similar ratios between mass outflow rates and nuclear accretion rates in AGN. 

In order to estimate the impact of the ionized gas outflow on the circumnuclear gas in the bulge of the host galaxy, it is also important to estimate its kinetic power and compare it with the accretion power ($L_{\mathrm{bol}}$). Assuming that the relatively large linewdith of
the outflowing gas reflects a turbulent motion that is present at all
locations in the outflow region, the kinetic power is
$\dot{E}_{\mathrm{out}}=1/2\,\dot{M}_{\mathrm{out}}\,(v_{\mathrm{max}}^2+\sigma^2)$. 
The resulting values are also listed in Table~\ref{table8}. The ratios between the estimated kinetic power and $L_{\mathrm{bol}}$ are in the range $10^{-4} < \dot{E}_{\mathrm{out}}/L_{\mathrm{bol}} < 5\times10^{-2}$.
We note that $\dot{E}_{\mathrm{out}}$ in half of the sample galaxies exhibiting radial outflow (NGC~1068, NGC~2992 and NGC~4151) fulfill the requirements of the two-stage feedback models to be thermally coupled to the circumnuclear gas ($\dot{E}_{\mathrm{out}}\approx 0.005L_{\mathrm{bol}}$). 
Furthermore, the fact that none of the measured parameters appear to systematically depend on the AGN luminosity (and therefore on the accretion rate), supports the hypothesis that the outflow is not a single wind originating in the AGN, but is actually a product of the interaction of an already accelerated wind with its ambient medium.
It is also likely that the kinetic energy of the outflow in these three objects blows away the molecular gas reservoir in the central 500 pc necessary to trigger AGN and starformation activity. Assuming that the typical lifetime of the outflow is $10^7$ yr, the total kinetic energy injected by the outflows to the ISM ranges between $3\times 10^{56}$ to $10^{57}$ erg. Assuming that these galaxies have a total mass $M_{\mathrm{total}}\sim10^{10}$ M$_\odot$ and a typical gas mass M$_{\mathrm{gas}}\sim5\times10{^8}$ contained within a radius of $500$ pc (Schinnerer et al. 2000; Garc\'ia-Burillo et al. 2005; Storchi-Bergmann et al. 2010; Friedrich et al. 2010), the gravitational binding energy of the gas is $E_{\mathrm{bind}}\approx G M_{\mathrm{total}} M_{\mathrm{gas}}/r\approx5\times10^{56}$ erg. Thus, the outflows may remove a considerable amount of warm/cool gas from the central regions of the host galaxy.     

The estimated kinetic power in the three remaining galaxies exhibiting outflows of coronal gas (NGC 3783, NGC 6814 and NGC 7469) is at least one order of magnitude less than might be expected on the basis of the quasar feedback models, even considering a two-stage feedback model. Furthermore, these outflows are also incapable of removing a significant fraction of the warm/cool gas from the central region since their kinetic energy is less than $3\times10^{55}$ erg. 
As can be seen in Table~\ref{table8}, and following the discussion presented in Section~\ref{acceleration}, our results suggest the existence of two types of outflows: fast and collimated with a relatively high $\dot{E}_{\mathrm{out}}$, and slow and less collimated with a much smaller $\dot{E}_{\mathrm{out}}$. Note that the outflow in NGC~2992 does not fit in these scenarios, since it presents characteristics similar to those observed in ULIRGs as discussed above. 
The two galaxies with the highest $\dot{E}_{\mathrm{out}}/L_{\mathrm{bol}}$ (NGC 1068 and NGC 4151) are also the only objects in which a clear interaction of the surrounding medium with the radio jet is observed. This may suggest that Seyfert galaxies with strong and collimated radio emission are also hosts of powerful outflows of ionized gas. 
Further radio and near-IR integral-field observations with high spatial resoultion would be required to confirm this hypothesis and
to gauge the real impact of ionized gas outflows on the evolution of Seyfert galaxies.  

\section{Summary and Conclusions} \label{conclusions}

SINFONI and OSIRIS near-IR AO-assisted integral field spectroscopy has been used in this work to probe the kinematics of the NLR and CLR in nearby AGNs. 
Thanks to our high S/N integral field data we were able to extract the 2D distribution and kinematics of Br$\gamma$ and the high-ionization line [Si~{\sc vi}] in the central $<300$ pc with spatial resolutions ranging from 4 to 36 pc. Our main results and conclusions can be summarized as follows:

\begin{itemize}
	\item In all cases the NLR and CLR are resolved (except in NGC~3783), and their morphologies are similar, indicating that they are being produced by the same ionizing source. The two regions present a bright compact core and extended emission. While the size of the NLR seems to be larger than our FOV in some objects, the CLR is more compact and could be well measured in the sample galaxies (except in Circinus) with sizes ranging from $r=80$ pc in NGC~6814 to $r=150$ pc in NGC~1068. 
	\item The kinematics of the NLR at these scales can be put into three groups: (a) velocity fields dominated by rotation, (b) disturbed rotational patterns, and (c) velocity fields dominated by non-circular motions. For the galaxies with velocity fields dominated by rotation, the NLR and stellar velocity fields are similar and indicate that the majority of the NLR gas is distributed in a disk.
	\item The velocity and dispersion maps of the NLR and CLR in the sample galaxies are similar, with a trend for the low-ionization gas to be a better tracer of the disk component, while the high-ionization gas presents more deviant behavior with velocity fields dominated by non-circular motions.  
	\item Several pieces of evidence suggest that the non-circular motions observed in the velocity fields of the NLR and CLR correspond to outflows. The main argument favouring this interpretation is the evidence for radial acceleration to a projected distance of $\sim100$ pc (followed by deceleration in some cases) observed in the velocity channel maps and the radial velocity maps. Other evidences are: (1) For the galaxies with existing radio images, the velocity channel maps reveal blueshifts in the direction of the radio jet pointing towards us and redshifts in the direction of the jet pointing away from us, (2) the velocity dispersion clearly increases at the nucleus and/or along the directions of the blueshifts and redshifts and (3) the non-circular velocities are too high to be explained by any sort of reasonable gravitational potential.
	\item Kinematic models consisting of rotation and/or biconical outflow have been successfully applied to the data. The models provide a good match to the overall velocity patterns, despite slight local variations in the velocity fields.
	\item Regarding the rotational component, the data clearly shows that the disks of low- and high-ionized gas are spatially coincident with the inner disks of molecular gas, suggesting that the three types of gas are spatially mixed. Since  
high excitation emission is observed outside the borders of the bicone, the torus must be a clumpy medium such that a fraction of ionizing photons escape in different directions and ionize the gas in the disk.
	\item The maximum velocity of the outflowing component ranges between 120 km s$^{-1}$ in NGC~6814 to 1900 km s$^{-1}$ in NGC~1068, and it is reached at a typical distance of $\sim180$ pc from the AGN. We observe two types of outflows based on their opening angle: collimated with full opening angles $<70\degr$ and less collimated with full opening angles $>90\degr$.
	\item We find an anticorrelation between the outflow velocity ($v_{\mathrm{max}}$) and the half-opening angle of the bicone ($\theta_{\mathrm{out}}$), and another anticorrelation between $\theta_{\mathrm{out}}$ and the molecular gas mass ($M_{\mathrm{gas}}$) in $r<30$ pc. These anticorrelations suggest that galaxies with greater amount of gas in their centers have higher degree of collimation (small $\theta_{\mathrm{out}}$) and higher outflow velocities. This indicates that the torus invoked by unified models is indeed the structure responsible for the collimation of the outflows on small spatial scales and that the accumulation of gas clouds towards the nucleus (higher $M_{\mathrm{gas}}$) increments its collimation power and produces outflows with higher velocities (stops the outflow expanding sideways). 
	\item	The observed bipolar outflows are consistent with the torus model of AGN (except in NGC~3783). 
The five intermediate type Seyfert galaxies studied here (four Sy 1.5s and one Sy 1.9) are dominated by blueshifts on one side and redshifts on the other side of the AGN, consistent with an intermediate viewing angle, whereas the Seyfert 2 galaxy NGC 1068 shows blueshifts and redshifts on either side of the AGN, consistent with the unified model prediction of a bicone axis near the plane of the sky. Considering the uncertainties in the kinematic modeling, and possible misalignments in the torus-bicone system, it is likely that we have a direct view to the BLR in most of the intermediate type Seyferts (except in NGC~3783). We conclude that the majority of these objects (and possibly most Seyfert 1 galaxies) are actually seen close to the edge of the cone, rather than down the middle. Alternatively, a clumpy torus may explain the presence of broad lines even if the observer is not looking into the bicone, which is likely the case in NGC~3783.  
	\item We have estimated mass outflow rates in the range 1 to 10 M$_\odot$ yr$^{-1}$, which are $\sim10^2-10^3$ times the accretion rate necessary to feed the AGN. The estimated kinetic powers of the outflows are $10^2-10^4$ times smaller than the AGN bolometric luminosities. 
We note that $\dot{E}_{\mathrm{out}}$ in half of the galaxies presenting outflows (NGC~1068, NGC~2992 and NGC~4151) fulfill the requirements of two-stage feedback models to be thermally coupled to the circumnuclear gas ($\dot{E}_{\mathrm{out}}\approx 0.005L_{\mathrm{bol}}$), and match the $M_{BH}-\sigma^*$ relation. 
The two galaxies with the highest $\dot{E}_{\mathrm{out}}/L_{\mathrm{bol}}$ (NGC 1068 and NGC 4151) are also the only objects in which a clear interaction of the surrounding medium with the radio jet is observed. This may suggest that Seyfert galaxies with strong and collimated radio emission are also hosts of powerful outflows of ionized gas. 
	\item The high velocity components detected at the nucleus of some galaxies and the high $\dot{M}_{\mathrm{out}}/\dot{M}_{\mathrm{acc}}$ ratios indicate that the outflowing gas is not likely a single wind from the AGN, but the result of a two-stage process in which the circumnuclear ISM has been pushed away by an already accelerated AGN wind. The radio jet may also have a contribution to the formation of the outflows at small spatial scales, transfering momentum to the circumnuclear gas.    
\end{itemize}





\acknowledgments

The authors thank Jim Lyke and Randy Campbell for their support at the W. M. Keck Obsetvatory and all those who assisted in the observations at the ESO/VLT. We also thank the referee for a careful and critical reading of the manuscript. 




Facilities: \facility{Keck:II (OSIRIS)}, \facility{VLT:Yepun(SINFONI)}.



\appendix

\section{Description of NLR and CLR gas in individual objects} 

In this Appendix, we discuss the emission-line properties of each object, and give a brief review of some data in the literature. For each galaxy, we discuss the structure and kinematics of the NLR and CLR, and point out relations of these structures with the radio jet and [O~{\sc iii}] emission.

\subsection{Circinus} \label{circinus}

Circinus displays several characteristics of a typical Seyfert 2 nucleus: the observed line ratios (Oliva et al. 1994), an ionization cone observed in [O~{\sc iii}] (Marconi et al. 1994) with the corresponding countercone appearing in the [Si~{\sc vii}] line (Prieto et al. 2005), narrow prominent coronal lines in its optical/near-IR spectrum (Oliva et al. 1994; Maiolino et al. 2000), and broad (FWHM$>3300$ km s$^{-1}$) H$\alpha$ emission detected in polarized light (Oliva et al. 1998).

Br$\gamma$ and the coronal lines present a relatively symmetric flux distribution extending out to a radius of 8 pc along a PA of $0\degr$. Note that the complete extension of the NLR and CLR is not accesible in our observations due to the limited FOV of SINFONI. At larger scales, Circinus shows low- and high-ionized gas extending along the minor axis of the galaxy out to $r\sim30$ pc, with a morphology that is reminiscent of the ionization cones seen in other Seyfert galaxies (Marconi et al. 1994, Prieto et al. 2005). 

The Br$\gamma$ kinematics in the central 16 pc are consistent with ordered rotation in a disk with a PA of $18\degr$ and an inclination of $65\degr$. Within the errors, these parameters agree with the large-scale galactic disk (Freeman et al. 1977). As already pointed out by MS06, this suggests that a considerable amount of Br$\gamma$ in this galaxy must have a stellar origin. This is supported by the fact that there exist offsets between the photometric centers at different wavelengths and between the peaks of several emission lines (Marconi et al. 1994; MS06), indicative of the presence of central super star clusters (MS06; M\"uller-S\'anchez et al. 2010). The velocity field of [Si~{\sc vi}] is also relatively regular, with a best-fit PA of $0\degr$ and inclination of $57\degr$. However, the best-fit PA to the [Si~{\sc vi}] velocity map is not consistent with that of H$_2$ or Br$\gamma$. Furthermore, Circinus exhibits asymmetric and broadened spectral profiles, which can be well fitted by a strong narrow component and a weaker blueshifted broad component. These two facts are indicative of outflows. As the velocity field is affected by artifacts at $\sim0.25$ and $-0.25\arcsec$ caused probably by illumination effects on the SINFONI detector, we decided not to perform a kinematic modeling of the CLR in this galaxy. A study of the CLR kinematics in Circinus, based on integrated spectra, is presented in MS06.

\subsection{NGC~1068} \label{1068}

NGC~1068 has the most extensively studied NLR in the literature.
It was the first one to be shown to have a conically shaped NLR
(Pogge 1988a). The $HST$ [O~{\sc iii}] images of this galaxy were discussed
in detail by Evans et al. (1991), Capetti et al. (1997) and Bruhweiler et
al. (2001). These images show that most of the emission originates in a V-shaped region with opening angle of $50\degr$, extending for approximately 750
pc along PA $35\degr$, and 430 pc in the direction perpendicular
to the bicone axis. The direction of this emission is
coincident with the radio jet observed by Wilson \& Ulvestad (1982).
The radio emission is extended along PA $\sim33\degr$ on the larger scales (Wilson \& Ulvestad 1982) but is aligned along the N-S
direction in the nuclear region (Gallimore et al. 1996). 
The Br$\gamma$ flux map shown in Figure~\ref{fig3} has the same morphology of the  [O~{\sc iii}] emission in the central 300 pc. 

The coronal lines in this galaxy have also been extensively studied. Ground-based long-slit spectroscopy in the optical and near-IR revealed complex spectral shapes, dominated by extreme broadening in all high-ionization lines (Rodr\'iguez-Ardila et al. 2006). The HST/STIS spectra show several coronal lines in the optical, from 
[Ne~{\sc v}] and [Fe~{\sc viii}] lines, whose ionisation potential are $\sim100$ eV, to the extreme ionisation [S~{\sc xii}] line, with an ionisation potential of 504.7 eV \citep{mazzalay10}. This very high-ionization line has been reported only in NGC~1068 (Kraemer \& Crenshaw 2000). 

The strongest coronal lines in our sample are observed in NGC~1068. The three coronal lines contained in the spectral range of our observations, [Si~{\sc vi}], [Al~{\sc ix}] and [Ca~{\sc viii}], dominate the spectra in the central region. In our observations, the CLR tends to follow the same flux distribution as Br$\gamma$ and [O~{\sc iii}], but is slightly more compact. Both, the NLR and CLR emission are stronger towards the NE suggesting strong extinction towards the SW. The kinematics of the two regions are dominated by non-circular motions that can only be explained in terms of radial outflow. The evidences are the following: (1) the velocity channel maps reveal blueshifts in the direction of the radio jet pointing towards us and redshifts in the direction of the jet pointing away from us, (2) there are clear signatures of acceleration along a PA of $\sim30\degr$, and (3) the velocities are too high to be explained just by gravitational forces. 

The maximum velocity of the outflow $v_{\mathrm{max}}$ is reached at a projected distance of $\sim80$ pc. After this point, the gas appears to decelerate. Due to the limited FOV of SINFONI, we do not see the complete deceleration phase, but it has been observed that it reaches systemic velocity at $z_{\mathrm{max}}=400$ pc (Das et al. 2006).    
The method used to model the kinematics is presented in M\"uller-S\'anchez et al. (in prep.). The kinematics can be well modeled with radial outflow along a biconical geometry. The parameters of the bicone are listed in Table~\ref{table6}.    
Furtermore, the kinematics are consistent with the predictions of the Unified Model for a type 2 object. The inclination of the bicone axis is $9\degr$, nearly perpendicular to the LOS. Based on this value and the external and internal half-opening angles of the cone ($\theta_{\mathrm{out}}=27\degr$, $\theta_{\mathrm{in}}=14\degr$), blueshifted velocities are expected to dominate the velocity vectors in the NE simply because there are more clouds moving towards us in this region. Based on the same argument, redshifted clouds should dominate in the SW part of the galaxy. In addition, redshifted velocities should also be present in the NE corresponding to the back part of the outflow, and blueshifted velocities from the front part of the cone should appear in the SW. These properties are evident in the velocity tomography of NGC~1068 (Figure~\ref{fig11}). This galaxy exhibits the fastest and most collimated outflow of the galaxies in our sample ($\dot{E}_{\mathrm{out}}/L_{\mathrm{bol}}=0.05$) indicating that the energy released by the outflows can be thermally coupled with the ISM.        




\subsection{NGC~2992} \label{2992}

The host galaxy disk is highly inclined at $i\sim65\degr$, which makes it
suitable for extraplanar emission studies. The galaxy has a companion
NGC~2993, which is located $3\arcmin$ SE, connected
with a tidal bridge, and whose tidal forces might have an important
effect on the kinematics. The 6 cm radio emission is
concentrated in the central parts, forming a figure-of-eight structure
at PA$=160\degr$. (Ulvestad \& Wilson 1984). The [O~{\sc iii}] emission is aligned with the galactic minor axis (PA$=120\degr$) and approximately with the radio structure. Multi-component optical emission lines were detected, and the velocity field of H$\alpha$ was suggested to be a superposition of rotation and an outflow (Veilleux et al. 2001).

This galaxy has a compact CLR ($r=100$ pc) but an extended NLR ($r>280$) pc. The two are elongated along a PA$\sim30\degr$. It presents several similarities with the Circinus galaxy which include: the inclination of the galactic disk, a hard nuclear continuum which produces [Si~{\sc vi}], [Al~{\sc ix}] and [Ca~{\sc viii}] emission, the shape of the coronal lines profiles (asymmetric with blue wings, see Figure~\ref{fig1a}), extended  [O~{\sc iii}] and H$\alpha$ emission along a PA$\sim120\degr$ with a morphology that is reminiscent of ionization cones, inconsistency between the best-fit PA of the [Si~{\sc vi}] velocity field (when fitted with kinemetry) and the PA of the H$_2$ inner disk, and the presence of an outflow of molecular gas (MS06, Friedrich et al. 2010). All these facts suggest that although the [Si~{\sc vi}] velocity field appears rather regular, an outflowing component must be part of the kinematics of the NLR and CLR. The Br$\gamma$ and  [Si~{\sc vi}] velocity fields are best fitted with a model incorporating radial outflow plus rotation. The resulting parameters of the model are presented in Table~\ref{table7}. Interestingly, in this galaxy the outflow reaches its maximum velocity of 200 km s$^{-1}$ at $r=900$ pc, the largest distance among the galaxies in our sample. Furthermore, it also has the largest opening angle ($\theta_{\mathrm{out}}=57\degr$). The radius of the accelerating gas in this galaxy is consistent with the one of the figure-of-8 observed in radio observations \citep{ulvestad84} and slightly shorter than the one measured in morphological and spectroscopic studies of H$\alpha$ and  [O~{\sc iii}] emission \citep{veilleux01, garcialorenzo01}.  Therefore, the outflow in this galaxy presents different characteristics to the ones observed in the rest of the sample galaxies. We interpret these properties as evidence for an hybrid origin of the outflow (stellar and the AGN). This interpretation is supported by the fact that NGC~2992 is a combination between an intermediate Seyfert and a starburst galaxy (Glass 1997). 
Furthermore, the outtflow in NGC~2992 is similar to those observed in nearby starburst galaxies, which usually extend to radial distances of kpc (M82: Westmoquette et al. 2009, NGC~253: M\"uller-S\'anchez et al. in prep.). Aditionally, the large $\theta_{\mathrm{out}}$ may be an indication of a truncated cone instead of a cone having a sharp apex. The SFR in this galaxy is estimated to be in the range 2-5 M$_\odot$ yr$^{-1}$ \citep{friedrich10}. Thus, the mechanical power from the stars is $1-4\times10^{42}$ erg s$^{-1}$, which is of the order of the kinetic power of the outflow.

\subsection{NGC~3783} \label{3783}

An unresolved nuclear point source is detected in radio observations (Ulvestad \& Wilson 1984), ground-based observations in the optical (Winge et al. 1992), and $HST$  [O~{\sc iii}] images presented in Schmitt et al. (2003). The total extent of the  [O~{\sc iii}] emission is $r\sim0.6\arcsec$ (130 pc). Our Br$\gamma$ and  [Si~{\sc vi}] images also show an unresolved core but extended emission out to $r=140$ pc along a PA$\approx0\degr$. In $HST$ images, NGC~3783 presents the largest CLR, with more than
400 pc radius in the [Fe~{\sc vii}] line \citep{mazzalay10}. As pointed out by these authors, the compact [O~{\sc iii}] emission could be a result of an oversubtraction of the continuum image removing  [O~{\sc iii}] emission from the outer regions of the halo. 

The Br$\gamma$ morphology and velocity field are similar to that of  [Si~{\sc vi}], and rather different from the H$_2$ 1--0 S(1) (see Figure~\ref{fig10}). The kinematics of the two regions are dominated by non-circular motions consistent with radial outflow. In this case the evidences are the following: (1) the radial velocity maps and the velocity channel maps show redshifted velocities in the north part of the galaxy accelerating away from the nucleus out to $\sim0.35\arcsec$ and then appear to decelerate, (2) the maximum velocity of $\sim200$ km s$^{-1}$ observed at $\sim0.35\arcsec$ is too high to be explained by any reasonable gravitational potential and (3) the velocity dispersion of the ionized gas increases in the direction of the acceleration. A kinematic model incorporating rotation plus radial outflow provides a good fit to the data. 
The parameters of the model are presented in Table~\ref{table6}. The best fit PA$_{\mathrm{disk}}=-15\degr$ and $i_{\mathrm{disk}}=38\degr$ are consistent with those of the H$_2$ kinematics. The half-opening angle of the cone is $\theta_{\mathrm{out}}=34\degr$, $\theta_{\mathrm{in}}=27\degr$ and its axis is inclined $-30\degr$. Therefore the northern cone is almost entirely redshifted and the southern cone is in its majority blueshifted, consistent with an intermediate viewing angle. Note that the blueshifted part of the bicone is in its majority hidden by the plane of the galaxy. The SW side of the galaxy is closer to us, and the southern cone therefore experiences more extinction. This is consistent with the Br$\gamma$ and  [Si~{\sc vi}] images which show weaker, and in some place absent, emission SW of the location of the SMBH (Fig.~\ref{fig12}). The case of NGC~3783 provides strong evidence against the hypothesis that the radio jet is driving the bipolar outflow of ionized gas, since there is no collimated radio emission in the central parsecs of this galaxy \citep{orienti09} so the clouds seem to accelerate even in the absence of any radio material. Furthermore, it is also the only Seyfert galaxy in our sample exhibiting broad lines and a line of sight far outside of the bicone ($\alpha=26\degr$). This gives support to a clumpy torus such that our view to the BLR is almost independent of the inclination and opening angles of the bicone. Within this context, \citet{vollmer09} found that NGC~3783 has the most transparent torus of the galaxies in our sample.





\subsection{NGC~4151} \label{4151}

Pogge (1989) detected [O~{\sc iii}] emission in ground-based images that is extended by $20\arcsec$ along PA$=50\degr$. The [O~{\sc iii}] emission was later confirmed with $HST$ to have a biconical morphology along PA$=55\degr$ (Evans et al. 1993). The total extent of the emission is $5.5\arcsec$ (350 pc). At radio wavelengths a complex structure of knots has been identified within the central few arcseconds along a PA of $77\degr$ (Mundell et al. 2003). 

Br$\gamma$ and  [Si~{\sc vi}] show a marginally resolved core as well as extended
emission out to 80 pc along a PA=$50\degr$. Note that the complete extension of the NLR and CLR is not accesible in our observations due to the limited FOV of OSIRIS.
The Br$\gamma$ velocity field is similar to that of [Si~{\sc vi}], and rather different from the H$_2$ 1-0 S(1). The kinematics of the two regions are dominated by non-circular motions consistent with radial outflow. In this case the evidences for outflow are the following: (1) the radial velocity maps and the velocity channel maps show redshifted velocities in the NE part of the galaxy and blueshifted velocities in the SW, which are accelerating away from the nucleus along a PA$\sim-130\degr$, (2) the redshifted velocities in the NE follow the direction of the radio jet pointing away from us and the blueshifts in the SW are observed in the direction of the jet pointing towards us, (3) the maximum velocity of $\sim400$ km s$^{-1}$ observed at $\sim0.5\arcsec$ is too high to be explained by any reasonable gravitational potential and (4) the velocity dispersion of the ionized gas increases along a PA$\sim-130\degr$ consistent with the extended V-shape morphology observed in [O~{\sc iii}] images. A kinematic model incorporating rotation plus radial outflow provides a good fit to the data. The parameters of the model are presented in Table~\ref{table6}. The underlying rotation has a best fit PA$_{\mathrm{disk}}=175\degr$ and $i_{\mathrm{disk}}=45\degr$, which is consistent with the kinematics of the molecular gas in the nuclear region and roughly consistent with the large scale galactic rotation. The bicone has an external half-opening angle of $29\degr$, an internal angle $\theta_{\mathrm{in}}=15\degr$ and is inclined $i_{\mathrm{bicone}}=-45\degr$. Therefore the NE cone is entirely redshifted and the SW cone is entirely blueshifted, consistent with the prediction of the Unified Model for a type 1.5 object. The misalignment of the NLR and the radio jet in NGC~4151 suggest that the radio jet is not the principal driving force on the outflowing NLR clouds in this galaxy.

\subsection{NGC~6814} \label{6814}

The [O~{\sc iii}] image of this galaxy \citep{schmitt96} exhibits an unresolved core and some extended emission along PA$\sim-15\degr$. The total extent of the emission is $r=0.6\arcsec$ (55 pc). The Br$\gamma$ and [Si~{\sc vi}] emission are consistent with this mophology, but in our OSIRIS data, the core is marginally resolved. This in fact is the most compact CLR in our sample. There is no emission related to the western radio extension found by Ulvestad \& Wilson (1984), but the overall distributions of Br$\gamma$ and  [Si~{\sc vi}] emission in the central $\sim100$ pc seem to follow that of the radio emission (see Figure~\ref{fig15}). 

The spatial resolution of the SINFONI data is estimated from the broad Br$\gamma$ emission to be $\sim0.57\arcsec$ (FWHM), which is $\sim60$ pc at the distance of NGC~6814. The azimuthal averages of the SINFONI observed distributions (Br$\gamma$ and  [Si~{\sc vi}]) have an HWHM of 30 pc consistent with the PSF. The OSIRIS data has a higher spatial resolution of $\sim0.17\arcsec$ (FWHM), equivalent to 18 pc, and for this reason the properties measured from the OSIRIS data are taken to be more accurate, although, the measured and derived properties from both data sets are consistent, particularly the kinematics (see Figures~\ref{fig3b}, \ref{fig4}, \ref{fig6} and \ref{fig7}).  
Both, the Br$\gamma$ and  [Si~{\sc vi}] velocity fields, exhibit a perturbed rotational pattern. The best fit PA and $i$ of the rotational component are consistent with those of the molecular gas kinematics. The velocity residuals map obtained after subtracting the rotational component to the [Si~{\sc vi}] (or Br$\gamma$) velocity map reveal peculiar velocity gradients with blueshifts and redshifts on each side of the nucleus. These approaching and receding regions are also evident in the velocity channel maps (Figure~\ref{fig14}), and indicate that material is accelerating away from the nucleus. Further evidence for outflows comes from the collimated radio emission which is spatially coincident with the coronal gas (see Figure~\ref{fig15}). Assuming that the radio jet is approaching in the south part of the galaxy, the blueshifted velocities seen in [Si~{\sc vi}] in this region would be outflowing. A kinematic model incorporating rotation plus radial outflow provides a good fit to the data. The parameters of the model are presented in Table~\ref{table6}. Note that this model provides a good fit to both, the OSIRIS and the SINFONI datasets. The underlying rotation has a best fit PA$_{\mathrm{disk}}=145\degr$ and $i_{\mathrm{disk}}=50\degr$, which is consistent with the kinematics of the molecular gas in the nuclear region. The bicone has a half-opening angle $\theta_{\mathrm{out}}$ of $47\degr$, $\theta_{\mathrm{in}}=25\degr$ and is inclined $i_{\mathrm{bicone}}=-34\degr$. Therefore the northern cone is almost entirely redshifted and the southern cone is almost entirely blueshifted, consistent with an intermediate viewing angle. Within the uncertainties, a direct view to the BLR is possible.

\subsection{NGC~7469} \label{7469}

A variety of observations at many wavelengths indicate that the active nucleus is surrounded by a $r=1.5\arcsec$ (480 pc) ring of starburst activity (Wilson
et al. 1991; Mauder et al. 1994; Genzel et al. 1995; Soifer et al. 2003; Davies et al. 2004). The [O~{\sc iii}] emission is contained in a $r=2.5\arcsec$ region having an almost circular morphology. 
Within the SINFONI and OSIRIS FOV, which fall inside the ring, Br$\gamma$ and highly ionized emission line gas, [Si~{\sc vi}] and [Ca~{\sc viii}], are detected.
In addition to the diffuse radio emission associated with the nuclear
starburst ring, an E-W extended nuclear source
is also detected (Thean et al. 2001), which, at high spatial
resolution, separates into five sources along an E-W
direction within $0.1\arcsec$ of the nucleus (Lonsdale et al.
2003). There is no indication in the Br$\gamma$ flux distribution of a counterpart to this radio emission. However, the  [Si~{\sc vi}] emission seems to follow the direction of the nuclear radio sources. [Si~{\sc vi}] is extended along a PA=$90\degr$ out to $r=90$ pc. The OSIRIS and SINFONI data are consistent with these measurements. Although the OSIRIS data have a slightly better spatial resolution than the SINFONI data of this galaxy, we decided to perform the kinematic analysis using the SINFONI datacube, because its FOV covers better the region where the CLR is extended (E-W).    

The Br$\gamma$ emission detected in NGC~7469 is nearly symmetric, extends out to $r=125$ pc and peaks at $\sim0.1\arcsec$ (32 pc) west from the AGN location.  [Si~{\sc vi}] presents a similar morphology, but it is slightly more extended in the E-W and peaks $\sim32$ pc East from the AGN. These offsets may indicate that the Br$\gamma$ emission is associated with star formation rather than the AGN. This is supported by the facts that the Br$\gamma$ velocity dispersion is rather low in the central region and its velocity field is dominated by rotation. 
The velocity field of Br$\gamma$ in the SINFONI and OSIRIS datasets is well fitted by a rotating disk with PA$=143\degr$ and $i=45\degr$. These values are consistent with the PA and $i$ of the molecular gas disk. The [Si~{\sc vi}] velocity field, however, is rather different from the Br$\gamma$ kinematics. It is dominated by a non-circular component exhibiting increments of velocity as a function of distance in the E-W direction. The magnitudes of the velocities reach a maximum at $r\sim0.16\arcsec$ and then appear to decrement. These are signatures of acceleration followed by deceleration and therefore indicate that the CLR is in an outflow. This phenomenon is observed in both, the OSIRIS and SINFONI datasets (see Figures~\ref{fig3b}, \ref{fig4}, \ref{fig6} and \ref{fig7}). We performed a kinematic modeling of the CLR including rotation and radial outflow and found that the kinematics are best fitted with just an outflowing component. This can be clearly seen in the lower panels of Figure \ref{fig20}, where we show the LOS velocities inside the bicone model as it would appear in the plane of the sky. The velocities outside this region do not show obvious regular patterns. The model parameters are presented in Table~\ref{table6}. The half-opening angle of the bicone is $\theta_{\mathrm{out}}=45\degr$,the internal opening angle is $\theta_{\mathrm{in}}= 28$ and its axis is inclined $-33\degr$. Therefore the eastern cone is almost entirely redshifted and the western cone is in its majority blueshifted, consistent with an intermediate viewing angle. As in the rest of intermediate type Seyferts of our sample (except NGC~3783), within the uncertainties, a direct view to the BLR is possible. 

\clearpage



\clearpage

\clearpage

\begin{figure}
\epsscale{.99}
\plotone{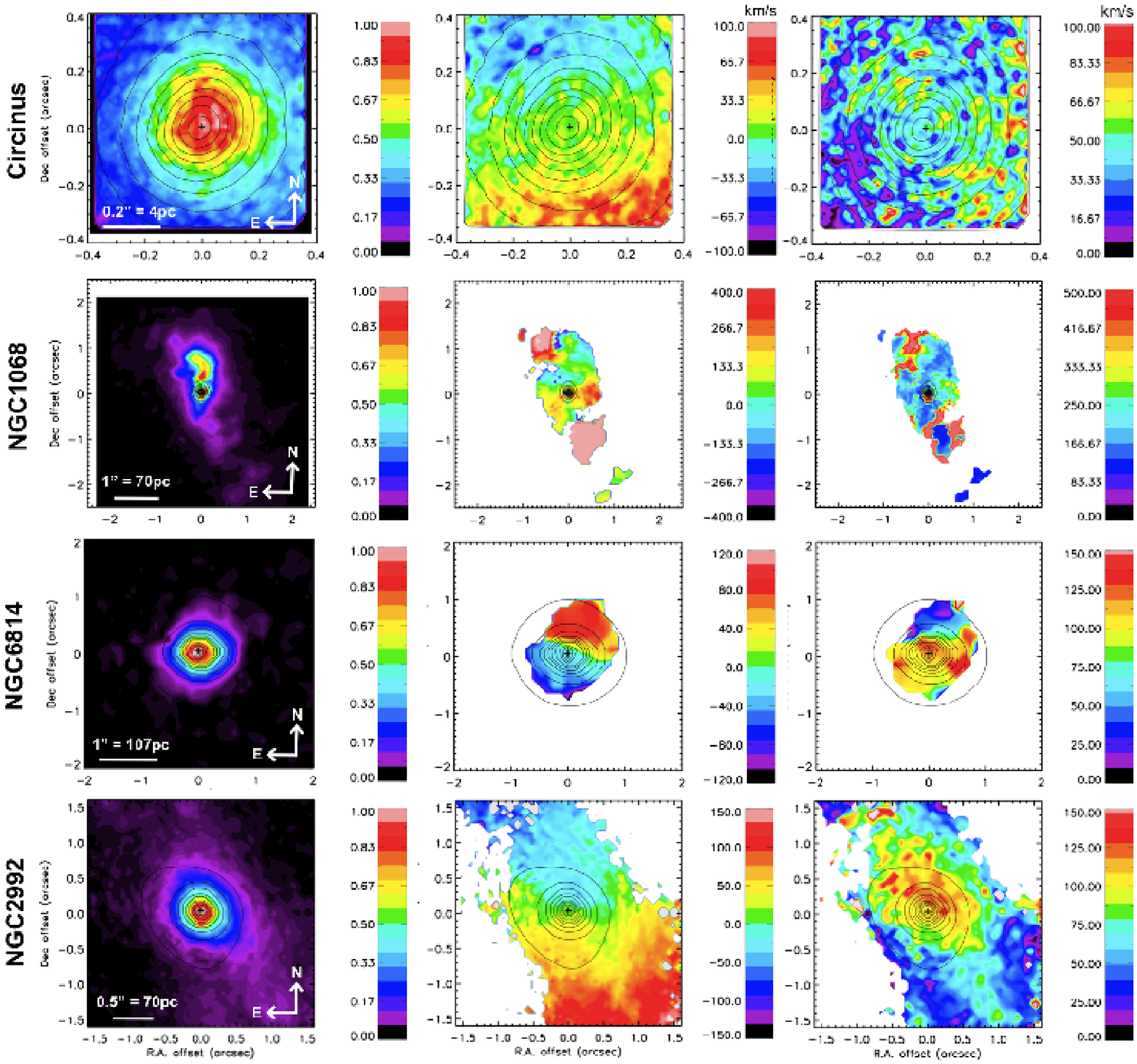}
\caption{Two-dimensional maps of the Br$\gamma$ flux distribution, velocity, and velocity dispersion (from left to right), for each galaxy observed with SINFONI. The galaxies are sorted by distance. The contours delineate the $K-$band continuum emission and the position of the AGN (the peak of non-stellar continuum emission at $2.2\mu$m) is marked with a cross. 
Regions in white correspond to pixels where the line properties are uncertain and thus were masked out. These rejected pixels in the velocity and dispersion maps are those with a flux density lower than $5\%$ of the peak of Br$\gamma$ emission.
\label{fig3}}
\end{figure}

\clearpage

\begin{figure}
\epsscale{.99}
\plotone{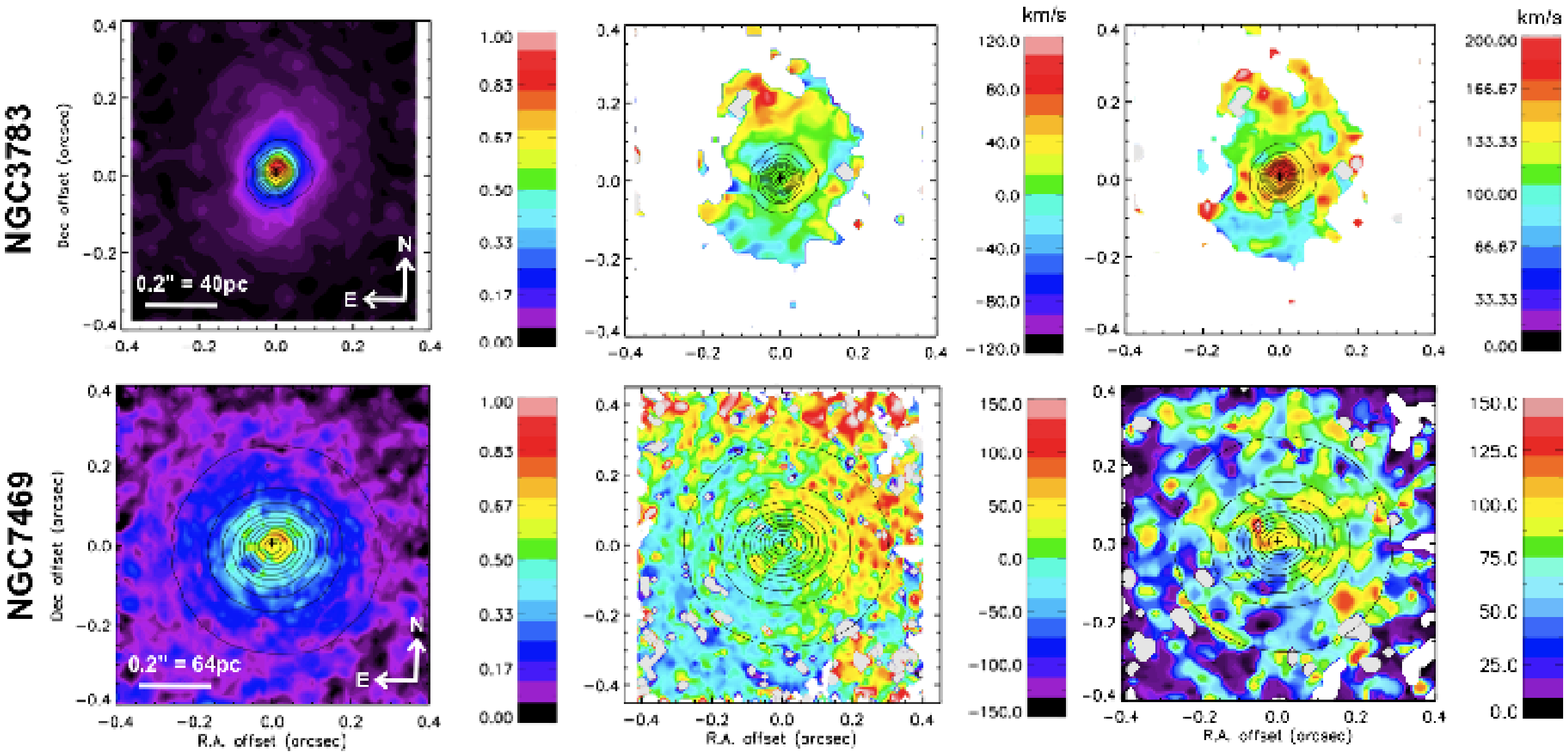}
\caption{Continuation of Figure~\ref{fig3}.
\label{fig3b}}
\end{figure}

\clearpage

\begin{figure}
\epsscale{.99}
\plotone{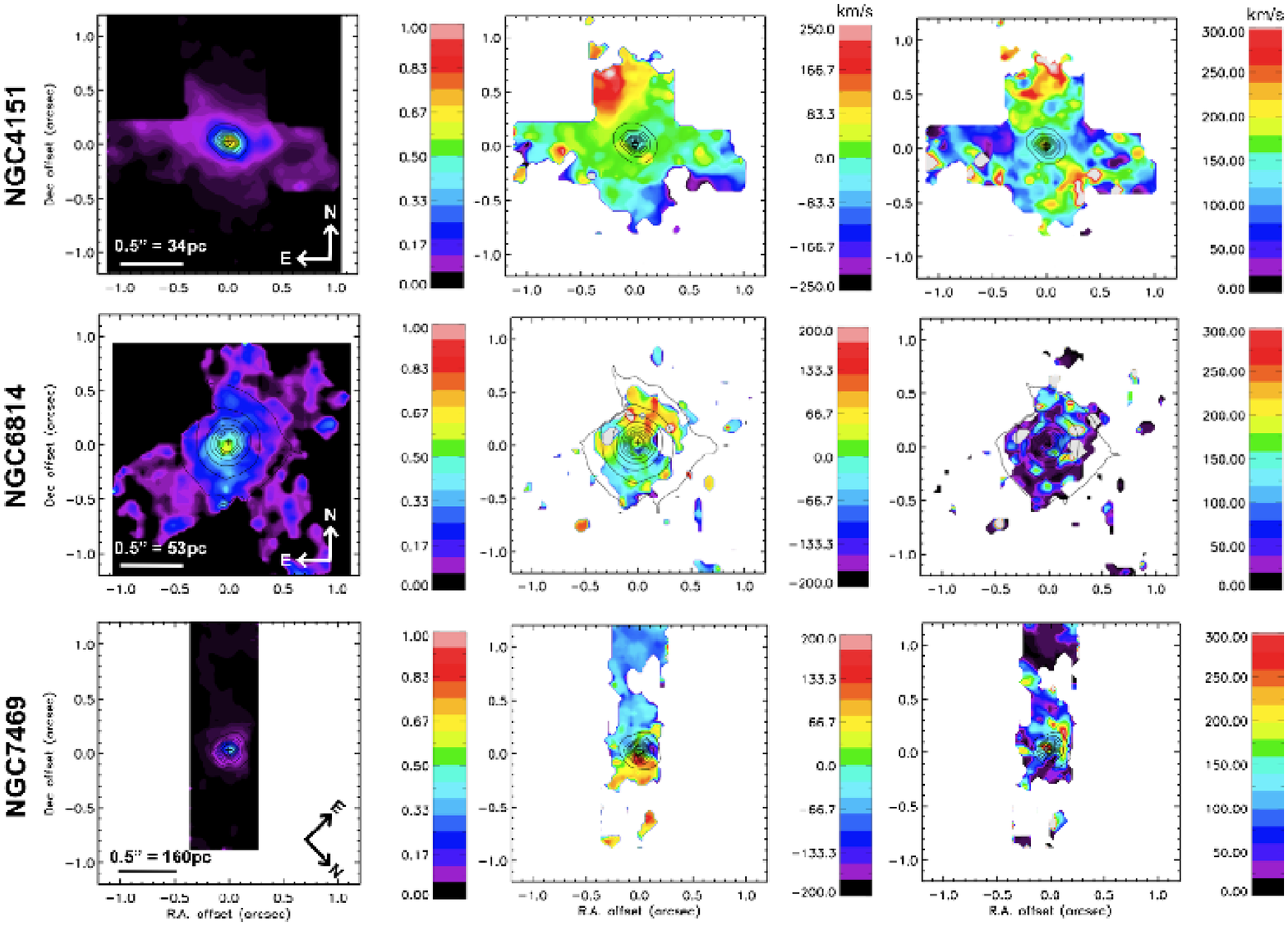}
\caption{Same as Figure~\ref{fig3}, but for the galaxies observed with OSIRIS. 
\label{fig4}}
\end{figure}

\clearpage

\begin{figure}
\epsscale{.99}
\plotone{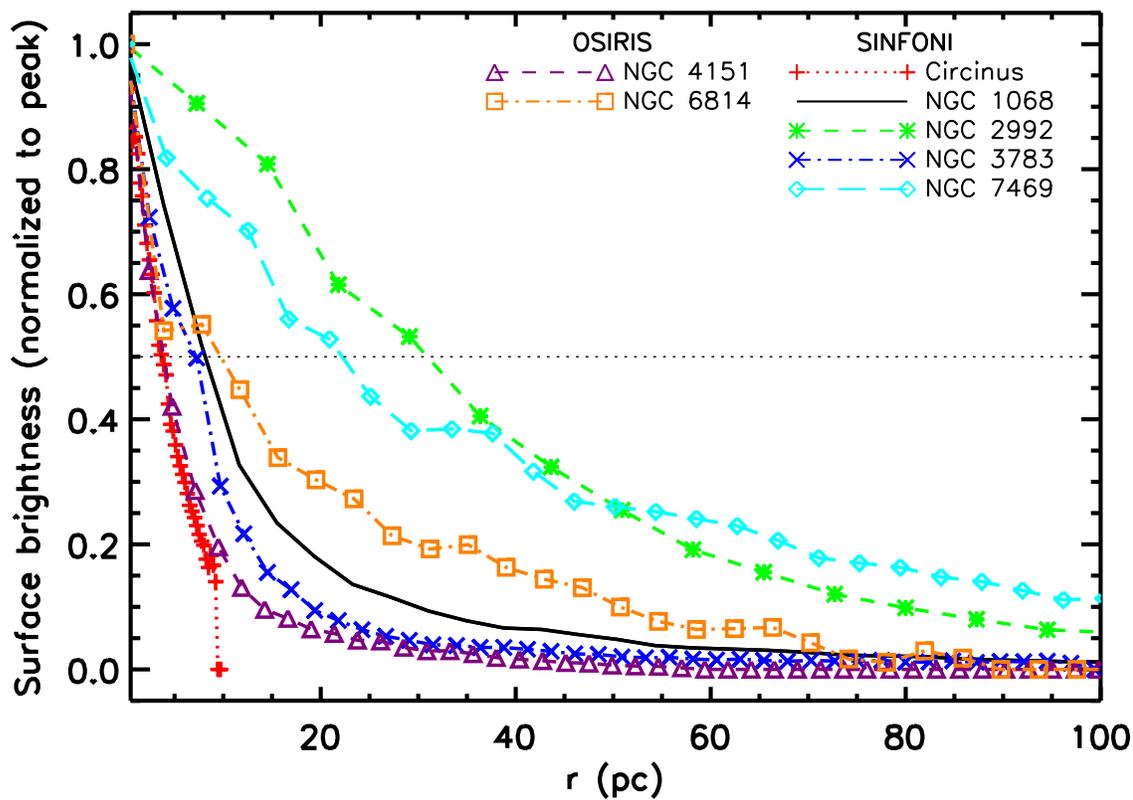}
\caption{Azimuthally-averaged surface-brightness radial profiles of the Br$\gamma$ flux distribution for the galaxies observed with SINFONI and OSIRIS as indicated in the legend. The horizontal dashed line indicates the HWHM$_{\mathrm{Br\gamma}}$. Tipically, the standard deviation of the azimuthally averaged flux is $5\%$. 
\label{fig5}}
\end{figure}

\clearpage

\begin{figure}
\epsscale{.99}
\plotone{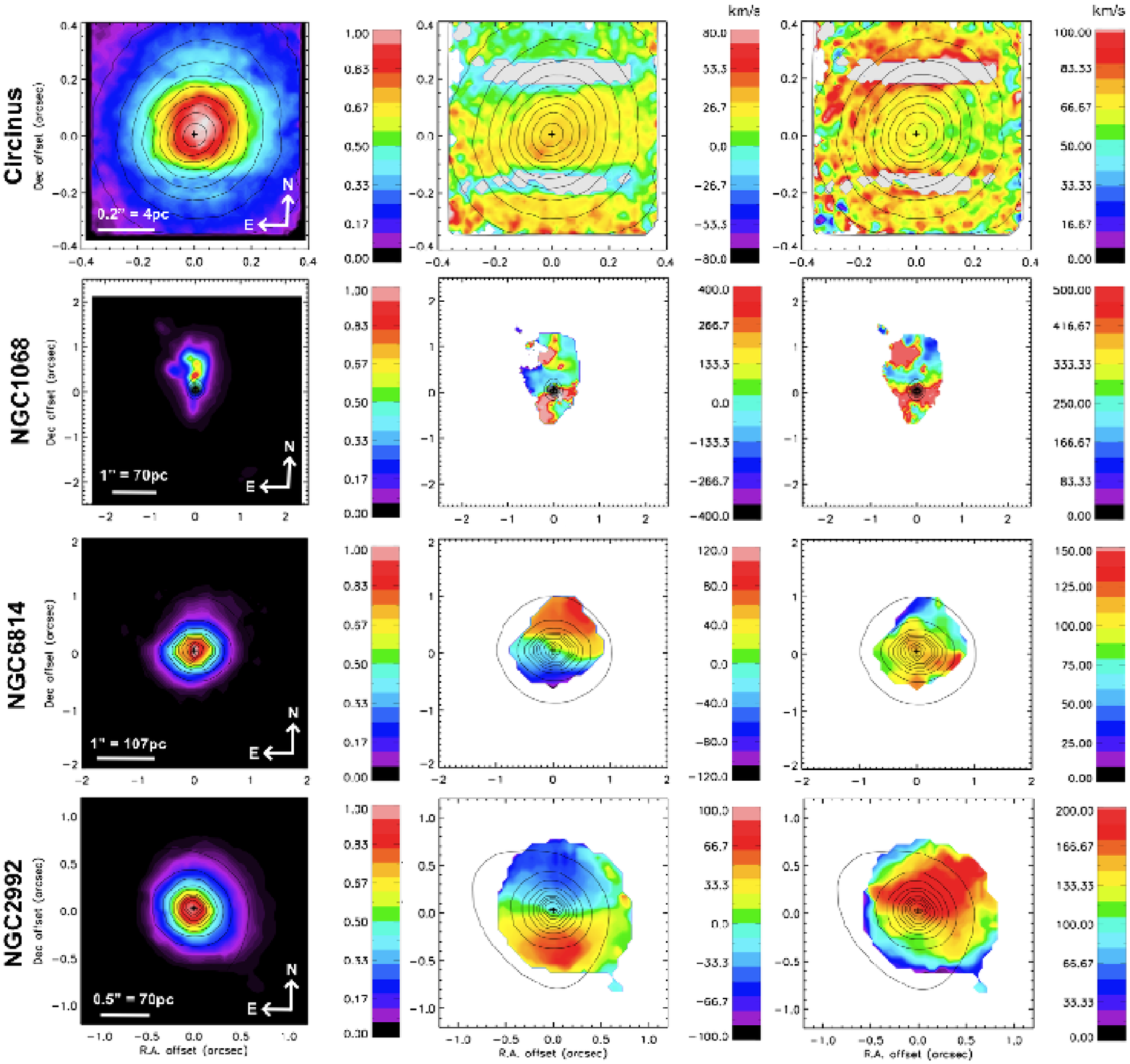}
\caption{Two-dimensional maps of the [Si~{\sc vi}] flux distribution, velocity, and velocity dispersion (from left to right), for each galaxy observed with SINFONI. 
The galaxies are sorted by distance. The contours delineate the $K-$band continuum emission and the position of the AGN (the peak of non-stellar continuum emission at $2.2\mu$m) is marked with a cross. 
Regions in white correspond to pixels where the line properties are uncertain and thus were masked out. These rejected pixels in the velocity and dispersion maps are those with a flux density lower than $5\%$ of the peak of [Si~{\sc vi}] emission.
\label{fig6}}
\end{figure}

\clearpage

\begin{figure}
\epsscale{.99}
\plotone{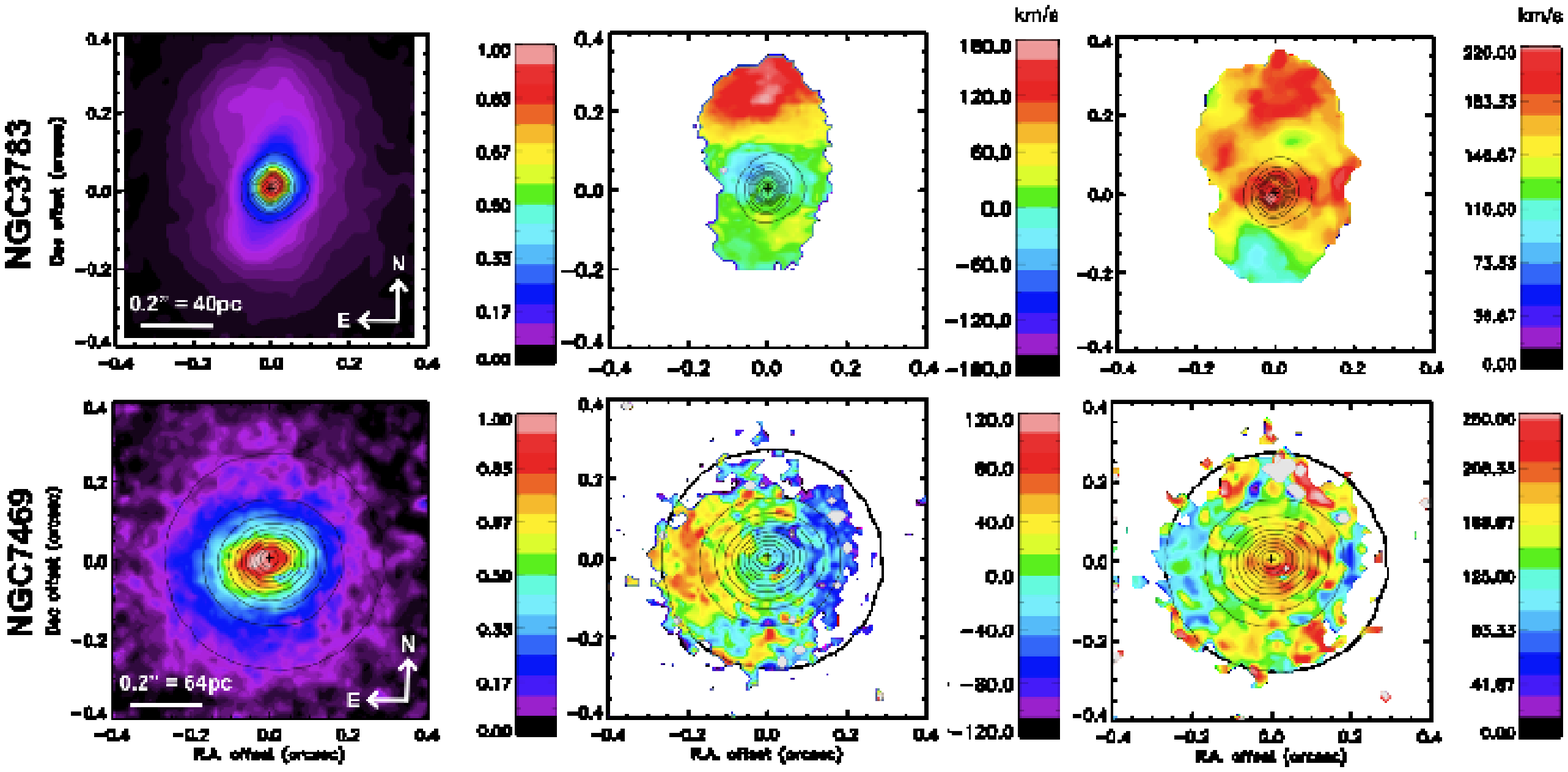}
\caption{Continuation of Figure~\ref{fig6}.
\label{fig6b}}
\end{figure}

\clearpage

\begin{figure}
\epsscale{.99}
\plotone{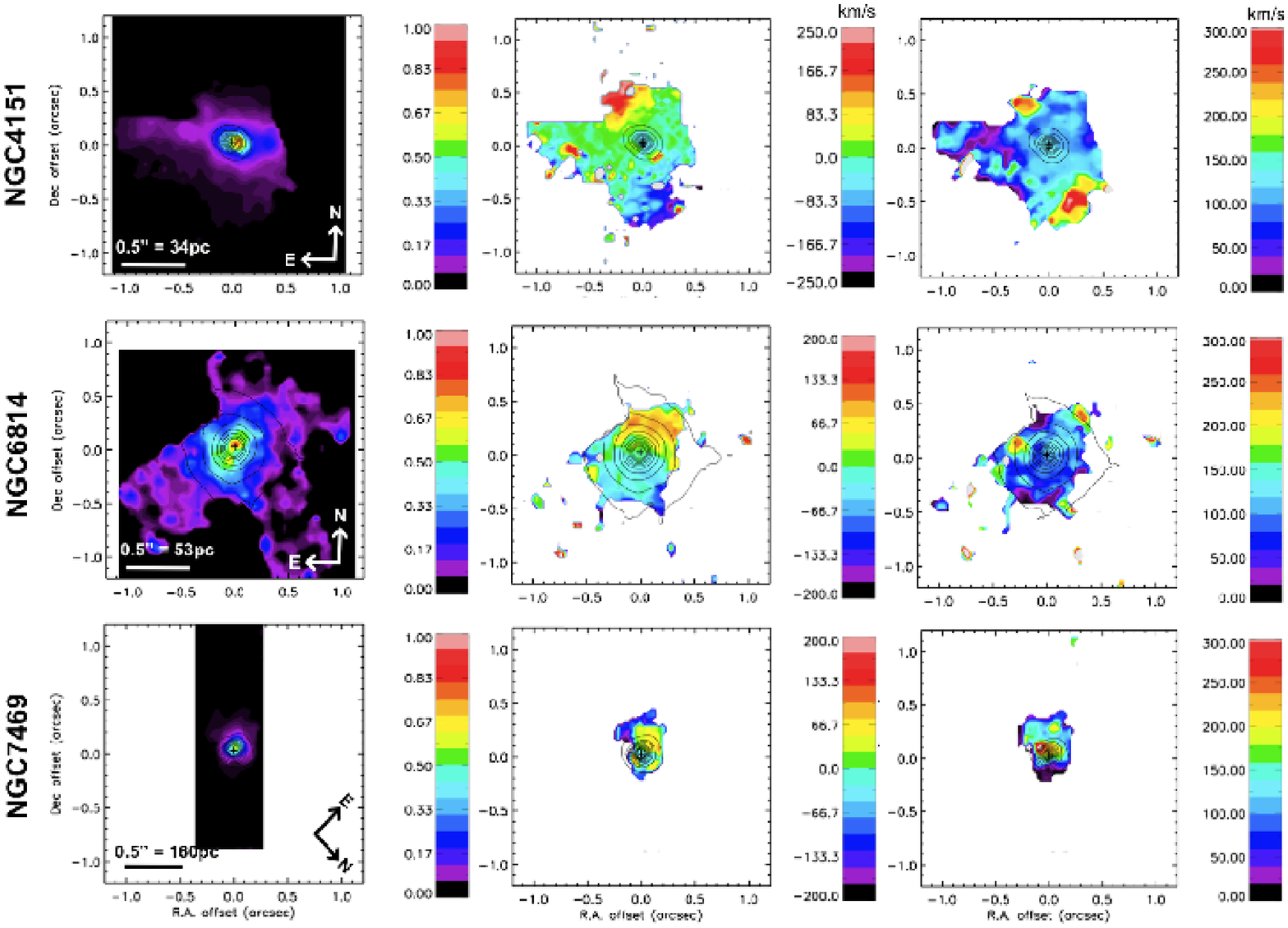}
\caption{Same as Figure~\ref{fig6}, but for the galaxies observed with OSIRIS.
\label{fig7}}
\end{figure}

\clearpage

\begin{figure}
\epsscale{.99}
\plotone{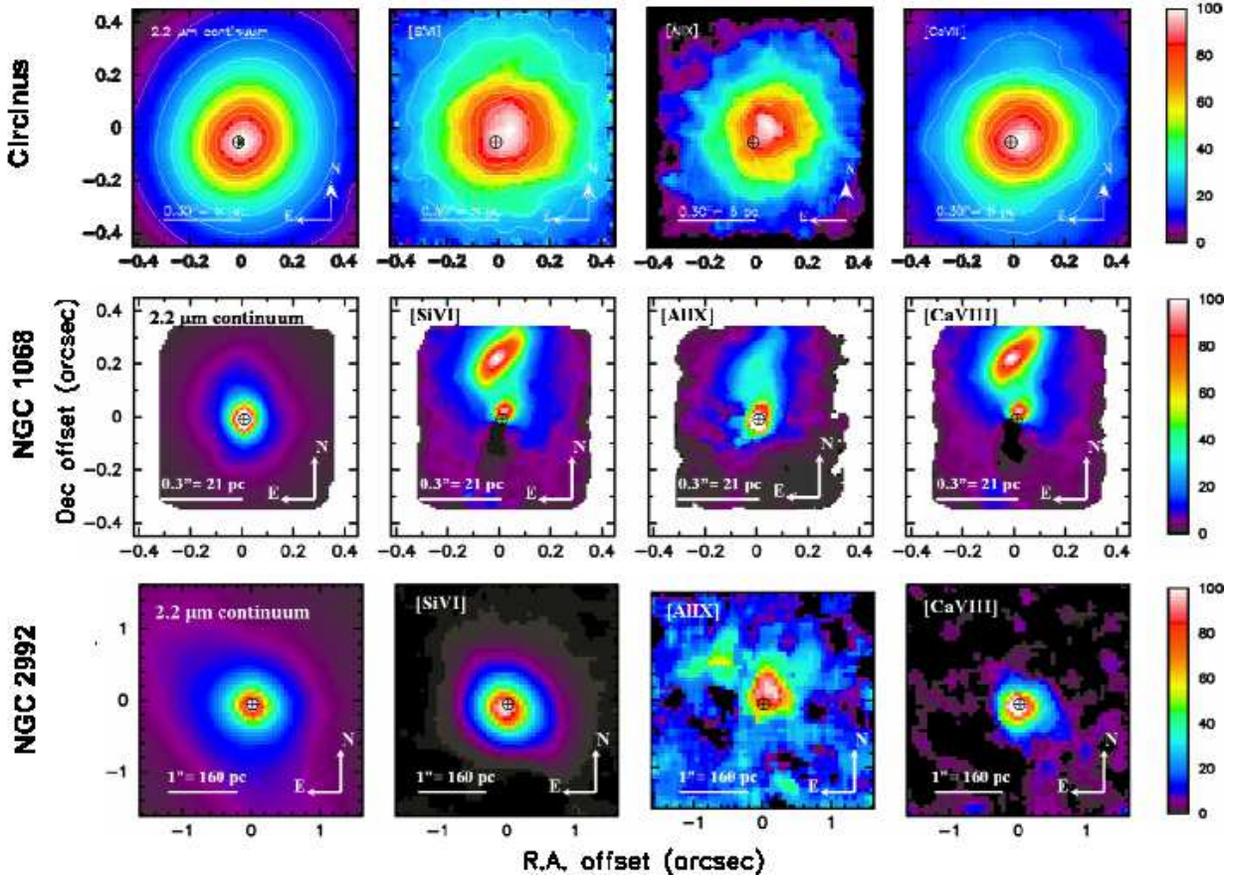}
\caption{Flux maps of  [Si~{\sc vi}], [Al~{\sc ix}] and [Ca~{\sc viii}] emission (second, third and fourth columns) in Circinus, NGC~1068 and NGC~2992. The first column represents the continuum emission at $2.2 \mu$m. These are the only three galaxies in our sample exhibiting these three near-IR coronal lines. Note the similarities in their morphologies and that [Al~{\sc ix}] presents always the most compact emission. See text for details.  
\label{fig8}}
\end{figure}

\clearpage

\begin{figure}
\epsscale{.99}
\plotone{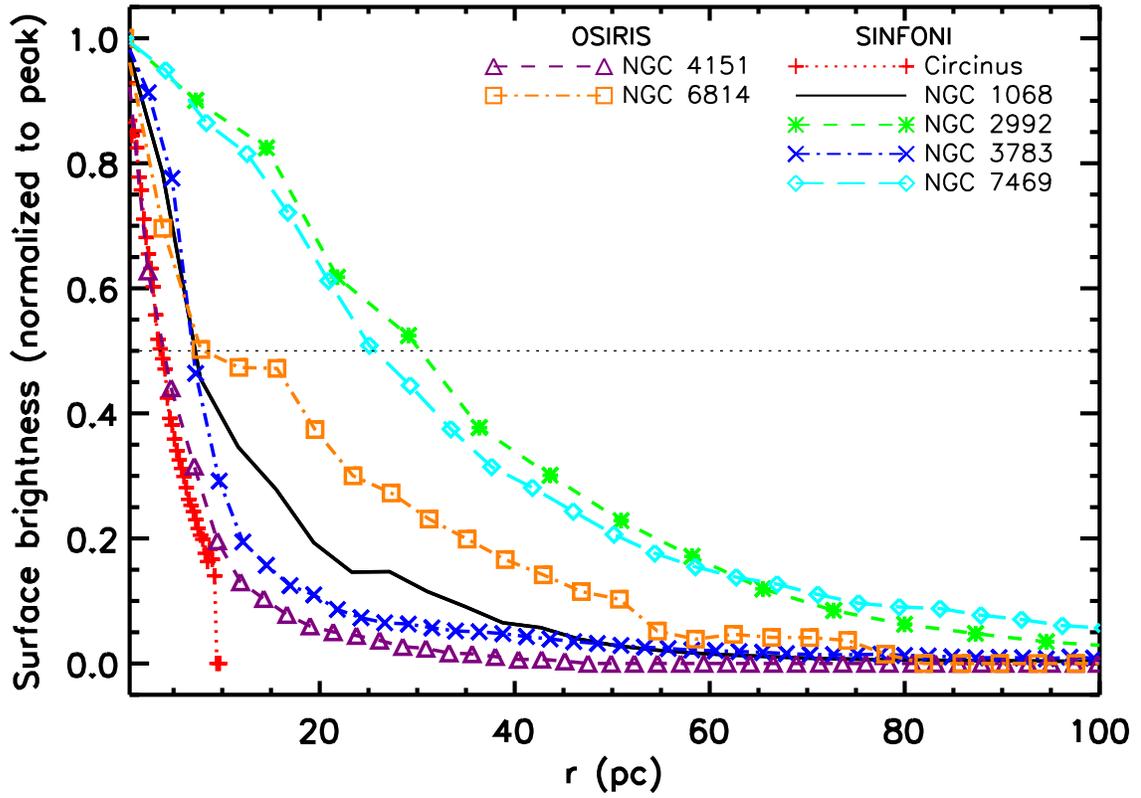}
\caption{Azimuthally-averaged surface-brightness radial profiles of the [Si~{\sc vi}] flux distribution for the galaxies exhibiting [Si~{\sc vi}] emission. The horizontal dashed line indicates the HWHM$_{\mathrm{[Si~{vi}]}}$. 
Tipically, the standard deviation of the azimuthally averaged flux is $5\%$.
\label{fig9}}
\end{figure}

\clearpage

\begin{figure}
\epsscale{.99}
\plotone{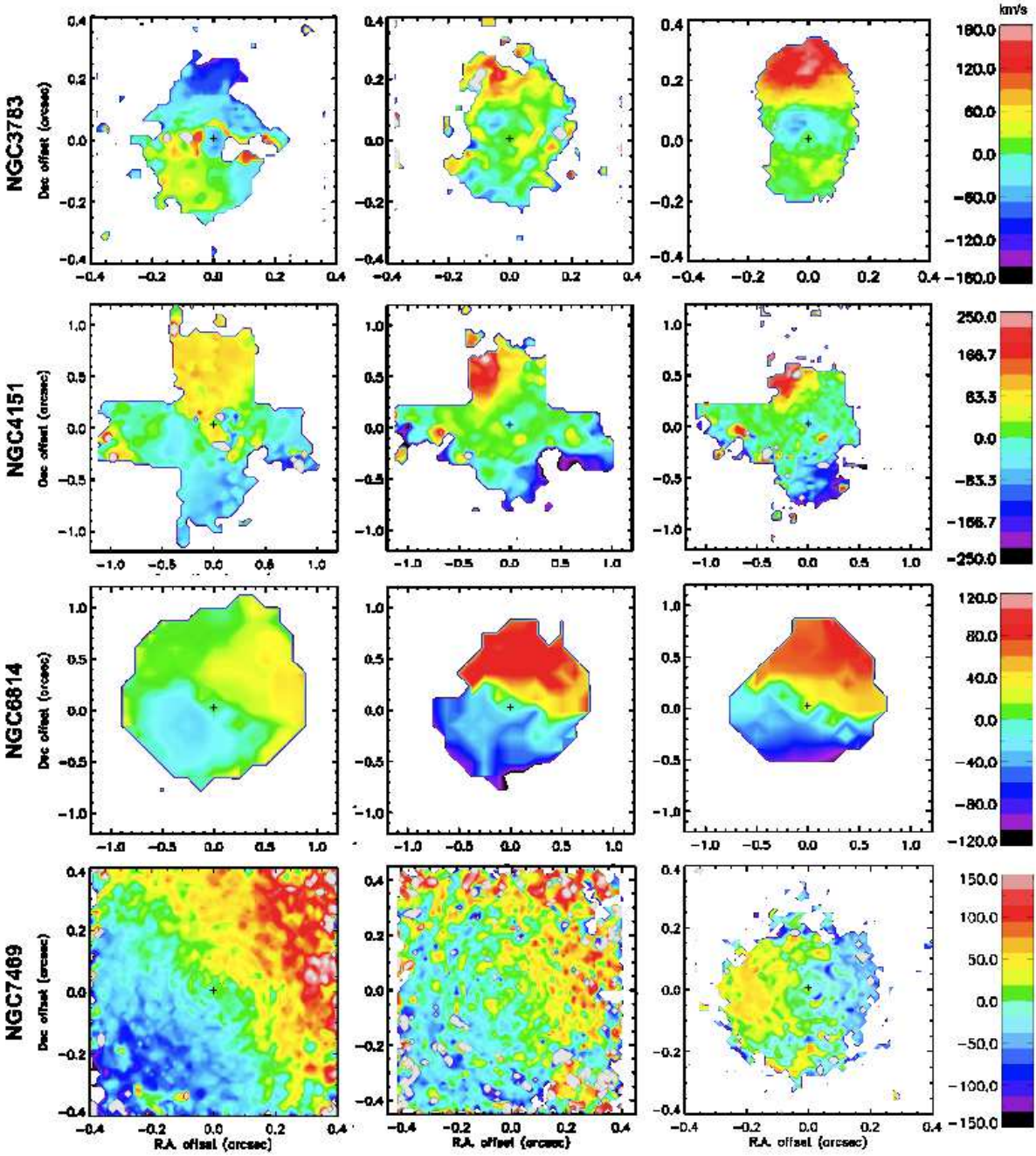}
\caption{Velocity fields of H$_2$ (left), Br$\gamma$ (middle) and [Si~{\sc vi}] (right) emission in the central regions of NGC~3783 (FOV$\sim160$ pc), NGC~4151 (FOV$\sim160$ pc), NGC~6814 (FOV$\sim250$ pc) and NGC~7469 (FOV$\sim250$ pc). The [Si~{\sc vi}] velocity fields show increments of velocity with distance in a different direction to the H$_2$ rotation indicative of an outflow of $>100$ km s$^{-1}$. See text for details. 
\label{fig10}}
\end{figure}

\clearpage

\begin{figure}
\epsscale{.99}
\plotone{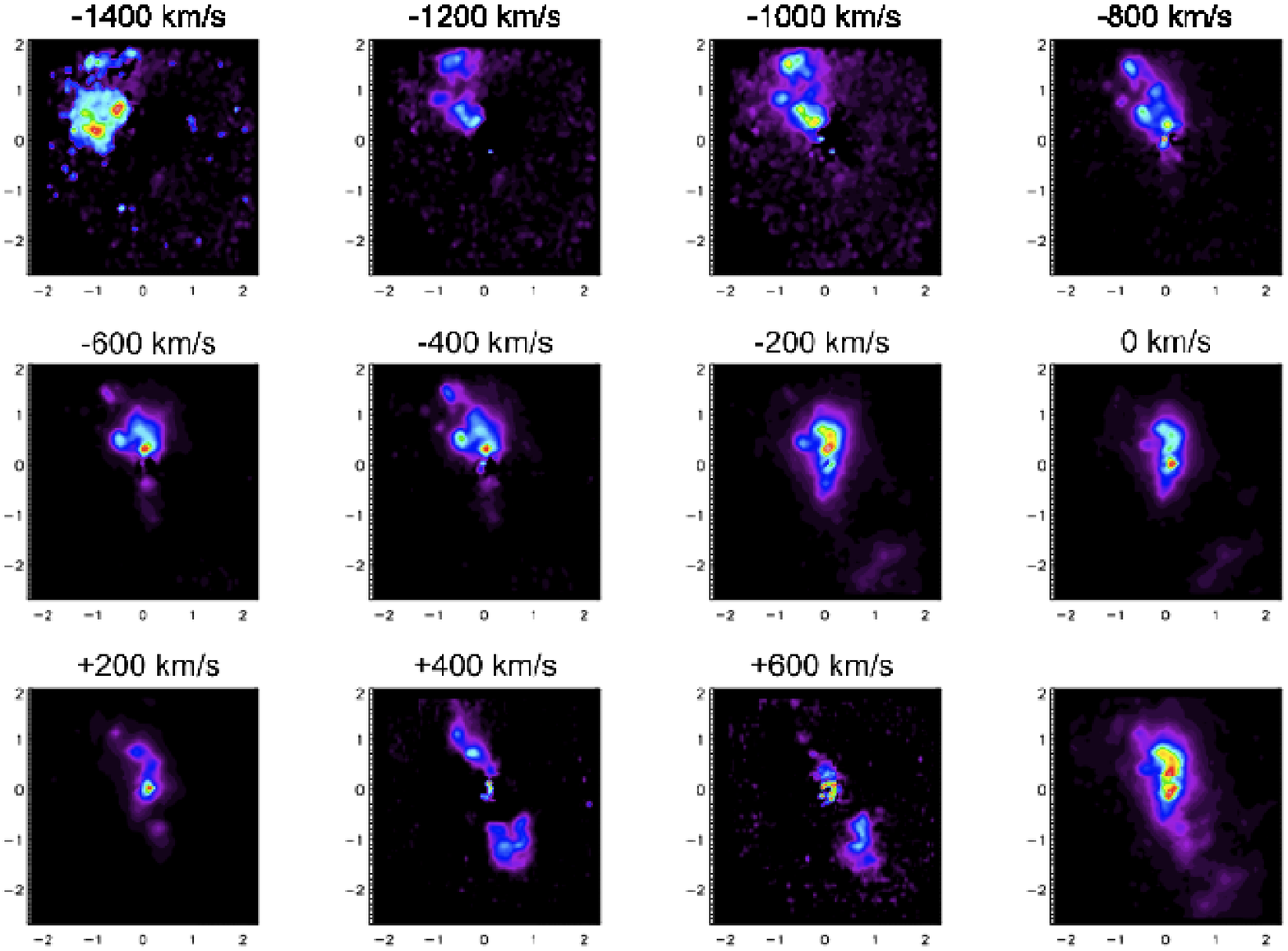}
\caption{Velocity tomography of the [Si~{\sc vi}] emission-line in the Seyfert 2 galaxy NGC 1068. The channel maps were obtained by integrating the flux within velocity bins of 200 km s$^{-1}$ along the  [Si~{\sc vi}] emission-line profile. The
number in the upper part of each panel correspond to the central velocity of the bin in km s$^{-1}$ relative to systemic. The bottom right panel shows the integrated intensity image by summing the velocity slices. 
\label{fig11}}
\end{figure}

\clearpage

\begin{figure}
\epsscale{.99}
\plotone{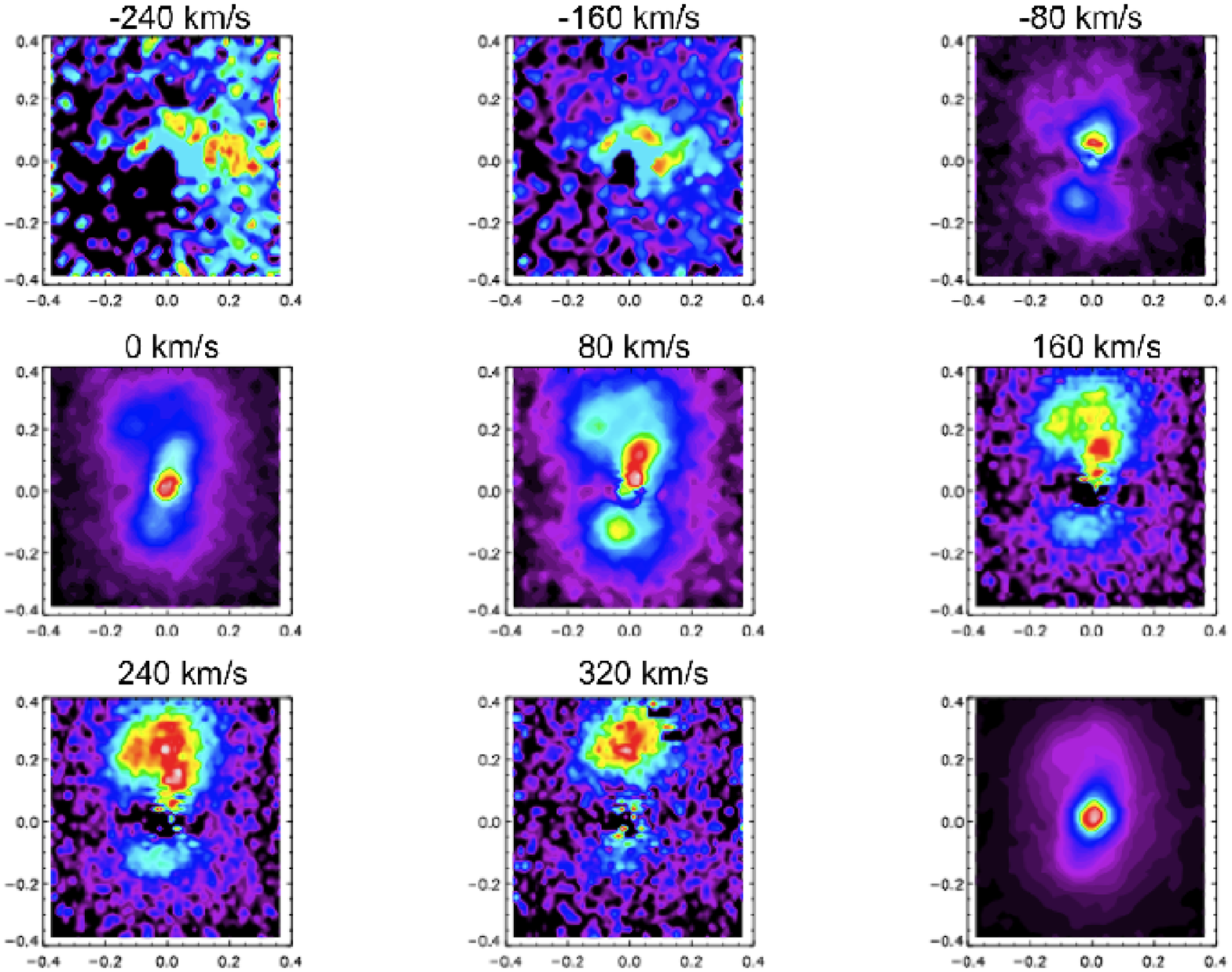}
\caption{Velocity tomography of the [Si~{\sc vi}] emission-line in the Seyfert 1.5 galaxy NGC 3783. The channel maps were obtained by integrating the flux within velocity bins of 80 km s$^{-1}$ along the  [Si~{\sc vi}] emission-line profile. The
number in the upper part of each panel correspond to the central velocity of the bin in km s$^{-1}$ relative to systemic. The bottom right panel shows the integrated intensity image by summing the velocity slices. Note that the acceleration in this galaxy is mostly observed in the north part. See text for details. 
\label{fig12}}
\end{figure}

\clearpage

\begin{figure}
\epsscale{.99}
\plotone{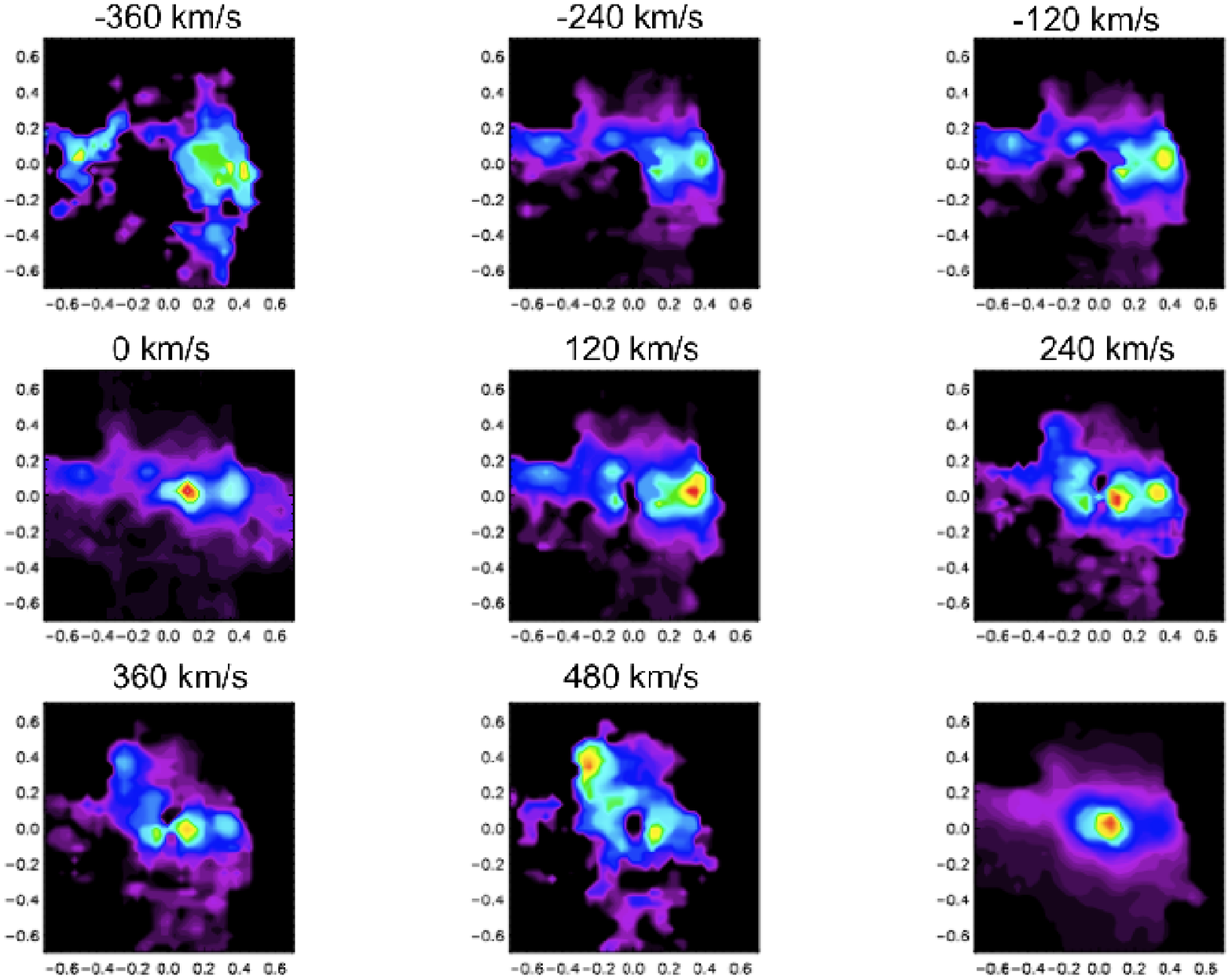}
\caption{Velocity tomography of the  [Si~{\sc vi}] emission-line in the Seyfert 1.5 galaxy NGC 4151. The channel maps were obtained by integrating the flux within velocity bins of 120 km s$^{-1}$ along the  [Si~{\sc vi}] emission-line profile. The number in the upper part of each panel correspond to the central velocity of the bin in km s$^{-1}$ relative to systemic. The bottom right panel shows the integrated intensity image by summing the velocity slices. 
\label{fig13}}
\end{figure}

\clearpage

\begin{figure}
\epsscale{.99}
\plotone{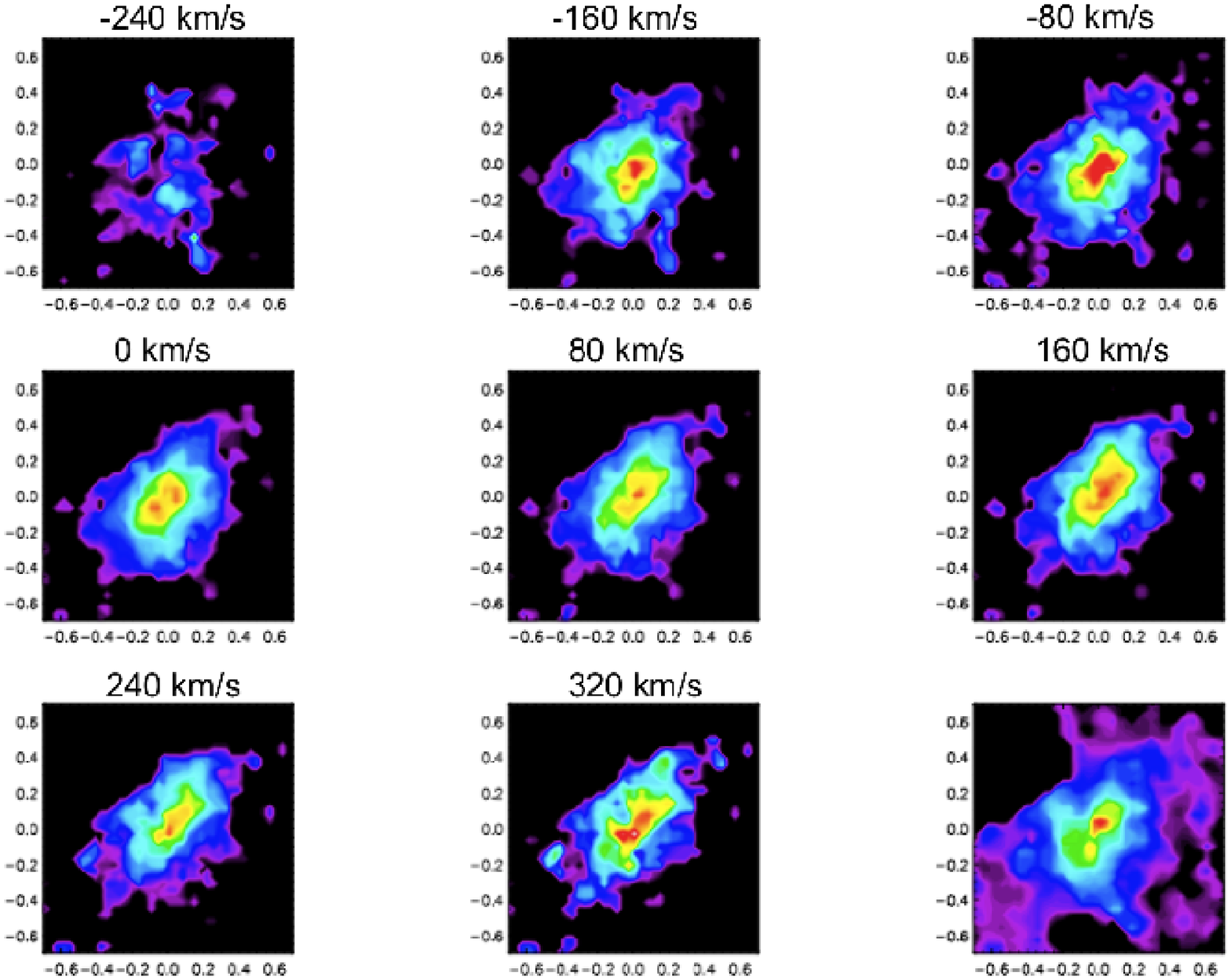}
\caption{Velocity tomography of the [Si~{\sc vi}] emission-line in the Seyfert 1.5 galaxy NGC 6814. The channel maps were obtained by integrating the flux within velocity bins of 80 km s$^{-1}$ along the [Si~{\sc vi}] emission-line profile. The
number in the upper part of each panel correspond to the central velocity of the bin in km s$^{-1}$ relative to systemic. The bottom right panel shows the integrated intensity image by summing the velocity slices. 
\label{fig14}}
\end{figure}

\clearpage

\begin{figure}
\epsscale{.99}
\plotone{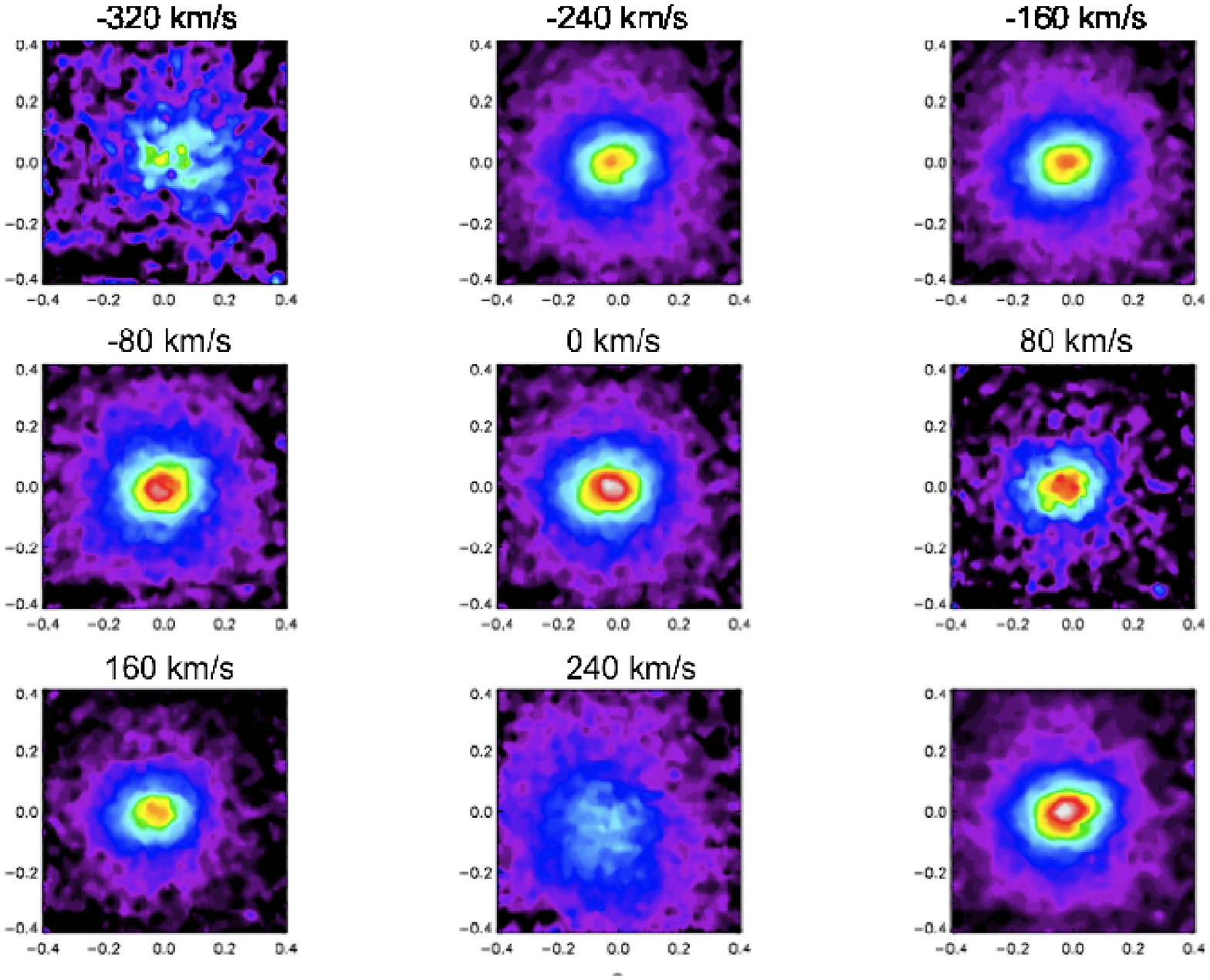}
\caption{Velocity tomography of the [Si~{\sc vi}] emission-line in the Seyfert 1.5 galaxy NGC 7469. The channel maps were obtained by integrating the flux within velocity bins of 80 km s$^{-1}$ along the [Si~{\sc vi}] emission-line profile. The
number in the upper part of each panel correspond to the central velocity of the bin in km s$^{-1}$ relative to systemic. The bottom right panel shows the integrated intensity image by summing the velocity slices. 
\label{fig14a}}
\end{figure}

\clearpage

\begin{figure}
\epsscale{.99}
\plotone{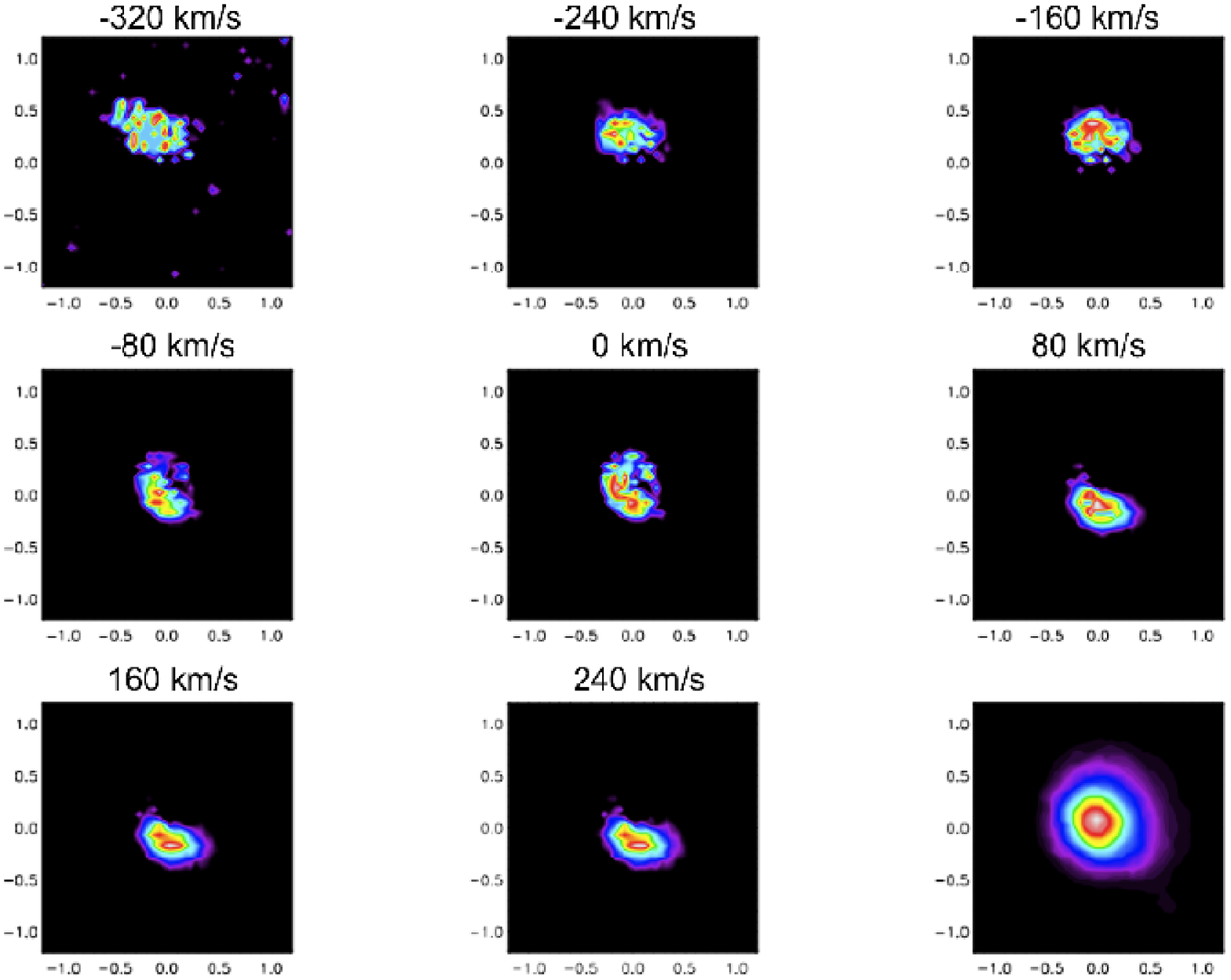}
\caption{Velocity tomography of the [Si~{\sc vi}] emission-line in the Seyfert 1.9 galaxy NGC 2992. The channel maps were obtained by integrating the flux within velocity bins of 80 km s$^{-1}$ along the [Si~{\sc vi}] emission-line profile. The
number in the upper part of each panel correspond to the central velocity of the bin in km s$^{-1}$ relative to systemic. The bottom right panel shows the integrated intensity image by summing the velocity slices. 
\label{fig14b}}
\end{figure}

\clearpage

\begin{figure}
\epsscale{.99}
\plotone{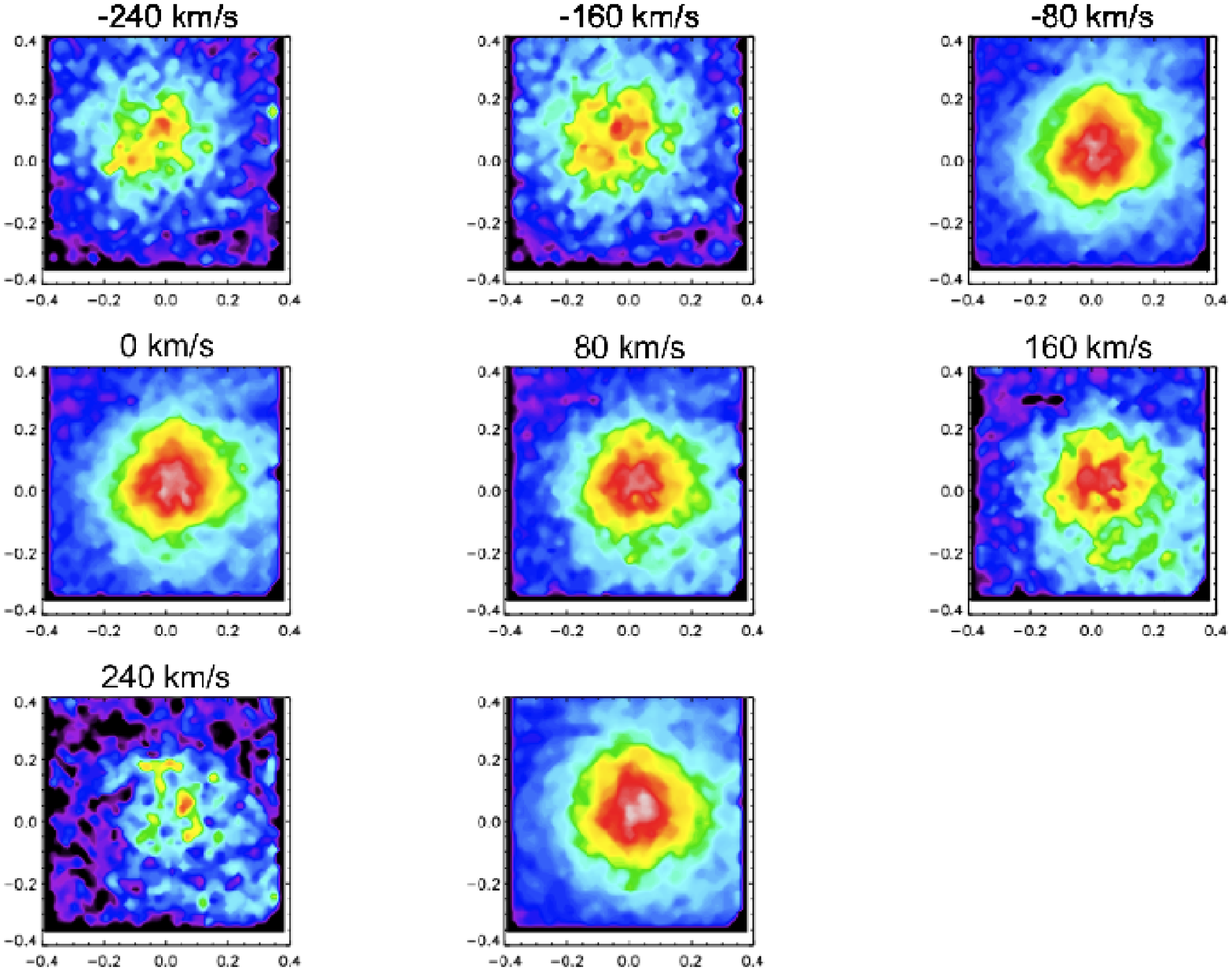}
\caption{Velocity tomography of the Br$\gamma$ emission-line in the Seyfert 2 galaxy Circinus. The channel maps were obtained by integrating the flux within velocity bins of 80 km s$^{-1}$ along the Br$\gamma$ emission-line profile. The
number in the upper part of each panel correspond to the central velocity of the bin in km s$^{-1}$ relative to systemic. The bottom right panel shows the integrated intensity image by summing the velocity slices. 
\label{fig14c}}
\end{figure}

\clearpage

\begin{figure}
\epsscale{.99}
\plotone{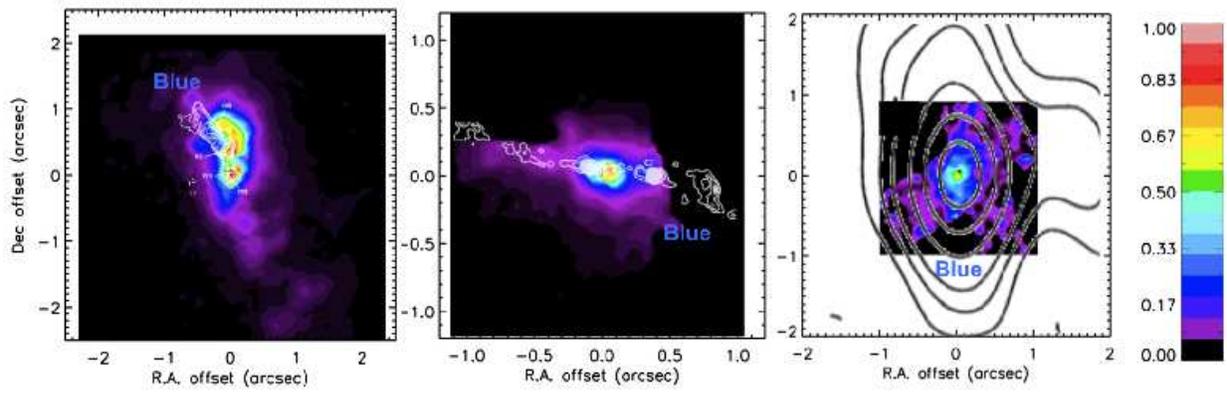}
\caption{Contour plots of radio continuum emission overlaid on the [Si~{\sc vi}] flux maps. The overlays correspond, from left to right, to NGC~1068 (radio image from Gallimore et al. 1996), NGC~4151 (radio image from Mundell et al. 2003) and NGC~6814 (radio image from Ulvestad \& Wilson 1984). The legend ``blue'' indicates the direction of the radio jet coming towards us as well as the regions where the blueshifts in the velocities of the coronal gas are observed. 
\label{fig15}}
\end{figure}

\clearpage

\begin{figure}
\epsscale{.99}
\plotone{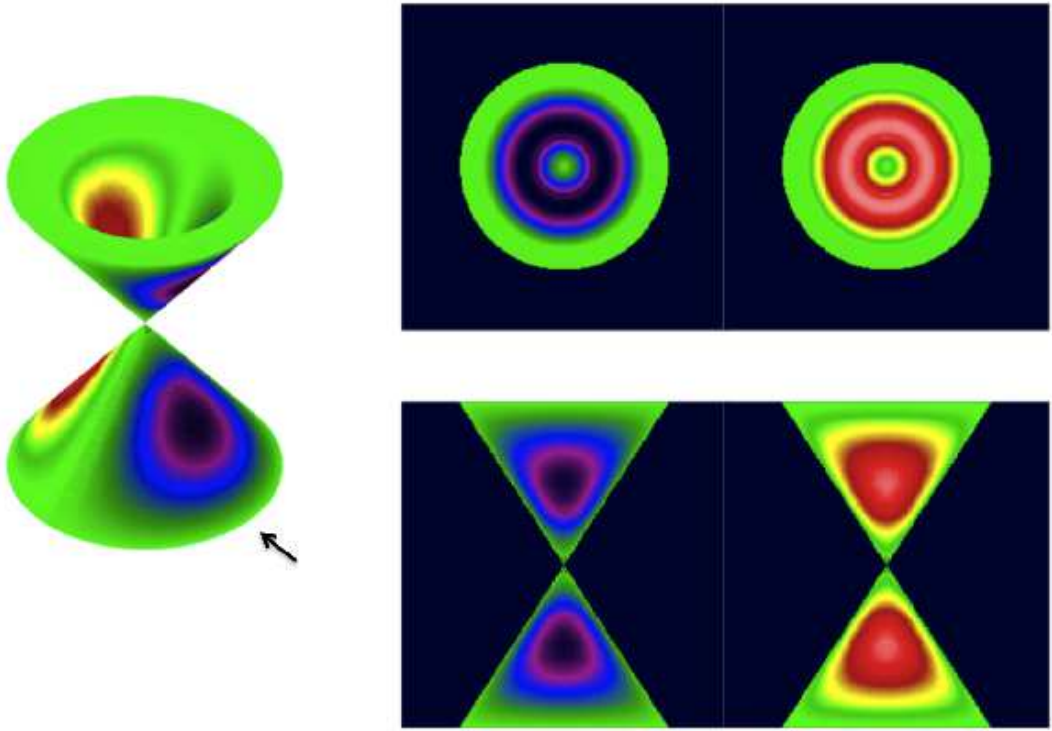}
\caption{Example of a bicone model incorporating radial acceleration and deceleration for two test galaxies having different inclinations and PA$_{\mathrm{bicone}}=0\degr$. \textit{Left:} Three-dimensional structure of the model with the arrow indicating the LOS. \textit{Top Right:} Front and back projections of a galaxy having $i_{\mathrm{bicone}}=90\degr$. \textit{Bottom right:} Front and back projections of a galaxy having $i_{\mathrm{bicone}}=0\degr$. The velocities are assigned a colour based on the amount of redshift (warm colours) or blueshift (cold colours). Green represents approximately zero velocity.
\label{fig16}}
\end{figure}

\clearpage

\begin{figure}
\epsscale{.99}
\plotone{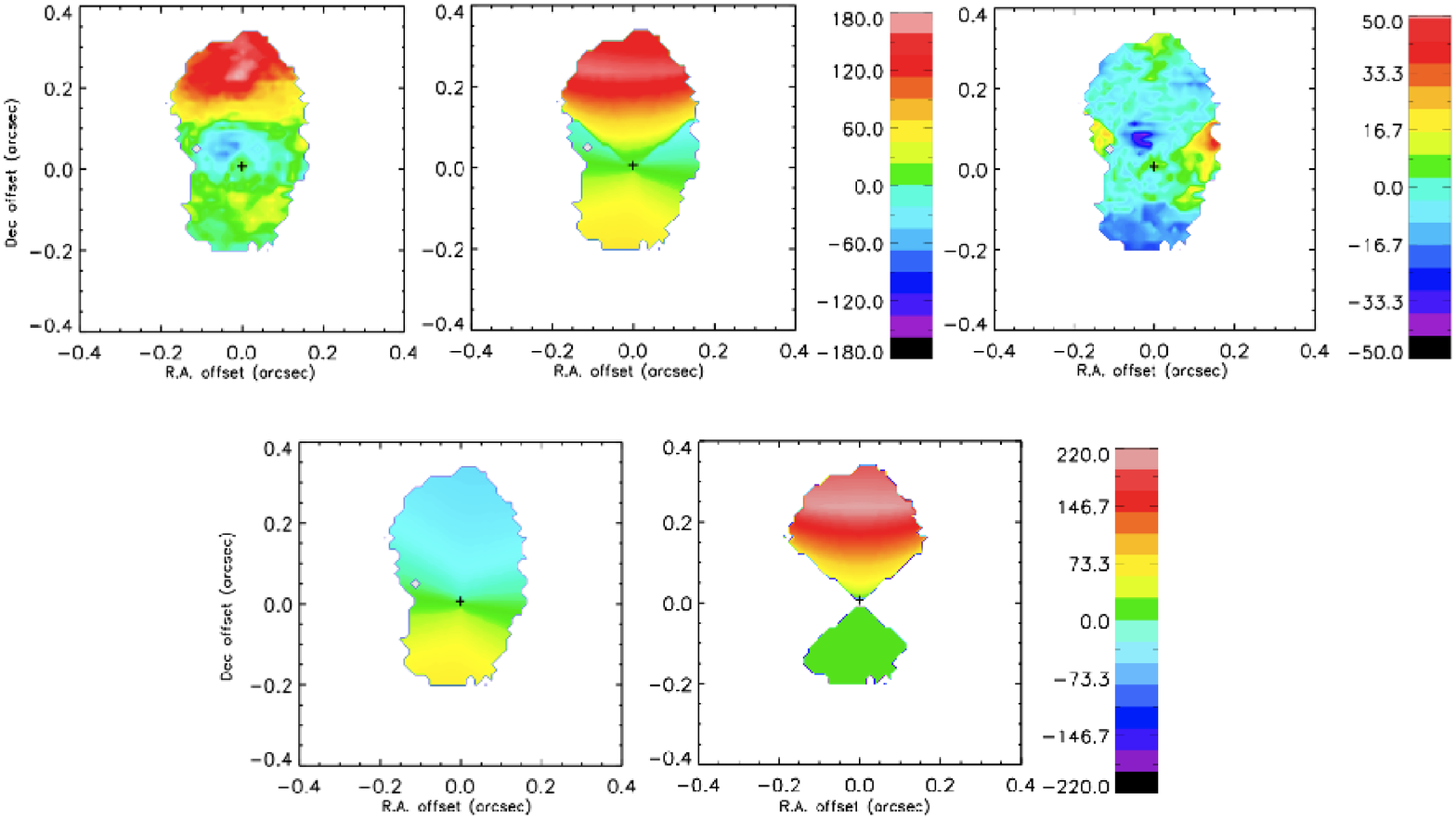}
\caption{Maps showing the data and the best-fit kinematic model used to reproduce the [Si~{\sc vi}] kinematics in NGC~3783. The maps are centered at the position of the AGN (black cross), and the scales are in km s$^{-1}$. \textit{Top Left:} [Si~{\sc vi}] velocity field used for the modeling, \textit{Top Middle:} Best-fit kinematic model incorporating rotation and biconical outflow, \textit{Top Right:} Residuals (data-model) to the fit, \textit{Bottom Left:} Rotational component of the combined kinematic model shown in the top middle panel, \textit{Bottom Right:} Outflowing component of the combined kinematic model shown in the top middle panel.
\label{fig17}}
\end{figure}

\begin{figure}
\epsscale{.99}
\plotone{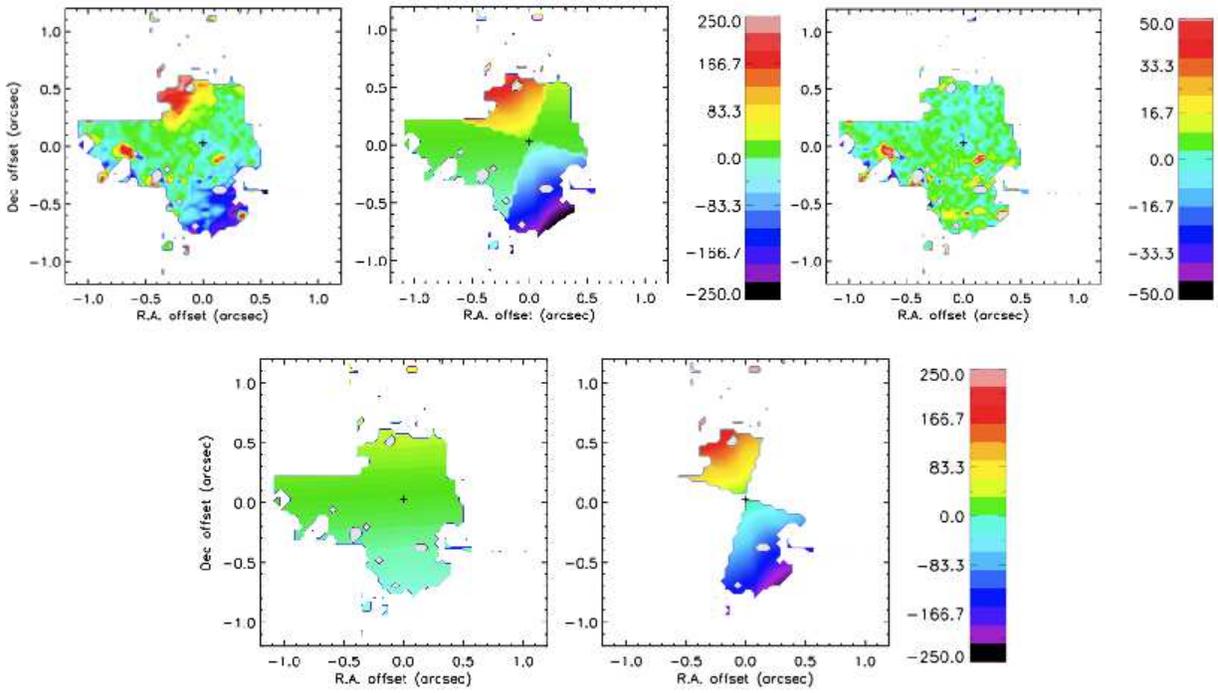}
\caption{Same as Figure~\ref{fig17} but for NGC~4151.
\label{fig18}}
\end{figure}

\begin{figure}
\epsscale{.99}
\plotone{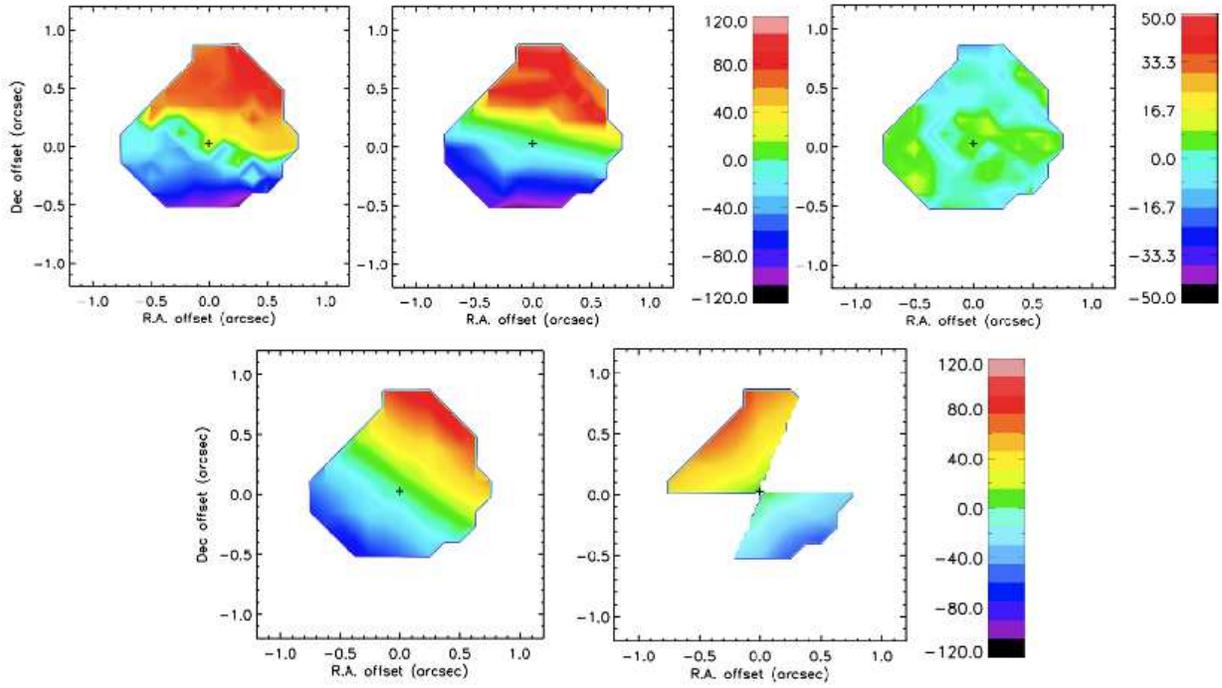}
\caption{Same as Figure~\ref{fig17} but for NGC~6814.
\label{fig19}}
\end{figure}

\begin{figure}
\epsscale{.99}
\plotone{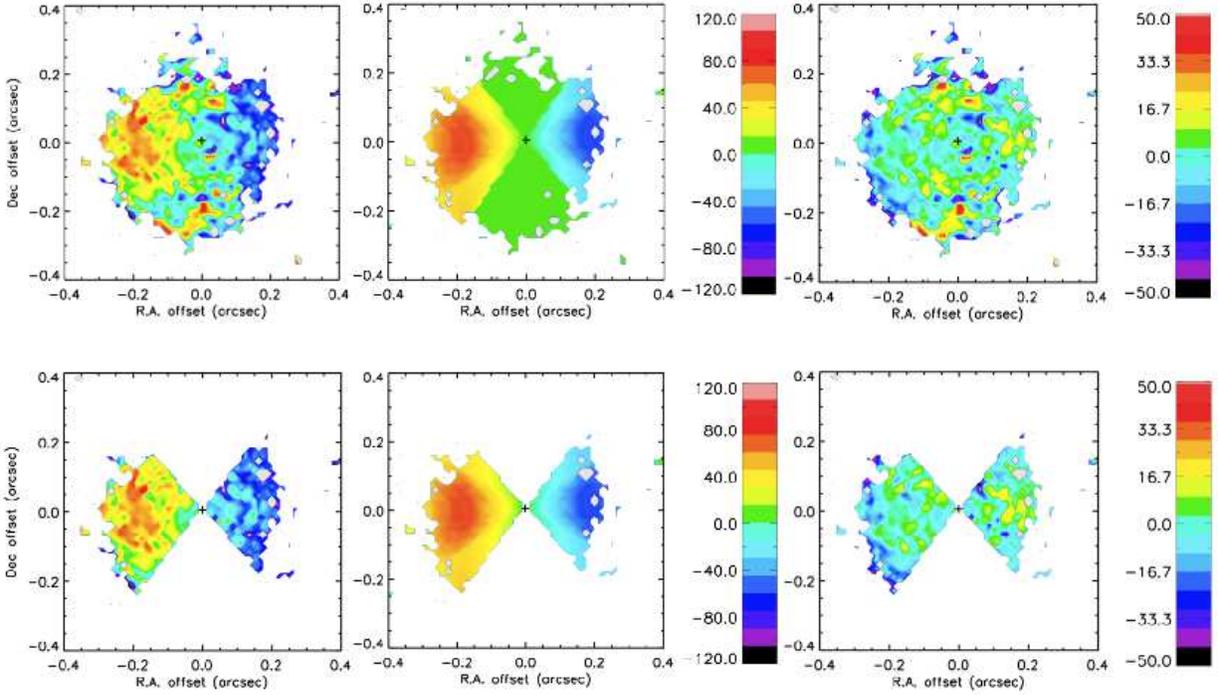}
\caption{Maps showing the data and the best-fit kinematic model used to reproduce the [Si~{\sc vi}] kinematics in NGC~7469. The maps are centered at the position of the AGN (black cross), and the scales are in km s$^{-1}$. While the top row shows the LOS velocities in the entire FOV, the bottom row shows only the LOS velocities inside the bicone model. \textit{Left column:} [Si~{\sc vi}] velocity field used for the modeling, \textit{Middle column:} Best-fit kinematic model corresponding only to biconical outflow, \textit{Right column:} Residuals (data-model) to the fit.
\label{fig20}}
\end{figure}

\begin{figure}
\epsscale{.99}
\plotone{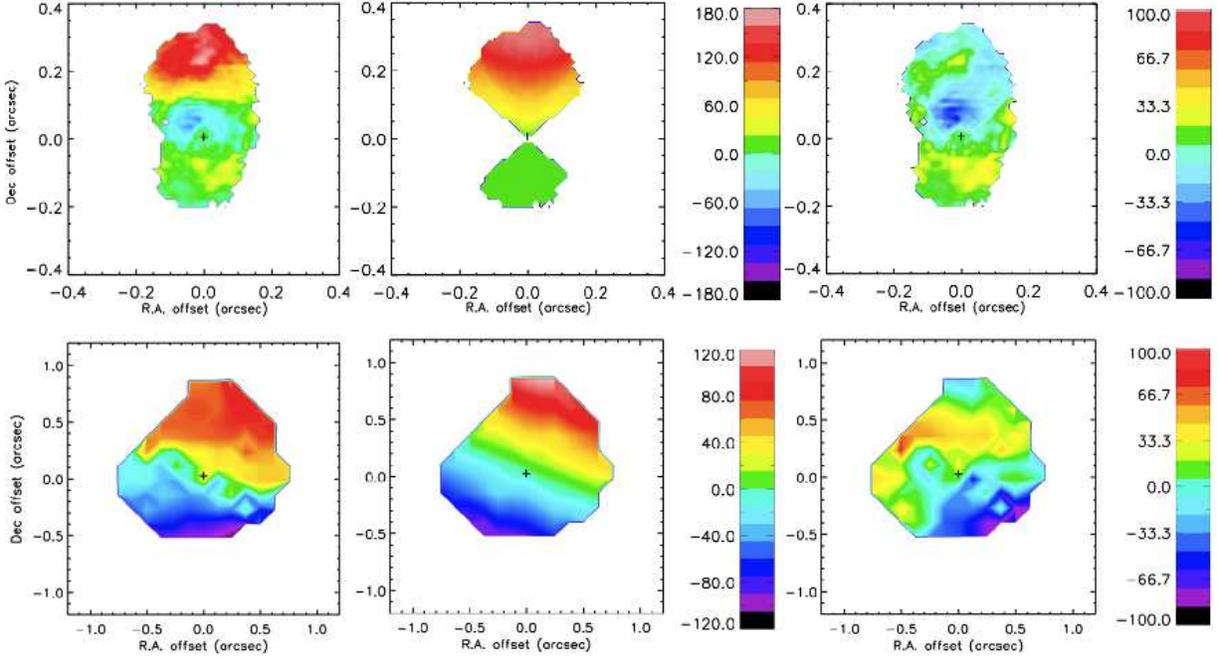}
\caption{Maps showing the data and the best-fit with pure rotation for a rotation-dominated object (NGC~6814, bottom), and the best fit with pure outflow for an outflow-dominated case (NGC~3783, top). The maps are centered at the position of the AGN (black cross), and the scales are in km s$^{-1}$. \textit{Left column:} [Si~{\sc vi}] velocity fields used for the modeling, \textit{Middle column:} Best-fit incorporating only one kinematical component (top-outflow, bottom-rotation), \textit{Right column:} Residuals (data-model) to the fit. In the two cases, the quality of the fit decreases when compared to the best fits incorporating rotation plus biconical outflow.
\label{fig20b}}
\end{figure}


\begin{figure}
\epsscale{.99}
\plotone{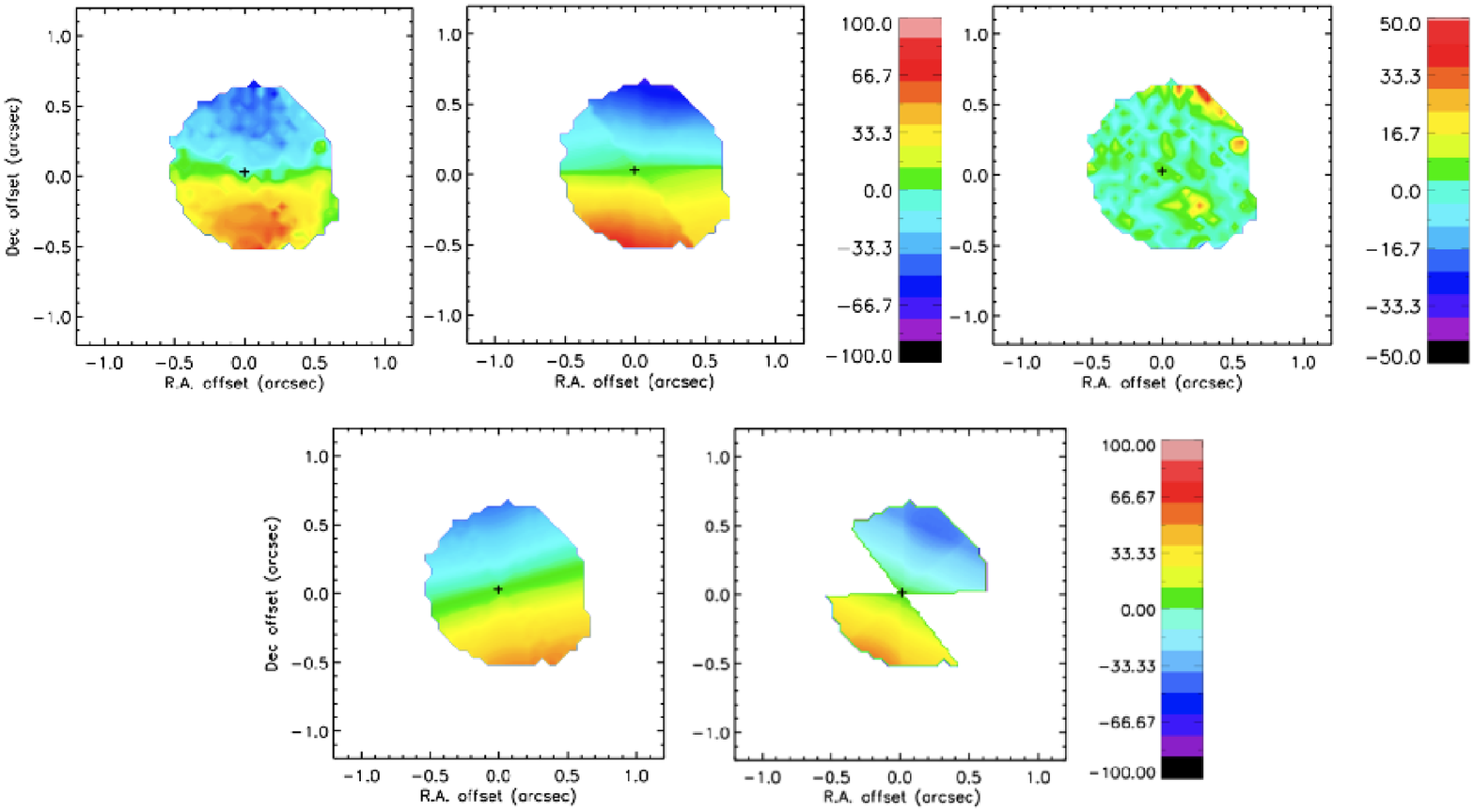}
\caption{Maps showing the data and the best-fit kinematic model used to reproduce the [Si~{\sc vi}] kinematics in NGC~2992. The maps are centered at the position of the AGN (black cross), and the scales are in km s$^{-1}$. \textit{Top Left:} [Si~{\sc vi}] velocity field used for the modeling, \textit{Top Middle:} Best-fit kinematic model incorporating rotation and biconical outflow, \textit{Top Right:} Residuals (data-model) to the fit, \textit{Bottom Left:} Rotational component of the combined kinematic model shown in the top middle panel, \textit{Bottom Right:} Outflowing component of the combined kinematic model shown in the top middle panel.
\label{fig22}}
\end{figure}

\begin{figure}
\epsscale{.99}
\plotone{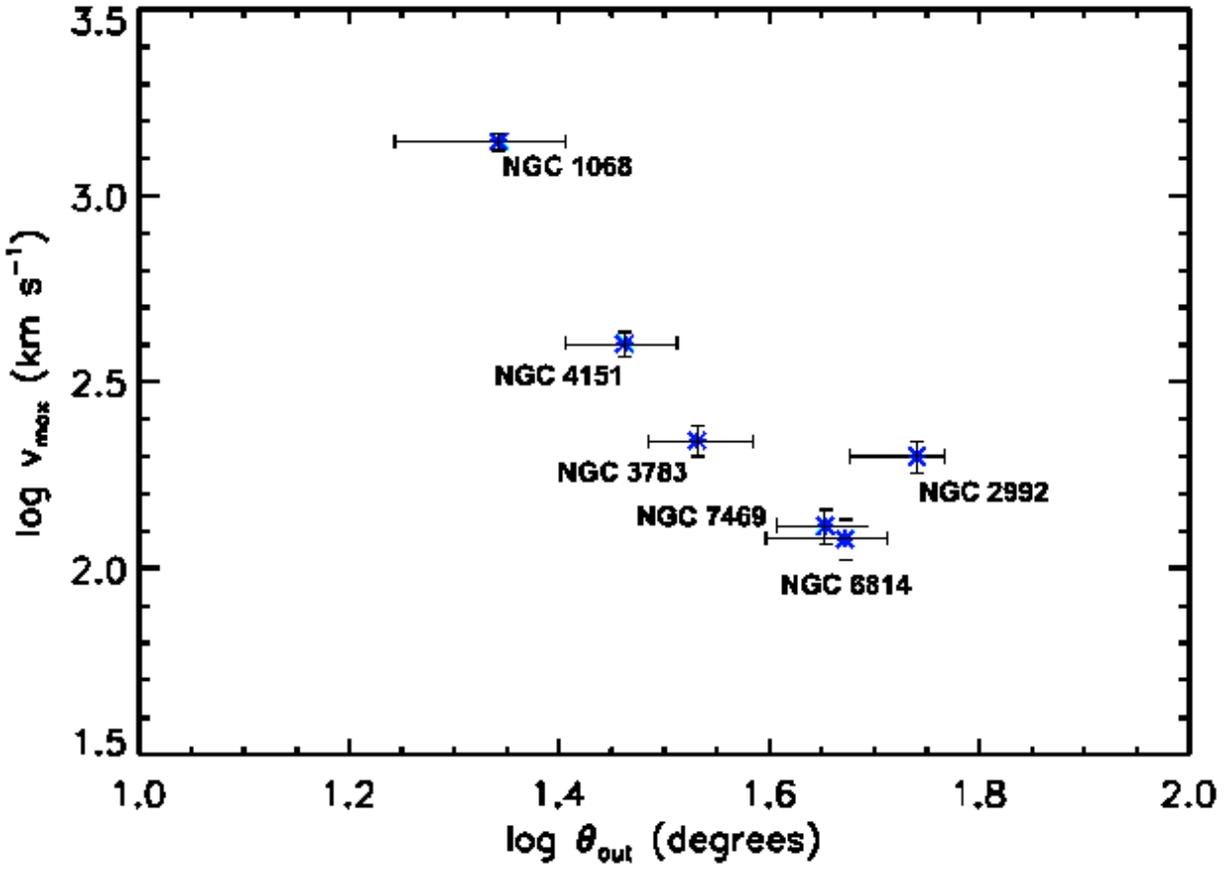}
\caption{Graph showing the anticorrelation found between the half opening angle of the outflow ($\theta_{\mathrm{out}}$) and its maximum velocity ($v_{\mathrm{max}}$). The error bars correspond to the uncertainties in the estimation of the best-fit parameters (Tables \ref{table6} and \ref{table7}). 
\label{fig23}}
\end{figure}

\begin{figure}
\epsscale{.99}
\plotone{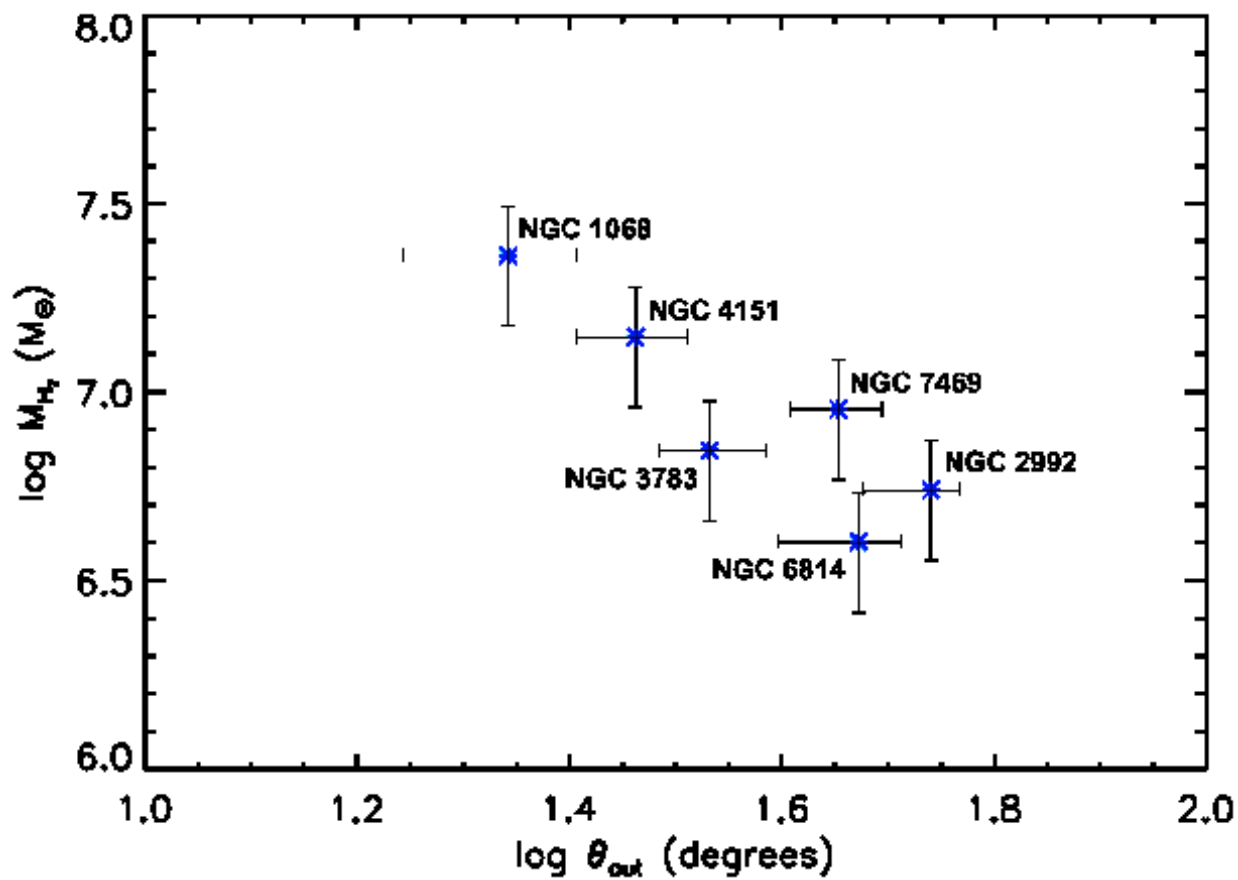}
\caption{Graph showing the anticorrelation found between the half opening angle of the outflow ($\theta_{\mathrm{out}}$) and the molecular gas mass ($M_{\mathrm{H_2}}$) in the central $r<30$ pc. The values of $M_{\mathrm{H_2}}$ were obtained from H09 (except for NGC~2992 which was taken from Friedrich et al. 2010). The vertical error bars represent the typical $35\%$ error of the gas mass estimates (H09). The horizontal error bars correspond to the uncertainties in the estimation of $\theta_{\mathrm{out}}$ (Tables \ref{table6} and \ref{table7}).   
\label{fig24}}
\end{figure}

\begin{figure}
\epsscale{.99}
\plotone{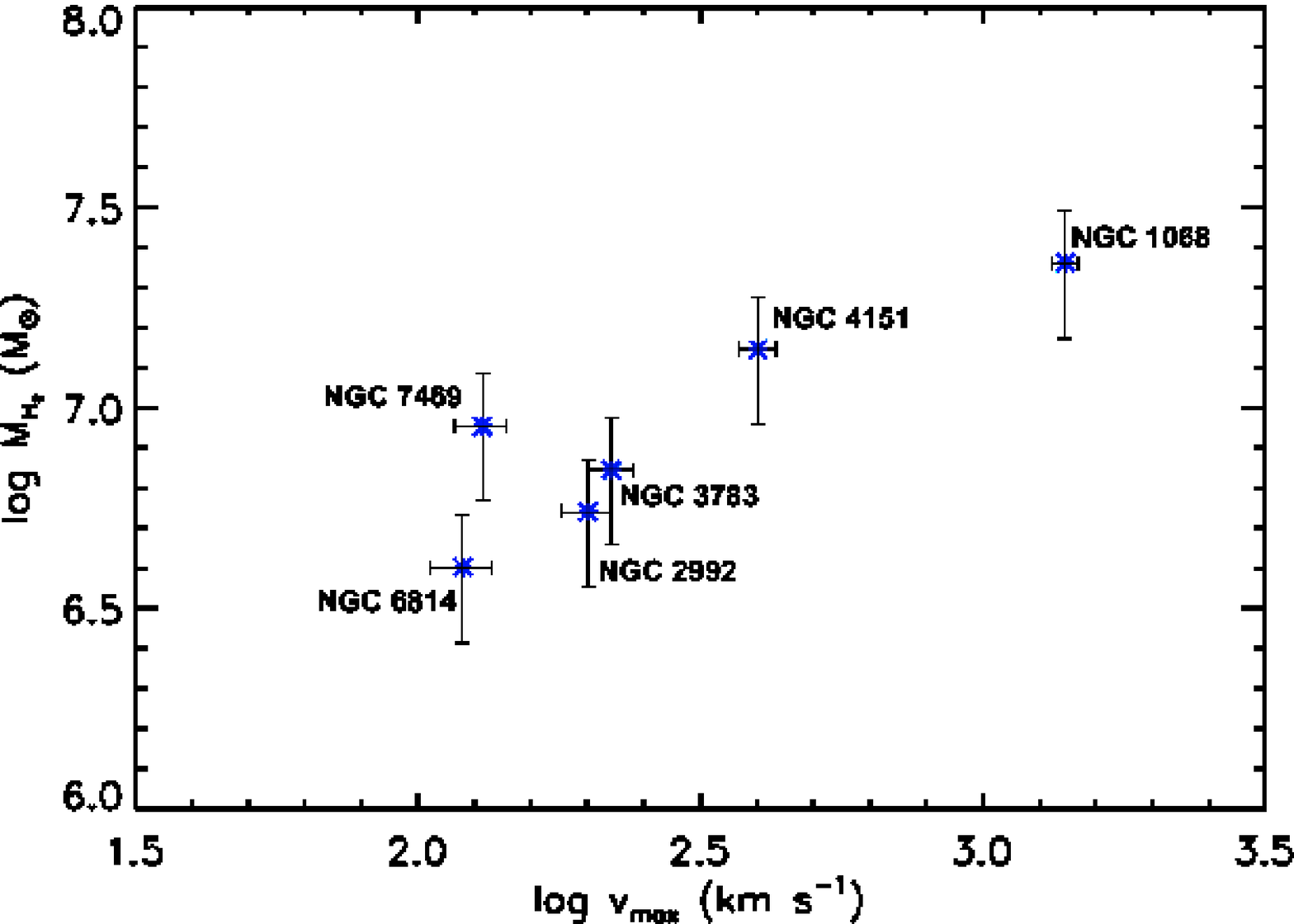}
\caption{Graph showing the correlation found between the maximum velocity of the outflow ($v_{\mathrm{max}}$) and the molecular gas mass ($M_{\mathrm{H_2}}$) in the central $r<30$ pc. The values of $M_{\mathrm{H_2}}$ were obtained from H09 (except for NGC~2992 which was taken from Friedrich et al. 2010). 
The vertical error bars represent the typical $35\%$ error of the gas mass estimates (H09). The horizontal error bars correspond to the uncertainties in the estimation of $v_{\mathrm{max}}$ (Tables \ref{table6} and \ref{table7}).
\label{fig25}}
\end{figure}

\clearpage 

\begin{figure}
\epsscale{.7}
\plotone{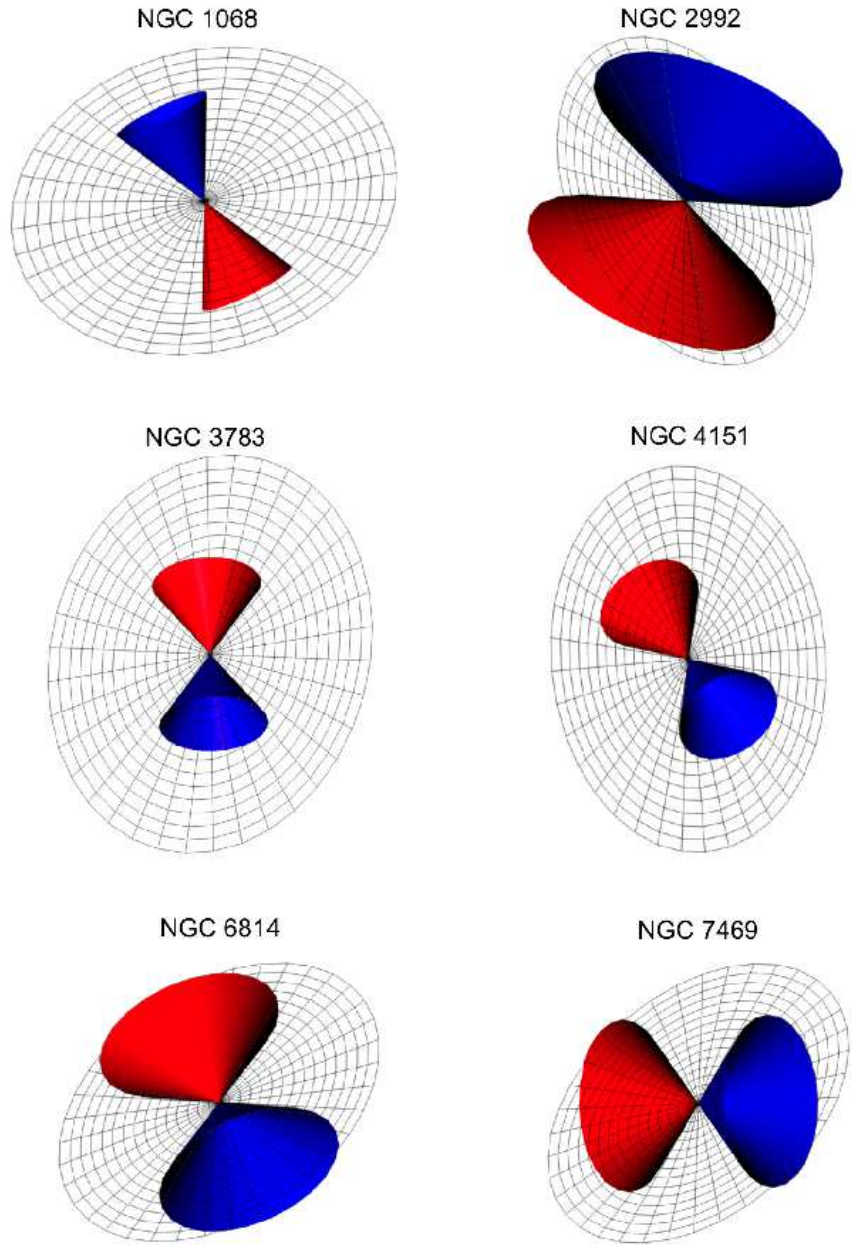}
\caption{Geometric models of the NLR/CLR bicones and the inner galactic disks in NGC 1068 (top left), NGC 2992 (top right), NGC 3783 (middle left), NGC 4151 (middle right), NGC 6814 (bottom left) and NGC 7469 (bottom right) based on the parameters listed in Tables~\ref{table6} and ~\ref{table7}. The color blue indicates that the majority of the radial velocities in the cone are blueshifted. Accordingly, a red cone indicates that the majority of the radial velocities are redshifted. See Section~\ref{orientation} for details.
\label{fig26}}
\end{figure}

\clearpage

\begin{landscape}
\begin{table}
\begin{center}
{\scriptsize
\begin{tabular}{l c c c c c c c c c c c c c}
\hline
\hline \noalign{\smallskip}
Object & Classification\tablenotemark{a} & D\tablenotemark{a} & Scale & Instrument & Band & 
FWHM\tablenotemark{b} & FOV & Date & Broad Br$\gamma$\tablenotemark{c} & 
Narrow Br$\gamma$\tablenotemark{c} & [Si~{\sc vi}]\tablenotemark{c} & 
[Al~{\sc ix}]\tablenotemark{c} & [Ca~{\sc viii}]\tablenotemark{c}\\
& & (Mpc) & pc$/\arcsec$ & & & ($^{\prime\prime}$) & (pc) & & $2.16\mu$m & $2.16\mu$m & 
$1.96\mu$m & $2.04\mu$m & $2.32\mu$m\\ 
\hline \noalign{\smallskip}
Circinus & Sy2 & 4 & 20 & SINFONI & $K$ & 0.22 (4.4) & $16\times16$ & Jul 2004 & ND &D&D&D&D \\
NGC~1068 & Sy2 & 14 & 70 & SINFONI & $H+K$ & 0.08 (5.6) & $225\times225$ & Oct 2005, Nov 2006 & ND & D & D & D & D\\
NGC~2992 & Sy1.9 & 33 & 140 & SINFONI & $K$ & 0.3 (42) & $450\times450$ & Mar 2005 &D&D&D&D&D\\
NGC~3783 & Sy1.5 & 42 & 200 & SINFONI & $H+K$ & 0.085 (17) & $160\times160$ & Mar 2005 & D & D & D & ND & D\\
 &  &  &  & SINFONI & $H+K$ & 0.17 (34) & $640\times640$ & Mar 2005 & D & D & D & ND & D \\
NGC~4151 & Sy1.5 & 14 & 68 & OSIRIS & $K$ & 0.11 (7.5) & $160\times160$\tablenotemark{d} & Mar 2006 &D&D&D&ND&D \\
NGC~6814 & Sy1.5 & 22 & 107 & SINFONI & $K$ & 0.57 (61) & $850\times850$ & Apr, Jun 2009 & D & D & D & ND & MD \\
 &  &  &  & OSIRIS & $K$ & 0.17 (18) & $250\times250$\tablenotemark{d} & Apr, Sep 2006 & D & D & D & ND & MD \\
NGC~7469 & Sy1.5 & 66 & 320 & SINFONI & $K$ & 0.14 (44) & $250\times250$ & Jul 2004 &D&D&D&ND&MD\\
 &  &  & & OSIRIS & $K$ & 0.11 (35) & $192\times640$ & Sep 2006 &D&D&D&ND&MD\\
\hline
\hline
\end{tabular}
}
\tablenotetext{a}{Classification is taken from the catalogue of \citet{veron06}, and Distance (D) from the NASA/IPAC Extragalactic Database.}
\tablenotetext{b}{Spatial resolution (FWHM) estimated from the data itself, using the broad Br$\gamma$ emission and the non-stellar continuum, except for Circinus (see Section 2.2). The numbers in parentheses indicate the FWHM in pc.}
\tablenotetext{c}{Column indicating if the line was Detected (D), Marginally Detected (MD) or Not Detected (ND) in the spectra.}
\tablenotetext{d}{The FOV in these observations has a cross shape.}

\end{center}
\caption[The Nearby AGN Sample]{Summary of AGN data included in this study\label{table1}}
\end{table}
\end{landscape}

\clearpage

\begin{landscape}
\begin{table}
\begin{center}
{\scriptsize
\begin{tabular}{lcccccccccc}
\hline
\hline \noalign{\smallskip}
Galaxy & F(Br$\gamma$)$_{\mathrm{nuc}}$\tablenotemark{a} & 
F(Br$\gamma$)$_{\mathrm{int}}$\tablenotemark{b} & FWHM$_{\mathrm{Br\gamma}}$ & R$_{\mathrm{Br\gamma}}$\tablenotemark{c} & PA$_{\mathrm{Br\gamma}}$ &
$\sigma_{\mathrm{mean}}$ & $\sigma_{\mathrm{max}}$ & 
R$_{\mathrm{[O~{iii}]}}$\tablenotemark{d} & PA$_{\mathrm{[O~{iii}]}}$ & Ref\tablenotemark{e}\\
 & ($10^{-18}$ W m$^{-2}$) & ($10^{-18}$ W m$^{-2}$) & (pc) & (pc) & 
(\degr) & (km s$^{-1}$) &  (km s$^{-1}$) & (pc) & (\degr) & \\
\hline \noalign{\smallskip}
Circinus & 7 (8.8) & 10 & 8 & $8:$ & 10 & $45\pm5$ & 110 & 350 & 10 (-20)\tablenotemark{f} & 1\\
NGC~1068 & 29 (11) & 80 & 16 & $160:$ & 30 & -- & -- & 375 & 35 & 2\\
NGC~2992 & 2 (84) & 4 & 60 & $280:$ & 30 & $84\pm10$ & 140 & 2000 & 28 & 3\\
NGC~3783 & 3.3 (34) & 5 & 16 & 140\tablenotemark{g} & -4\tablenotemark{g} & $120\pm7$ & 190 & 155 & -10 & 2\\
NGC~4151 & 9 (15) & 13 & 8.4 & $80:$ & 50 & $180\pm15$ & 300 & 190 & 55 & 4\\
NGC~6814 & 1.3 (36) & 2 & 20 & 75 & -30 & $80\pm10$ & 140 & 55 & -40 & 4\\
NGC~7469 & 2.8 (88) & 4 & 46 & 130 & 10 & 90$\pm9$ & 150 & 180\tablenotemark{h} & 10\tablenotemark{h} & 5\\
\hline
\hline
\end{tabular}
}
\tablenotetext{a}{Nuclear Br$\gamma$ flux integrated in a circular aperture of $2\times$FWHM (two times the resolution) and centered at the nucleus. The numbers in parentheses indicate the diameter of the aperture in pc.}
\tablenotetext{b}{Total Br$\gamma$ flux integrated in a rectangular aperture covering the observed emission (inside $2\times$R$_{\mathrm{Br\gamma}}$).}
\tablenotetext{c}{Photometric semimajor axis of Br$\gamma$ emission measured using as reference the contours corresponding to $5\%$ of the peak of Br$\gamma$ flux. Lower limits are indicated by a colon.}
\tablenotetext{d}{Photometric semimajor axis of [O~{\sc iii}] emission}
\tablenotetext{e}{References for R$_{\mathrm{[O~{iii}]}}$ and PA$_{\mathrm{[O~{iii}]}}$: (1) Marconi et al. (1994), (2) \citet{schmitt03b}, (3) Garc\'ia-Lorenzo et al. (2001), (4) \citet{schmitt96}, (5) This work.}
\tablenotetext{f}{The PA in this galaxy changes from $\sim10\degr$ in the central 50 pc to $\sim110\degr$ at larger scales.}
\tablenotetext{g}{Measured in the $640\times640$ pc SINFONI datacube (see Fig. 27 of D07.)}
\tablenotetext{h}{Size and PA of the compact bright core. The radius of the starformation ring is $\sim1300$ pc. 
}
\tablecomments{Uncertainities in the flux measurements are approximately $5\%$}
\end{center}
\caption{Flux, size and velocity dispersion of the NLR\label{table2}}
\end{table}
\end{landscape}

\clearpage

\begin{table}
\begin{center}
{\scriptsize
\begin{tabular}{lccccccc}
\hline
\hline \noalign{\smallskip}
Galaxy & F([Si~{\sc vi}])$_{\mathrm{nuc}}$\tablenotemark{a} & 
F([Si~{\sc vi}])$_{\mathrm{int}}$\tablenotemark{b} & FWHM$_{\mathrm{[Si~{vi}]}}$ &  
R$_{\mathrm{[Si~{vi}]}}$\tablenotemark{c} & PA$_{\mathrm{[Si~{vi}]}}$ & 
$\sigma_{\mathrm{mean}}$ & $\sigma_{\mathrm{max}}$ \\
 & ($10^{-18}$ W m$^{-2}$) & ($10^{-18}$ W m$^{-2}$) & (pc) & (pc) & 
(\degr) & (km s$^{-1}$) &  (km s$^{-1}$) \\
\hline \noalign{\smallskip}
Circinus & 33 (8.8) & 46 & 7.6 & $8:$ & 0 & 70$\pm7$ & 100 \\
NGC~1068 & 180 (11) & 390 & 14 & 150 & 30 & -- & -- \\
NGC~2992 & 7 (84) & 10 & 60 & 100 & 27 & 120$\pm10$ & 220 \\
NGC~3783 & 14 (34) & 22 & 16 & 140\tablenotemark{d} & 0\tablenotemark{d} & 160$\pm13$ & 250 \\
NGC~4151 & 13 (15) & 19 & 8 & 80 & 50 & 150$\pm15$ & 320 \\
NGC~6814 & 2.2 (36) & 4 & 19 & 80 & -21 & 140$\pm14$ & 300 \\
NGC~7469 & 6 (88) & 10 & 50 & 90 & 90 & 150$\pm15$ & 300 \\
\hline
\hline
\end{tabular}
}
\tablenotetext{a}{Nuclear [Si~{\sc vi}] flux integrated in a circular aperture of $2\times$FWHM (two times the resolution) and centered at the nucleus. The numbers in parentheses indicate the diameter of the aperture in pc.}
\tablenotetext{b}{Total [Si~{\sc vi}] flux integrated in a rectangular aperture covering the observed emission (inside $2\times$R$_{\mathrm{[Si~{vi}]}}$).}
\tablenotetext{c}{Photometric semimajor axis of [Si~{\sc vi}] emission measured using as reference the contours corresponding to $5\%$ of the peak of [Si~{\sc vi}] flux. Lower limits are indicated by a colon.}
\tablenotetext{d}{Measured in the $640\times640$ pc SINFONI datacube (see Fig. 27 of D07.)}
\tablecomments{Uncertainities in the flux measurements are approximately $5\%$}
\end{center}
\caption{Flux, size and velocity dispersion of the CLR\label{table3}}
\end{table}

\clearpage

\begin{table}
\begin{center}
\begin{tabular}{lcccccccc}
\tableline
\tableline 
Galaxy & PA$_{\mathrm{Br\gamma}}$ & $i_{\mathrm{Br\gamma}}$ & 
$\mu_{\mathrm{Br\gamma}}$\tablenotemark{a} & 
PA$_{\mathrm{[Si~{vi}]}}$ & $i_{\mathrm{[Si~{vi}]}}$ & 
$\mu_{\mathrm{[Si~{vi}]}}$\tablenotemark{a} & 
PA$_{\mathrm{H_2}}$\tablenotemark{b} & $i_{\mathrm{H_2}}$\tablenotemark{b} \\
 & ($\degr$) & ($\degr$) & (km s$^{-1}$) & ($\degr$) & ($\degr$) & (km s$^{-1}$) & ($\degr$) & ($\degr$) \\
\tableline
Circinus & 18 & 65 & 12 & 0\tablenotemark{c} & 57\tablenotemark{c} & 16\tablenotemark{c} & 30 & 55\\
NGC~1068 & -- & -- & -- & -- & -- & -- & 85 & 40\\
NGC~2992 & 38\tablenotemark{d} & 55\tablenotemark{d} & 14 & 0 & 45 & 16 & 34 & 60 \\
NGC~3783 & 180 & 50 & $>20$ & 177 & 60 & $>20$ & -15 & 35\\
NGC~4151 & -125 & 45 & $>20$ & -130 & 45 & $>20$ & -170 & 45\\
NGC~6814 & 147 & 50 & $>20$ & 144 & 53 & $>20$ & 140 & 55\\
NGC~7469 & 143 & 45 & 9 & -89 & 40 & $>20$ & 140 & 50\\
\tableline
\tableline
\end{tabular}
\tablenotetext{a}{Mean value of the absolute values in the velocity residuals map. $\mu>20$ km s$^{-1}$ indicates a bad fit.}
\tablenotetext{b}{Values obtained from H09, except for NGC~2992 \citep{friedrich10}. These values are found to be consistent with those of the large scale galactic disk. This was verified by H09 and \citet{friedrich10}.}
\tablenotetext{c}{Values affected by artifacts in the SINFONI detector (see Figure~\ref{fig3}).}
\tablenotetext{d}{Results obtained keeping PA and $i$ constant during the fit, although the data indicate a change in the PA from $\sim0\degr$ in the central $r<0.7\arcsec$ to $\sim40\degr$ in the outer region \citep{friedrich10}.}
\end{center}
\caption{Best-fit kinematic parameters of a rotating disk\label{table4}}
\tablecomments{The typical estimated error of the kinematic parameters is $\pm 12 \degr$}
\end{table}

\clearpage

\begin{table}
\begin{center}
\begin{tabular}{lccccc}
\tableline\tableline
Parameter & NGC~1068\tablenotemark{a} & NGC~3783 & NGC~4151\tablenotemark{b} & NGC~6814\tablenotemark{b} & NGC~7469 \\
\tableline
$z_{\mathrm{max}}$ (pc)& 280$\pm15$ & 400$\pm18$ & $>230$ & $>330$ & 380$\pm25$ \\ 
$\theta_{\mathrm{in}}$ ($\degr$) & 14$\pm7$ & 27$\pm6$ & $15\pm6$ & $25\pm7$ & 28$\pm9$ \\
$\theta_{\mathrm{out}}$ ($\degr$) & 27$^{+5}_{-7}$ & $34^{+8}_{-5}$ & $30\pm7$ & $47^{+8}_{-7}$ & $45\pm6$ \\
$i_{\mathrm{cone}}$ ($\degr$) & $9\pm4$ & $-30\pm8$ & $-45\pm8$ & $-34^{+6}_{-8}$ & $-33^{+7}_{-9}$ \\
PA$_{\mathrm{cone}}$ ($\degr$) & 28$^{+4}_{-6}$ & $-177^{+6}_{-3}$ & $-145\pm5$ & $-146\pm8$ & $-88\pm6$ \\
$v_{\mathrm{max}}$\tablenotemark{c} (km s$^{-1}$) & 1900$\pm 80$ & 220$\pm20$ & $450\pm30$ & $120\pm15$ & $130\pm14$ \\
$r_t$\tablenotemark{d} (pc) & $130\pm15$ & 150$\pm18$ & $230\pm14$ & $330\pm20$ & $220\pm25$ \\
$v_c$\tablenotemark{e} (km s$^{-1}$) & -- & $40$ & $80$ & $60$ & -- \\
$i_{\mathrm{disk}}$ ($\degr$) & -- & 38$\pm12$ & 45$^{+7}_{-10}$ & 49$^{+6}_{-11}$ & -- \\
PA$_{\mathrm{disk}}$ ($\degr$) & -- & $-10\pm7$ & $-175\pm5$ & $145\pm8$ & -- \\
$\alpha$ \tablenotemark{f} ($\degr$) & 54 & 26 & 15 & 9 & 12 \\
$\beta$ \tablenotemark{g} ($\degr$) & 48 & 14 & 50 & 15 & 49 \\
$\big(\chi^2_{\mathrm{rot}}\big)_{\mathrm{min}}$\tablenotemark{h} & -- & 778 & 908 & 321 & 808 \\
$\big(\chi^2_{\mathrm{out}}\big)_{\mathrm{min}}$\tablenotemark{i} & -- & 707 & 819 & 400 & 680 \\
$\big(\chi^2_{\mathrm{rot+out}}\big)_{\mathrm{min}}$\tablenotemark{j} & -- & 616 & 756 & 271 & 727 \\
$N$\tablenotemark{k} & -- & 450 & 525 & 180 & 400 \\
\tableline
\tableline
\end{tabular}
\tablenotetext{a}{In this galaxy the fitting procedure incorporates only radial outflow and is described in a separate publication (M\"uller-S\'anchez et al. in prep.).}
\tablenotetext{b}{No deceleration observed.}
\tablenotetext{c}{Full module of the
outflow velocity vector, i.e. relative to the nucleus not to the observer.}
\tablenotetext{d}{Radial distance from the AGN to the location where $v_{\mathrm{max}}$ is observed.}
\tablenotetext{e}{Rotational velocity measured at $r=50$ pc.}
\tablenotetext{f}{Angle between the outer edge of the bicone and the LOS.}
\tablenotetext{g}{Relative tilt of the bicone axis to the disk. In NGC~1068 and NGC~7469 this angle is estimated using PA$_{\mathrm{H_2}}$ and $i_{\mathrm{H_2}}$ (see Table~\ref{table4}).}
\tablenotetext{h}{Minimum $\chi^2$ values for models incorporating only rotation.}
\tablenotetext{i}{Minimum $\chi^2$ values for models incorporating only biconical outflow.}
\tablenotetext{j}{Minimum $\chi^2$ values for models incorporating rotation and biconical outflow.}
\tablenotetext{k}{Number of data points in the velocity fields, equivalent to the number of degrees of freedom.}
\end{center}
\caption{Results of kinematic modeling of the CLR in five Seyfert galaxies. \label{table6}}
\tablecomments{The quoted parameters correspond to the model presenting the smallest $\big(\chi^2\big)_{\mathrm{min}}$ and the uncertainties refer to $3\sigma$ confidence level. The derived values are also the best-fit parameters for the NLR, except in NGC~7469.}
\end{table}

\clearpage

\begin{table}
\begin{center}
\begin{tabular}{lc}
\tableline\tableline
Parameter & NGC~2992\tablenotemark{a} \\
\tableline
$z_{\mathrm{max}}$ (pc) & $>900$ \\ 
$\theta_{\mathrm{in}}$ ($\degr$) & 22$^{+11}_{-7}$ \\
$\theta_{\mathrm{out}}$ ($\degr$) & 57$^{+5}_{-9}$ \\
$i_{\mathrm{cone}}$ ($\degr$) & 26$^{+8}_{-6}$ \\
PA$_{\mathrm{cone}}$ ($\degr$) & $-25\pm10$ \\
$v_{\mathrm{max}}$\tablenotemark{b} (km s$^{-1}$) & 200$\pm20$ \\
$r_t$\tablenotemark{c} (pc) & 900$\pm70$ \\
$v_c$\tablenotemark{d} (km s$^{-1}$) & 60$\pm20$ \\
$i_{\mathrm{disk}}$ ($\degr$) & 53$^{+10}_{-9}$ \\
PA$_{\mathrm{disk}}$ ($\degr$) & 27$\pm9$ \\
$\alpha$ \tablenotemark{e} ($\degr$) & 7 \\
$\beta$ \tablenotemark{f} ($\degr$) & 55 \\
$\big(\chi^2_{\mathrm{rot}}\big)_{\mathrm{min}}$\tablenotemark{g} & 423 \\
$\big(\chi^2_{\mathrm{out}}\big)_{\mathrm{min}}$\tablenotemark{h} & 537 \\
$\big(\chi^2_{\mathrm{rot+out}}\big)_{\mathrm{min}}$\tablenotemark{i} & 390 \\
$N$\tablenotemark{j} & 330 \\
\tableline
\tableline
\end{tabular}
\tablenotetext{a}{No deceleration observed.}
\tablenotetext{b}{Full module of the
velocity vector, i.e. relative to the nucleus not to the observer.}
\tablenotetext{c}{Radial distance from the AGN to the location where $v_{\mathrm{max}}$ is observed.}
\tablenotetext{d}{Rotational velocity measured at $r=50$ pc.}
\tablenotetext{e}{Angle between the outer edge of the bicone and the LOS.}
\tablenotetext{f}{Relative tilt of the bicone axis to the disk.}
\tablenotetext{g}{Minimum $\chi^2$ value for models incorporating only rotation.}
\tablenotetext{h}{Minimum $\chi^2$ value for models incorporating only biconical outflow.}
\tablenotetext{i}{Minimum $\chi^2$ value for models incorporating rotation and biconical outflow.}
\tablenotetext{j}{Number of data points in the velocity field, equivalent to the number of degrees of freedom.}
\end{center}
\caption{Kinematic model of the CLR in NGC 2992\label{table7}}
\tablecomments{The quoted parameters correspond to the model presenting the smallest $\big(\chi^2\big)_{\mathrm{min}}$ and the uncertainties refer to $3\sigma$ confidence level. The derived values are also the best-fit parameters for the NLR. }
\end{table}

\clearpage

\begin{table}
\begin{center}
\begin{tabular}{lcccccc}
\tableline\tableline
Galaxy & \multicolumn{3}{c}{NLR} & \multicolumn{3}{c}{CLR} \\
 & Rotation\tablenotemark{a} & Outflow\tablenotemark{a} & Best Fit\tablenotemark{b} & Rotation\tablenotemark{a} & Outflow\tablenotemark{a} & Best Fit\tablenotemark{b} \\
\tableline
Circinus & D$^*$ & NP$^*$ & Table~\ref{table4} & U & U & -- \\
NGC~1068 & NP$^*$ & D$^*$ & Table~\ref{table6} & NP & D & Table~\ref{table6} \\
NGC~2992 & D$^*$ & P$^*$ & Table~\ref{table7}  & D & P & Table~\ref{table7} \\
NGC~3783 & P & D & Table~\ref{table6}  & P & D & Table~\ref{table6} \\
NGC~4151 & P$^*$ & D$^*$ & Table~\ref{table6}  & P & D & Table~\ref{table6} \\
NGC~6814 & D & P & Table~\ref{table6}  & D & P & Table~\ref{table6} \\
NGC~7469 & D & NP & Table~\ref{table4} & NP & D & Table~\ref{table6} \\
\tableline
\tableline
\end{tabular}
\tablenotetext{a}{Column indicating if the corresponding kinematical component is Dominant (D), Present (P) or Not Present (NP) in the velocity field.``U'' stands for uncertain. The asterisk indicates that in these galaxies the size of the NLR, as traced by [O~{\sc iii}] emission, is larger than our FOVs. However, the model parameters of NGC~1068, NGC~2992 and NGC~4151 are consistent with those derived in previous kinematic studies of the NLR in these galaxies with larger FOVs. All CLRs are fully covered by the SINFONI and OSIRIS FOVs, except in the Circinus galaxy. See text for details.}
\tablenotetext{b}{Column indicating the table in the paper in which the best fit kinematic parameter values are presented.}
\end{center}
\caption{Summary of results of the kinematic analysis\label{table7a}}
\end{table}

\clearpage

\begin{table}
\begin{center}
\begin{tabular}{lcccccccc}
\hline
\hline
Galaxy & $A$\tablenotemark{a} & $\dot{M}_{\mathrm{out}}$ & $\dot{M}_{\mathrm{acc}}$ & $\dot{E}_{\mathrm{out}}$ & $L_{\mathrm{bol}}$ & $\dot{E}_{\mathrm{out}}/L_{\mathrm{bol}}$ & Ref\tablenotemark{b} \\
 & $10^4$ pc$^{2}$ & M$_\odot$ yr$^{-1}$ &
M$_\odot$ yr$^{-1}$ & 10$^{42}$ erg s$^{-1}$ & 10$^{42}$ erg s$^{-1}$ \\
\tableline
NGC~1068 & 2 & 9 & 0.015 & 5 & 88 & 0.05 & 1 \\
NGC~2992 & 200 & 120 & 0.015 & 2.5  & 85 & 0.029 & 2 \\
NGC~3783 & 4 & 2.5 & 0.03 & 0.07 & 180 & 0.0004 & 1 \\
NGC~4151 & 8 & 9 & 0.01 & 0.65 & 55 & 0.012 & 2 \\
NGC~6814 & 25 & 7.5 & 0.014 & 0.08 & 80 & 0.001 & 2 \\
NGC~7469 & 11 & 4 & 0.04 & 0.06 & 250 & 0.0002 & 1 \\
\hline
\hline
\end{tabular}
\tablenotetext{a}{ Lateral surface area of the biconical outflow obtained directly from the kinematic models, using $r_t$ and $\theta_{\mathrm{out}}$.}
\tablenotetext{b}{References for $L_{\mathrm{bol}}$: (1) Prieto et al. (2010), (2) \citet{woo02}.}
\end{center}
\caption{Mass outflow rates ($\dot{M}_{\mathrm{out}}$) and their corresponding kinetic powers ($\dot{E}_{\mathrm{out}}$) compared to mass accretion rates ($\dot{M}_{\mathrm{acc}}$) and accretion powers ($L_{\mathrm{bol}}$) \label{table8}}
\end{table}






\end{document}